\documentclass[final,authoryear,3p,times]{elsarticle}

\usepackage{graphicx}
\usepackage{subfigure}
\usepackage{float}
\usepackage{amsmath}
\usepackage{color}
\usepackage{multirow}

\usepackage{amssymb}

 \usepackage{lineno}

\journal{Ocean Modelling}

\begin{document}
\begin{frontmatter}



\cortext[cor1]{Corresponding author. Tel.:+1(540)231-7570; fax:+1(540)231-4574.}

\title{Approximate deconvolution large eddy simulation of a stratified two-layer quasigeostrophic ocean model}


\author{Omer San\corref{cor1}\fnref{address1}}
\ead{omersan@vt.edu}
\author{Anne E. Staples\fnref{address1}}
\author{Traian Iliescu\fnref{address2}}

\address[address1]{Department of Engineering Science and Mechanics, Virginia Tech, Blacksburg, VA, USA}

\address[address2]{Department of Mathematics, Virginia Tech, Blacksburg, VA, USA}

\begin{abstract}

We present an approximate deconvolution (AD) large eddy simulation (LES) model for the two-layer quasigeostrophic equations. We applied the AD-LES model to mid-latitude two-layer square oceanic basins, which are standard prototypes of more realistic stratified ocean dynamics models. Two spatial filters were investigated in the AD-LES model: a tridiagonal filter and an elliptic differential filter. A sensitivity analysis of the AD-LES results with respect to changes in modeling parameters was performed. The results demonstrate that the AD-LES model used in conjunction with the tridiagonal or differential filters provides additional dissipation to the system, allowing the use of a smaller eddy viscosity coefficient. Changing the spatial filter makes a significant difference in characterizing the effective dissipation in the model. It was found that the tridiagonal filter introduces the least amount of numerical dissipation into the AD-LES model. The differential filter, however, added a significant amount of numerical dissipation to the AD-LES model for large values of the filter width. All AD-LES models reproduced the DNS results at a fraction of the cost within a reasonable level of accuracy.

\end{abstract}


\end{frontmatter}

\linenumbers


\section{Introduction}
\label{sec:intro}

The investigation of characteristics of forced-dissipative general circulation models is of primary importance in developing our understanding of the large-scale nonlinear motions of geophysical flows. As one of the main circulation sources, winds drive the general circulation associated with the subtropical and subpolar gyres, which can be identified with the strong, persistent, sub-tropical and sub-polar western boundary currents in the North Atlantic Ocean (the Gulf Stream and the Labrador Current) and North Pacific Ocean (the Kuroshio and the Oyashio Currents) and sub-tropical counterparts in the southern hemisphere \citep{stommel1972gulf,mcwilliams2006fundamentals}. One of the major similarities between the various ocean basins is the asymmetry of the gyres: strong western boundary currents and weaker flow in the interior; weak and shallow eastern boundary currents. The most obvious motivation for being interested in forced-dissipative wind-driven ocean circulation is the connection between ocean currents and climate dynamics \citep{ghil2008climate}.

The wind-driven circulation in an enclosed, midlatitude rectangular or square basin is a classical problem, studied extensively by modelers \citep{allen1980models,holland1980example,griffa1989wind, vallis2006atmospheric,miller2007numerical}. Various models are derived from the full-fledged equations of geophysical flows, Boussinesq equations (BEs) or the primitive equations (PEs), to guide the theoretical studies on boundary currents, alternating zonal flows, or jet formations, as well as to identify some key issues related to the relative insensitivity of the model dynamics to the changes of parameters that is closely linked to a dynamical system point of view \citep{speich1995successive,meacham2000low,chang2001transition,nauw2004regimes,dijkstra2005nonlinear,dijkstra2005low}. The quasigeostropic (QG) model is a simplification of the primitive equation model that retains many of the essential features of geophysical fluid flows. Details of the mathematical and physical approximations may be found in standard textbooks on geophysical fluid dynamics, such as \cite{pedlosky1987geophysical}, \cite{vallis2006atmospheric}, and \cite{mcwilliams2006fundamentals}. The main assumptions that go into the QG models are: the hydrostatic balance, the $\beta$-plane approximation, the geostrophic balance, and the eddy viscosity parameterization.

The one-layer QG model, sometimes called the barotropic vorticity equation (BVE), represents one of the most commonly used mathematical models for these types of geostrophic flows with various dissipative and forcing terms \citep{majda2006non, vallis2006atmospheric,nadiga2001dispersive}. In reality, the ocean is a stratified fluid on a rotating Earth driven from its upper surface by patterns of momentum and buoyancy fluxes \citep{marshall1997hydrostatic}. While the barotropic model is not stratified, it exhibits many of the features that are observed in the stratified case.
To explore some of the effects of the stratification, the one-layer barotropic equation can be extended to the 1.5-layer model, also called the reduced gravity QG model \citep{özgökmen2001connection}. There are two layers in this model, but the second layer is infinitely deep and at rest (passive), and the dynamics are effectively barotropic. The two-layer model takes the next step in increasing the complexity of stratification by adding a second dynamically active layer \citep{holland1978role,özgökmen1998emergence,berloff1999large,dibattista2001equilibrium,berloff2009mechanism}. The dynamics in this model include the first baroclinic modes. The complexity of the models could be increased by adding more active layers, resulting in the N-layer models \citep{siegel2001eddies}, which, in turn, yield the three dimensional primitive equations when N goes to infinity \citep{mcwilliams2006fundamentals}.
In this study, we use the {\it two-layer QG (QG2)} model.

Geophysical turbulence is strongly affected by the planetary vorticity, the variation of the Coriolis parameter with latitude, the so-called $\beta$ effect \citep{maltrud1991energy,smith2002turbulent,chen2003physical}. The inverse cascade typically occurring in pure two-dimensional turbulence, in this case preferentially transfers small-scale energy towards zonal modes; the resulting flow is then anisotropic and characterized by a strong interaction between waves and turbulence, and is known as the arrest of the inverse energy cascade \citep{rhines1975waves,sukoriansky2007arrest,espa2008quasi,san2012stationary}. \cite{rhines1975waves} explained the emergence of flow anisotropy and the organization of a banded pattern of alternating zonal currents, or jets, due to Rossby wave dynamics in terms of a competition between nonlinear and $\beta$ terms in the barotropic vorticity equation. Under the effects of planetary rotation, Rossby waves dominate turbulent motions prohibiting the triad interactions, and arrest the inverse energy cascade when the scale of motions becomes larger than a critical value, later known as the Rhines scale \citep{tanaka2010alternating}.

Along with the Rhines scale which is a measure of the strength of nonlinear interactions, another important scale for determining the dynamics of the large scale motions in the ocean is the Munk scale \citep{munk1950wind}, which corresponds to the dissipative behavior of the system and can be linked to the Reynolds number. Although the water molecular viscosity is around $10^{-6} \ m^2s^{-1}$, the one- and two-layer QG models use viscosities on the order of $10^{2} \ m^2s^{-1}$. This is called \emph{eddy viscosity (EV)} parameterization, and is used because the horizontal scale of the ocean basin is much larger than the effective scale for molecular diffusion. An impractically fine resolution would be necessary if the ocean models were to resolve the full spectra of turbulence down to the Kolmogorov scale. Thus, the viscosity coefficients employed in the QG models typically remain much greater than the molecular viscosity \citep{campin2011super}. The eddy viscosities generally used in the oceanic models are summarized in Table~\ref{tab:vis}. The eddy viscosity parameterization used in the QG models plays a crucial role in the dynamics of the problem. Indeed, \cite{berloff1999large} studied the wind-driven circulation in a three-layer QG model for varying values of the eddy viscosity coefficient in a square oceanic basin. For $\nu=1200 \ m^2s^{-1}$ an asymmetric steady state was found. When the eddy viscosity coefficient was decreased, the flow first displayed a variability characterized by the presence of interior Rossby waves. At $\nu = 1000 \ m^2s^{-1}$, the flow regime showed a quasi-periodic variability. At a smaller eddy viscosity coefficient, starting from $\nu = 800 \ m^2s^{-1}$, the flow regime was chaotic and showed a persistent eastward jet penetration by fluctuating between two preferred states, one of which corresponds to a low energy state and a long eastward jet, and the other to a high energy state and a short jet.
The study of \cite{berloff1999large} clearly shows that different EV coefficients  can result in different dynamics of the QG models.
Thus, a natural question is ``What EV coefficient should be used in the QG models?"
The EV coefficients summarized in Table~\ref{tab:vis} seem to convey, at first glance, a confusing message: they vary by as much as an order of magnitude.
At a closer look, however, Table~\ref{tab:vis} clarifies this issue:
With the ever increasing computational power, the mesh size used in numerical simulations with the QG models constantly decreases and allows the use of smaller EV coefficients.
The development of a rigorous, mathematical understanding and subsequent modeling strategy for the eddy viscosity coefficients (see Table~\ref{tab:vis}) is the ``elephant in the room," one of the major unsolved problems in ocean modeling \citep{visbeck1997specification,campin2011super,majda2006non,cushman2009introduction,vallis2006atmospheric}.
Although addressing this grand challenge is beyond the scope of this report, we do address the intimate relationship between the EV coefficients and the numerical resolution employed by the QG models.

\begin{table}[H]
\centering
\caption{The eddy viscosity coefficients used in QG models}
\label{tab:vis}       
\begin{tabular}{lll}
\hline\noalign{\smallskip}
Study & Range of $\nu \ (m^2s^{-1})$  & Resolution\\
\noalign{\smallskip}\hline\noalign{\smallskip}
\cite{bryan1963numerical} & 500 - 10000 & 40$\times$80 \\
\cite{gates1968numerical} & 6000 - 10000 & 74$\times$50 \\
\cite{holland1975generation} & 330 & 50$\times$50 \\
\cite{jiang1995multiple} & 300 & 50$\times$100 \\
\cite{özgökmen1998emergence} & 50 & 151$\times$151 \\
\cite{berloff1999large} & 400 - 1600 &  256$\times$256 \\
\cite{sura2001regime} & 200 & 120$\times$120 \\
\cite{berloff2009mechanism} & 100 & 512$\times$256 \\
\cite{tanaka2010alternating} & 100 & 500$\times$500 \\
\noalign{\smallskip}\hline
\end{tabular}
\end{table}

To capture the under-resolved flow, i.e., the flow in the regions where the grid size becomes greater than the specified Munk scale, {\it large eddy simulation (LES)} appears as a natural choice.
Most of the LES models have been developed for three-dimensional turbulent flows, such as those encountered in engineering applications \citep{sagaut2006large,berselli2006mathematics}.
These LES models fundamentally rely on the concept of the forward energy cascade and so their extension to geophysical flows is beset with difficulties.
The effective viscosity values in oceanic models are much greater than the molecular viscosity of seawater, hence a uniform eddy viscosity coefficient is generally used to parameterize the unresolved, subfilter-scale effects in most oceanic models \citep{mcwilliams2006fundamentals,vallis2006atmospheric}.
LES models specifically developed for two-dimensional turbulent flows, such as those in the ocean and atmosphere, are relatively scarce \citep{fox2008can,awad2009large,ozgokmen2009large,chen2011scale}, at least when compared to the plethora of LES models developed for three-dimensional turbulent flows.
\cite{holm2003modeling} combined the uniform eddy viscosity parameterization with the alpha regularization LES approach to capture the under-resolved flow where the grid length becomes greater than the specified Munk scale of the problem. In that work, the structural alpha parameterization was tested on the barotropic vorticity equation (BVE) in an ocean basin with double-gyre wind forcing, which displays a four-gyre mean ocean circulation pattern. It was found that the alpha models provide a promising approach to LES closure modeling of the
barotropic ocean circulation by predicting the correct four-gyre circulation structure for under-resolved flows.

\cite{san2011} put forth a new LES closure modeling strategy for two-dimensional turbulent geophysical flows. The new closure modeling approach utilizes {\em approximate deconvolution (AD)}, which is particularly appealing for geophysical flows because of no additional phenomenological approximations to the BVE.
Similar to the method suggested by \cite{holm2003modeling}, this framework also uses a Laplacian operator with a constant eddy viscosity coefficient to account for the dissipation mechanism. For a given system with eddy viscosity dissipation, the subfilter-scale contribution, however, is modeled by a non eddy viscosity AD closure approach.
The AD approach can achieve high accuracy by employing repeated filtering, which is computationally efficient and easy to implement. The AD method has been used successfully in LES of three-dimensional turbulent engineering flows \citep{stolz1999approximate,stolz2001approximate,stolz2001approximatec,stolz2004approximatei,domaradzki2002direct}
and even of small scale geophysical flows, such as the atmospheric boundary layer \citep{chow2005explicit,chow2009evaluation,duan2010bridging,zhou2011large}.
The AD methodology was also used in LES of large scale geophysical flows, such as the barotropic ocean circulation flow.
To assess the new AD closure modeling approach, \cite{san2011} tested it on the same two-dimensional barotropic flow problem as that employed in \cite{nadiga2001dispersive} and in \cite{holm2003modeling}.
It was shown that the new LES-AD model provides an accurate approximation for under-resolved subfilter-scale effects.




The main goal of this report is to extend the LES-AD approach used for the one-layer QG model \citep{san2011} to the two-layer QG model. A quantitative analysis of the effects of using the AD-LES model on QG2 models was performed in conjunction with the tridiagonal and differential filters. We investigated whether the combination of LES-AD modeling and a particular spatial filter can, in fact, account for some of the eddy viscosity parameterization used in practical QG numerical simulations. Our numerical experiments show that the AD-LES model does add numerical dissipation, but the exact amount and form still need to be determined. A sensitivity analysis was performed to find out how much of the dissipation the AD-LES model, equipped with various spatial filters, can account for. We demonstrated that the amount of the dissipation added to the system depends on the free modeling parameters. We emphasize that this issue is common to LES modeling in general. Indeed, not only is it hard to find the ``best" LES model, i.e., the model that produces the most accurate results at the lowest computational cost, but once this model is found, it is often hard to decide whether the success of the model is due to the actual closure model or the numerical discretization used \citep{berselli2006mathematics,sagaut2006large}. In an actual LES of turbulent flow there are several ingredients -- some are used at the continuum level (e.g., the closure model with its various parameters), and some are used at the discrete level (e.g., the temporal and spatial discretization or the linear solver). Often, it is hard to disentangle the modeling effects from the numerical discretization effects.
Our QG setting is no different in this respect. We plan to investigate this complex relationship in a future study, by performing extensive numerical experiments in simplified settings and by developing mathematical support for these numerical results.

The rest of the paper is organized as follows: Section \ref{sec:model} presents the two-layer QG equations for large-scale geophysical flows. The proposed AD methodology, which yields the mathematical model used in this report, is presented in Section \ref{sec:method}. The numerical methods used in our simulations are briefly discussed in Section \ref{sec:numerical_methods}. The results for the new AD model are presented in Section \ref{s_results}. Finally, the conclusions are summarized in Section \ref{sec:cons}.

\section{Governing equations}
\label{sec:model}
\subsection{The two-layer quasigeostrophic equations}
The two-layer quasigeostrophic model used in this study is one of the simplified forced-dissipative oceanic models that considers baroclinic effects. The stratified ocean is partitioned into two isopycnal layers, each of constant depth, density and temperature. The governing quasigeostrophic potential vorticity equations for the two dynamically active layers are \citep{pedlosky1987geophysical,salmon1998lectures,mcwilliams2006fundamentals}
\begin{eqnarray}
\frac{\partial q_1}{\partial t} + J(\psi_1,q_1) &=& D_1 + F_1, \\
\frac{\partial q_2}{\partial t} + J(\psi_2,q_2) &=& D_2 + F_2,
\label{eq:ge}
\end{eqnarray}
where the layer index starts from top, $q_i$ represents potential vorticities, and $\psi_i$ denotes for streamfunctions.
The Jacobian operator is defined as $J(a,b)=\frac{\partial a}{\partial x}\frac{\partial b}{\partial y} - \frac{\partial a}{\partial y}\frac{\partial b}{\partial x}$. The dissipation and forcing (Ekman pumping) terms are represented by $D_i$, and $F_i$, respectively. The potential vorticities for each layer are related to the velocity streamfunctions through the following elliptic coupled system of equations:
\begin{eqnarray}
q_1 &=& \nabla^2\psi_1 + \beta y + \frac{f_{0}^{2}}{g'H_1}(\psi_2-\psi_1), \\
q_2 &=& \nabla^2\psi_2 + \beta y + \frac{f_{0}^{2}}{g'H_2}(\psi_1-\psi_2).
\label{eq:pv}
\end{eqnarray}
The isopycnal flow velocity components can be found from the velocity streamfunctions:
\begin{equation}
u_i = -\frac{\partial \psi_i}{\partial y}; \quad v_i = \frac{\partial \psi_i}{\partial x}.
\label{eq:vel}
\end{equation}
The two symbols $\beta$ and $f_0$ are parts of the linearized $\beta$-plane approximation to the Coriolis parameter $f=f_0 + \beta y$. Here $f_0=2 \,  \Omega \, \mbox{sin}(\phi_0)$ is the local rotation rate at $y=0$, where $\Omega$ is the rotational speed of the earth and $\phi_0$ is the latitude at $y=0$. This is equivalent to approximating the spherical Earth with a tangent plane at $y=0$. Stratification is represented by two stacked isopycnal layers with thicknesses $H_1$ and $H_2$, starting from the top, and $g'=g \frac{\Delta \rho}{\rho_1}$ is reduced gravity associated with the density jump between the two layers in which $\Delta \rho$ is the density difference between the two layers, $\rho_1$ is the reference (upper layer) density, and $g$ is the gravitational acceleration. The inertial radius of deformation between layers, a measure of stratification strength, is defined as the Rossby deformation radius $R_d=\sqrt{\frac{g' H_1H_2}{f_{0}^2 H}}$, where $H=H_1 + H_2$.
In this study, the top and bottom layers of the ocean are forced by an Ekman pumping of the form
\begin{eqnarray}
F_1 &=& \frac{1}{\rho_1 H_1} \hat{k} \cdot \nabla \times \vec{\tau} ,  \\
F_2 &=& -\gamma \nabla^2\psi_2 ,
\label{eq:force}
\end{eqnarray}
where $\vec{\tau}=(\tau^{(x)},\tau^{(y)})$ is the stress vector for surface wind forcing, and $\hat{k}$ is unit vector in vertical direction. In the present model, we use a double-gyre wind forcing only for zonal direction: $\tau^{(x)}=\tau_0 \,  \mbox{cos}\bigg(\frac{2\pi}{L} y\bigg)$, where $L$ is the meridional length of the ocean basin centered at $y=0$, and $\tau_0$ is the maximum amplitude of the wind stress. This form of wind stress represents the meridional profile of easterly trade winds, mid-latitude westerlies, and polar easterlies from South to North. The bottom Ekman layer is parameterized by a linear bottom friction with coefficient $\gamma$.  In the equations above, $\nabla$ and $\nabla^2$ are the gradient and Laplacian operators, respectively.
For the dissipation terms, the following EV parameterizations are used:
%
\begin{eqnarray}
D_1 &=& \nu \nabla^4\psi_1 , \\
D_2 &=& \nu \nabla^4\psi_2 ,
\label{eq:dis}
\end{eqnarray}
where $\nu$ is eddy viscosity coefficient.

\subsection{Governing equations in dimensionless form}
The governing equations can be written in dimensionless form by using the Sverdrup balance to set the velocity scale of the form
\begin{equation}
V = \frac{2 \pi \tau_0}{\rho_1 H_1 \beta L }.
\label{eq:sverdrup}
\end{equation}
The dimensionless variables (denoted by tilde) are defined as
\begin{equation}
\tilde{x} = \frac{x}{L}; \ \tilde{y} = \frac{y}{L}; \ \tilde{t} = \frac{t}{L/V}; \ \tilde{q} = \frac{q}{\beta L}; \ \tilde{\psi} = \frac{\psi}{V L} .
\label{eq:sverdrup}
\end{equation}
Then the two-layer quasigeostrophic equations in dimensionless form become
\begin{eqnarray}
\frac{\partial \tilde{q}_1}{\partial \tilde{t}} + J(\tilde{\psi}_1,\tilde{q}_1) &=&\tilde{D}_1 + \mbox{sin}(2\pi \tilde{y}) , \\
\frac{\partial \tilde{q}_2}{\partial \tilde{t}} + J(\tilde{\psi}_2,\tilde{q}_2) &=&\tilde{D}_2 - \sigma \tilde{\nabla}^2\tilde{\psi}_2 ,
\label{eq:ge-non}
\end{eqnarray}
in which the dissipative terms can be written as
\begin{eqnarray}
\tilde{D}_1 &=& A \tilde{\nabla}^4\tilde{\psi}_1  \\
\tilde{D}_2 &=& A \tilde{\nabla}^4\tilde{\psi}_2.
\label{eq:dis}
\end{eqnarray}
In dimensionless form, the kinematic relationships between potential vorticities and streamfunctions become:
\begin{eqnarray}
\tilde{q}_1 &=& \mbox{Ro} \tilde{\nabla}^2\tilde{\psi}_1 + \tilde{y} + \frac{\mbox{Fr}}{\delta}(\tilde{\psi}_2-\tilde{\psi}_1) , \\
\tilde{q}_2 &=& \mbox{Ro} \tilde{\nabla}^2\tilde{\psi}_2 + \tilde{y} + \frac{\mbox{Fr}}{1-\delta}(\tilde{\psi}_1-\tilde{\psi}_2) .
\label{eq:pv-non}
\end{eqnarray}
For clarity of exposition, in the remainder of the paper we will drop the tilde symbol used for the dimensionless variables.
In the two-layer QG model, $\delta=\frac{H_1}{H}$ is the aspect ratio of vertical layer thicknesses, $\mbox{Ro}$ is the Rossby number, $\mbox{Fr}$ is the Froude number, $A$ is the lateral eddy viscosity coefficient, and $\sigma$ is the Ekman bottom later friction coefficient. The definitions of these dimensionless parameters are:
\begin{equation}
\mbox{Ro}=\frac{V}{\beta L^2}; \  \mbox{Fr}=\frac{f_{0}^{2}V}{g'\beta H}; \ \mbox{Re} =\frac{V L}{\nu}; \ A=\frac{\nu}{\beta L^3}; \ \sigma =\frac{\gamma}{\beta L}.
\label{eq:ndim}
\end{equation}

The following three length scales are useful for setting the problem parameters: (i) the Munk scale, $\delta_M=\bigg(\frac{\nu}{\beta}\bigg)^{1/3}$, for the viscous boundary layer; this is related to the smaller scale dissipation; (ii) the Stommel scale, $\delta_S=\frac{\gamma}{\beta}$, for the bottom boundary layer thickness; this is accounting for larger scale damping; and (iii) the Rhines scale, $\delta_I=\bigg(\frac{V}{\beta}\bigg)^{1/2}$, for the inertial boundary layer; this is measuring the strength of the nonlinearity.

In order to complete the mathematical model, boundary and initial conditions should be prescribed. In many theoretical studies of ocean circulation, the modelers either use free-slip boundary conditions or no-slip boundary conditions. Following \cite{cummins1992inertial,özgökmen1998emergence}, we use free-slip boundary conditions
for the velocity for both isopycnal layers, which translates into homogenous Dirichlet boundary conditions for the vorticity (Laplacian of streamfunction):
$\nabla^2 \psi |_{\Omega} = 0$. The impermeability boundary condition is imposed as $\psi |_{\Omega} = 0$. We start from a rest state ($\psi=0$),  integrate the model until a statistically steady state is obtained, and continue for several decades to compute time-averaged results.

\section{Approximate deconvolution method}
\label{sec:method}
The goal in AD is to use repeated filtering in order to obtain approximations of the unfiltered unresolved flow
variables when approximations of the filtered resolved flow variables are available.
These approximations of the unfiltered flow variables are then used in the subfilter-scale terms
to close the LES system.
To derive the new AD model, we start by denoting by $G$ the spatial filtering operator: $Gu=\bar{u}$, $G\bar{u}=\bar{\bar{u}}$ and so on, where $u$ represents any flow variable (i.e., potential vorticity and the streamfunction in this study).
Since $G=I-(I-G)$, an inverse to $G$ can be written formally as the non-convergent Neumann series:
\begin{equation}
G^{-1} \sim \sum_{i=0}^{\infty}(I-G)^i .
\label{eq:2}
\end{equation}
Truncating the series gives the van Cittert approximate deconvolution operator, $Q_N$.
We truncate the series at $N$ and obtain $Q_N$ as an approximation of $G^{-1}$:
\begin{equation}
Q_N = \sum_{i=1}^{N}(I-G)^{i-1} ,
\label{eq:3}
\end{equation}
where $I$ is the identity operator.
The approximations $Q_N$ are not convergent as $N$ goes to infinity, but rather are asymptotic
as the filter radius, $\Delta$, approaches zero \citep{berselli2006mathematics}.
An approximate deconvolution of any variable $u$ can now be obtained as follows:
\begin{equation}
u^*= Q_N u .
\label{eq:4}
\end{equation}
For higher values of $N$, we get increasingly more accurate approximations of $u$:
\begin{eqnarray}
Q_1 &=& I \\
Q_2 &=& 2I -G \\
Q_3 &=& 3I-3G + G^2 \\
Q_4 &=& 4I-6G + 4G^2 -G^3 \\
Q_5 &=& 5I-10G + 10G^2 - 5G^3 + G^4 \\
\vdots \nonumber
\label{eq:q5}
\end{eqnarray}

Following the same approach as that used in \cite{dunca2006stolz}, one can prove that these
models are highly accurate ($O(\Delta^{2N+2})$ modeling consistency error) and stable.
For example, if we choose $N=5$, we can find an AD approximation of the resolved variable $q$ as
\begin{equation}
q^*=5 q - 10 \bar{q} + 10 \bar{\bar{q}} -5 \bar{\bar{\bar{q}}} + \bar{\bar{\bar{\bar{q}}}}
\label{eq:6}
\end{equation}
and, similarly, an AD approximation of the variable $\psi$ as
\begin{equation}
 \psi^*=5 \psi - 10 \bar{\psi} + 10 \bar{\bar{\psi}} -5 \bar{\bar{\bar{\psi}}} + \bar{\bar{\bar{\bar{\psi}}}} ,
\label{eq:7}
\end{equation}
where $q$ and $\psi$ are the resolved potential vorticity and streamfunction variables. We use a bar to denote the application of one filtering operation. Using \eqref{eq:6} and \eqref{eq:7}, we can now approximate the subfilter-scale contribution by applying a filter to the governing equation. This results in the following model:

\begin{eqnarray}
\frac{\partial q_1}{\partial t} + J(\psi_1,q_1) &=& A \nabla^4\psi_1 + \mbox{sin}(2\pi y) + S^{*}_{1} , \label{eq:adm-g1} \\
\frac{\partial q_2}{\partial t} + J(\psi_2,q_2) &=& A \nabla^4\psi_2 - \sigma \nabla^2\psi_2 + S^{*}_{2} ,
\label{eq:adm-g}
\end{eqnarray}
where $S^{*}_{i}$ is the subfilter-scale term for the $i^{th}$ layer, given by
\begin{equation}
S^{*}_{i} = -\overline{J(\psi_{i}^{*},q_{i}^{*})} + J(\psi_{i},q_{i}) ,
\label{eq:adm-s}
\end{equation}
where asterisk represents the approximated value for the unfiltered (unresolved) quantities. To completely specify the new AD model \eqref{eq:adm-g1}-\eqref{eq:adm-s}, we need to choose a computationally efficient filtering operator. In Section~\ref{s_results}, we will show that the selection of the filtering operator affects the dissipative behavior of the system.

\subsection{Tridiagonal filter}
\label{ss_tf}
Following \cite{stolz1999approximate}, we use the following discrete second-order {\it tridiagonal filter (TF)}:
\begin{equation}
\alpha \bar{f}_{i-1}
+ \bar{f}_{i}
+ \alpha \bar{f}_{i+1}
= \left(\frac{1}{2} + \alpha\right)\left(f_{i}+\frac{ f_{i-1} + f_{i+1}}{2}\right) ,
\label{eq:9}
\end{equation}
where $\bar{f}_{i}$ represents the filtered value of a discrete quantity $f_i$.
Here, the subscript $i$ is the spatial index in the $x$-direction. This results in a tridiagonal system of equations for each fixed value of $y$. After solving Eq.~\eqref{eq:9}, we use the same filter in the $y$-direction (i.e., we replace index $i$ with $j$) for each fixed value of $x$. The resulting tridiagonal system of equations is solved efficiently by using the well-known Thomas algorithm. Since the TF has been constructed in the physical space, a Fourier analysis is applied to study its characteristics in the wavenumber space. This analysis leads to the transfer function, $T(\omega)$, that correlates the Fourier coefficients of the filtered variable to those of the unfiltered variable as follows:
\begin{equation}
\hat{\bar{f}}_k = T(\omega) \hat{f}_k ,
\label{eq:traf}
\end{equation}
where $\hat{\bar{f}}_k$ and $\hat{f}_k$ are the Fourier coefficients of the filtered and unfiltered variables, respectively (i.e., $f_i = \sum \hat{f}_k e^{\imath k x_i}$ and $\bar{f}_i = \sum \hat{\bar{f}}_k e^{\imath k x_i}$, where $x_i=\Delta_x i$ and $\Delta_x$ is the grid spacing in the $x$-direction). Using the relation $\cos(\theta) = (e^{\imath \theta} + e^{-\imath \theta})/2$, the transfer function of the TF given in Eq.~\eqref{eq:9} can be written as
\begin{equation}
T^{(TF)}(\omega) = \left(\frac{1}{2}+\alpha\right) \, \frac{1+\cos(\omega)}{1+2\alpha\cos(\omega)},
\label{eq:tfun-1}
\end{equation}
where $\omega = k \Delta_x$ is the modified wavenumber in the $x$-direction.
The free parameter, $\alpha$, which is in the range $0 \leq |\alpha| \leq 0.5$, determines the filtering properties, with high values of $\alpha$ yielding less dissipative results.
If the transfer function of the filter used in the AD closure is positive, then the existence and uniqueness of strong solutions of the AD model can be proved \citep{stanculescu2008existence}. The transfer function corresponding to the TF becomes positive definite in the interval of $0 \leq |\alpha| \leq 0.5$. More details can be found in \cite{san2011}.

\begin{figure}[!t]
\centering
\includegraphics[width=0.7\textwidth]{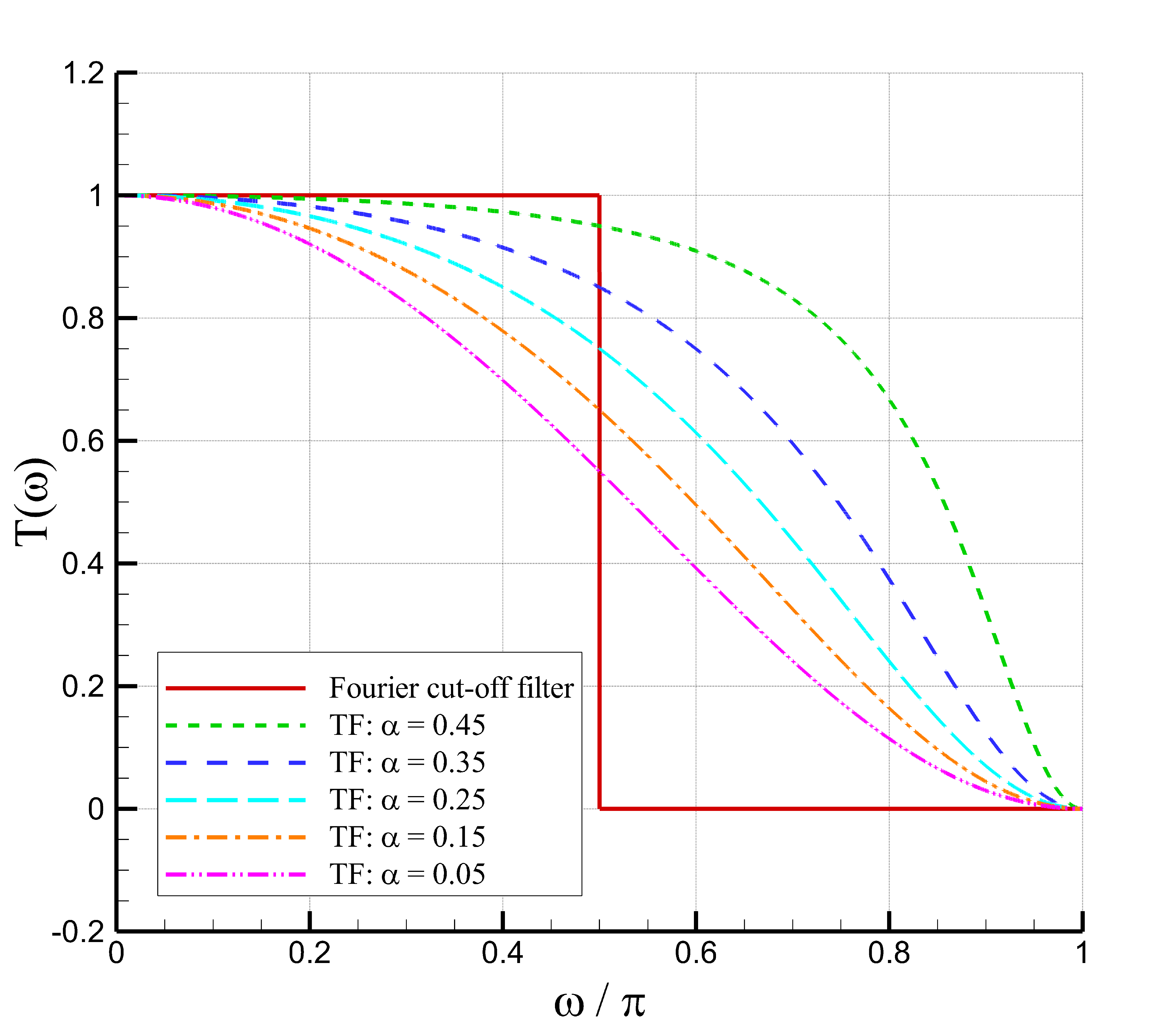}
\caption{
Transfer functions of the TF for different values of the parameter $\alpha$.
The transfer function of the Fourier cut-off filter is also included for comparison purposes.
}
\label{fig:tf}
\end{figure}

To show the characteristics of the TF in Eq.~\eqref{eq:9}, we plot in Fig.~\ref{fig:tf} its transfer function $T^{(TF)}(\omega)$ (which is given by Eq. \eqref{eq:tfun-1}) for different values of the free parameter $\alpha$. The transfer function of the Fourier cut-off filter is also shown for comparison purposes (see \cite{najjar1996study}). It is known that the Fourier cut-off filter removes the small scales with wavenumbers $\omega/\pi > 1/2$, while retaining the larger scales with wavenumbers $\omega/ \pi < 1/2$. It is clear from Fig.~\ref{fig:tf} and Eq. \eqref{eq:tfun-1} that $\alpha$ plays the role of a cut-off wavenumber for the TF: $\alpha = 0.5$ turns off the filter, whereas low $\alpha$ values result in more dissipation (i.e., high attenuation of all the wavenumber components).

\subsection{Elliptic differential filter}
\label{ss_df}
The second filter used in our numerical investigation is the elliptic {\it differential filter (DF)} \citep{germano1986differential,sagaut2006large,berselli2006mathematics}:
\begin{eqnarray}
\bar{f} - \lambda^2 \, \left(\frac{\partial^2 \bar{f}}{\partial x^2} + \frac{\partial^2 \bar{f}}{\partial y^2} \right) &=& f
\qquad \text{ in } \Omega ,
\label{eq:10} \\
\bar{f}
&=& f
\qquad \text{ on } \partial \Omega ,
\label{eq:10b}
\end{eqnarray}
where $\Omega$ is the computational domain and $\lambda$ is the Helmholtz length, which determines the effective width of the filter. The DF is also called Helmholtz filter. The filtered value $\bar{f}$ is obtained by applying the inverse Helmholtz operator to the unfiltered flow variable $f$. This inversion is done efficiently by using the fast Fourier transform (FFT) techniques \citep{press1992numerical}. Specifically, we use the fast sine transform to solve the discrete version of Eq.\eqref{eq:10}, which can be written as follows:
\begin{equation}
\bar{f}_{i,j} - \lambda^2 \left( \frac{\bar{f}_{i+1,j}-2\bar{f}_{i,j}+\bar{f}_{i-1,j}}{\Delta_{x}^{2}} + \frac{\bar{f}_{i,j+1}-2\bar{f}_{i,j}+\bar{f}_{i,j-1}}{\Delta_{y}^{2}}  \right) = f_{i,j}.
\label{eq:dishelm}
\end{equation}

\begin{figure}[!t]
\centering
\includegraphics[width=0.7\textwidth]{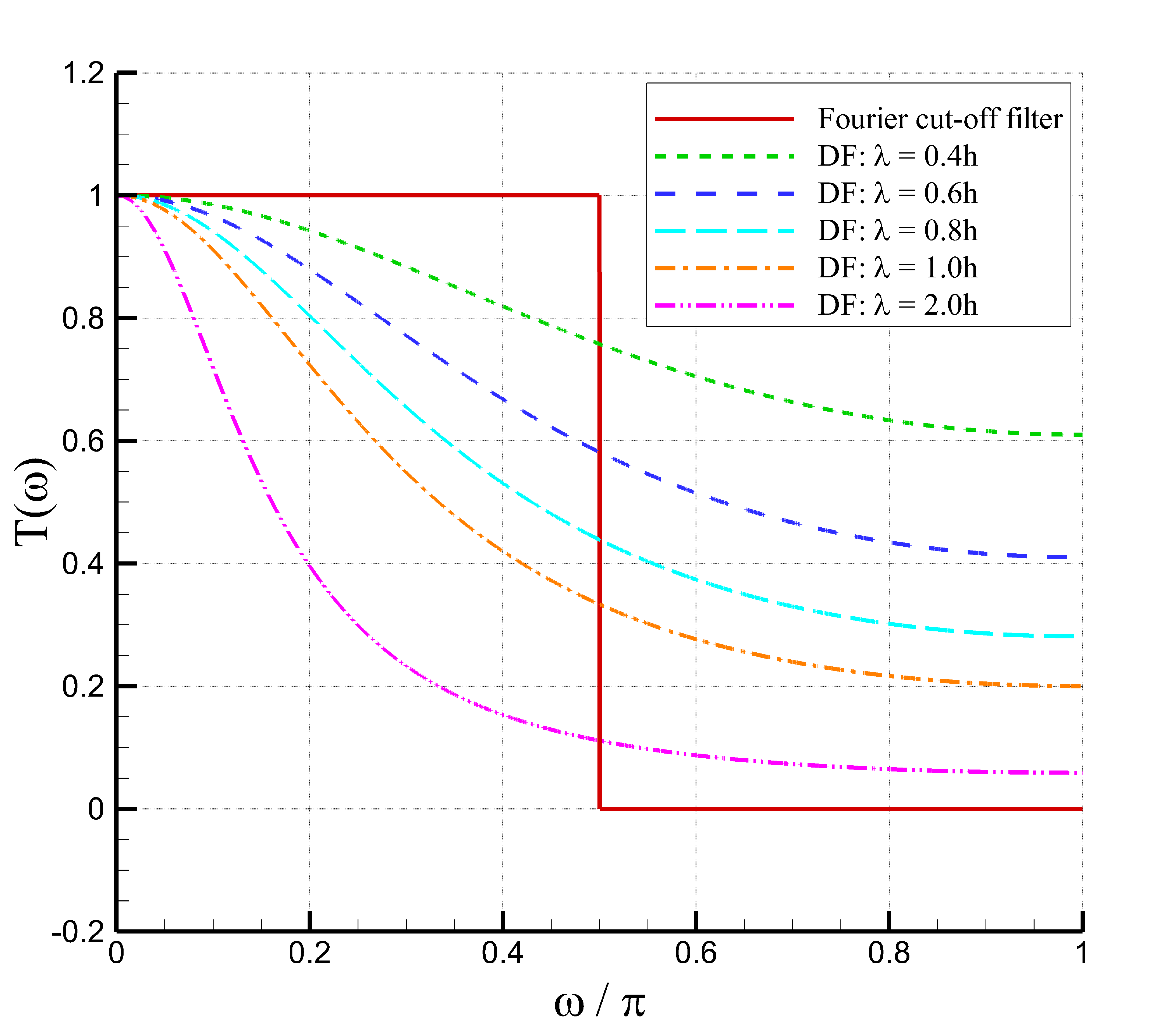}
\caption{
Transfer functions of the DF for different values of the Helmholtz length $\lambda$.
The transfer function of the Fourier cut-off filter is also included for comparison purposes.
}
\label{fig:df}
\end{figure}

The two-dimensional form of the DF in Eq. \eqref{eq:dishelm} is used throughout the paper.
In this section, however, to study the characteristics of the DF in the wavenumber space, we consider the one-dimensional version of the DF (in the $x$-direction)
\begin{equation}
\bar{f}_{i} - \lambda^2 \left( \frac{\bar{f}_{i+1}-2\bar{f}_{i}+\bar{f}_{i-1}}{\Delta_{x}^{2}} \right) = f_{i},
\label{eq:dishelmx}
\end{equation}
and perform a Fourier analysis similar to the analysis presented in Section~\ref{ss_tf}.
Thus, the transfer function of the DF becomes
\begin{equation}
T^{(DF)}(\omega)=\frac{1}{1-\frac{\lambda^2}{\Delta_{x}^2}(2\cos(\omega)-2)} .
\label{eq:tfunh}
\end{equation}
It is obvious that the transfer function $T^{(DF)}$ in Eq. \eqref{eq:tfunh} is positive, which ensures the well-posedness of the AD model (see \cite{stanculescu2008existence}).
To show the characteristics of the DF, we plot in Fig.~\ref{fig:df} its transfer function, $T^{(DF)}$, for different values of the parameter $\lambda$.
In this study, we parameterize the Helmholtz length $\lambda$ as a linear function of the grid spacing $h = \Delta_x = \Delta_y $.
Thus, increasing the value of $\lambda$ in Fig.~\ref{fig:df} amounts to increasing the filter width, while keeping the grid spacing fixed.
Fig.~\ref{fig:df} clearly shows that increasing $\lambda$ (i.e. increasing the filter width) results in a significant increase of the dissipation of the DF (the attenuation of the wavenumber components of the filtered variable).

The DF~\eqref{eq:10}-\eqref{eq:10b} was introduced in LES by \cite{germano1986differential}.
Since then, it has been successfully used in LES of three-dimensional engineering flows \citep{iliescu2003large} and small scale oceanic flows \citep{ozgokmen2009large}.
It has also been analyzed mathematically \citep{dunca2006stolz, layton2006residual,layton2007similarity,layton2012approximate, stanculescu2008existence, berselli2011convergence}. In this study, the DF is used in the LES of large scale oceanic flows.

\section{Numerical methods}
\label{sec:numerical_methods}
In many physically relevant situations, where the Munk and Rhines scales being close to each other, the solutions to oceanic models, such as the QG2 models, do not converge to a steady state as time goes to infinity \citep{medjo2000numerical}. Rather they remain time dependent by producing statistically steady state with one or multiple equilibria. Therefore, numerical schemes designed for numerical integration of such phenomena should be suited for such behavior of the solutions and for the long-time integration. In this study, the governing equations are solved by a fully conservative finite difference scheme along with a third-order Runge-Kutta adaptive time stepping algorithm. An efficient, linear-cost, fast sine transform method is utilized for solving the linear coupled inversion subproblem.

\subsection{Arakawa scheme for the Jacobian}
\cite{arakawa1966computational} suggested that the conservation of energy,
enstrophy, and skew-symmetry is sufficient to avoid computational instabilities stemming
from nonlinear interactions.
The second-order Arakawa scheme for the Jacobian (thenonlinear term in the governing equations) is
\begin{equation}
J(\psi,q) =\frac{1}{3}\bigl(J_{1}(\psi,q)+J_{2}(\psi,q)+J_{3}(\psi,q)\bigr) ,
\label{eq:ja1}
\end{equation}
where the discrete Jacobians have the following forms:
\begin{eqnarray}
J_{1}(\psi,q) &=& \frac{1}{4 \, \Delta_x \, \Delta_y} \bigl[-(q_{i+1,j}-q_{i-1,j})(\psi_{i,j+1}-\psi_{i,j-1}) \nonumber \\
& &   +(q_{i,j+1}-q_{i,j-1})(\psi_{i+1,j}-\psi_{i-1,j}) \bigr] ,
\label{eq:j1} \\
J_{2}(\psi,q) &=& \frac{1}{4 \, \Delta_x \, \Delta_y} \bigl[-q_{i+1,j}(\psi_{i+1,j+1}-\psi_{i+1,j-1})
+q_{i-1,j}(\psi_{i-1,j+1}-\psi_{i-1,j-1}) \nonumber \\
& &  +q_{i,j+1}(\psi_{i+1,j+1}-\psi_{i-1,j+1})
-q_{i,j-1}(\psi_{i+1,j-1}-\psi_{i-1,j-1}) \bigr] ,
\label{eq:j2} \\
J_{3}(\psi,q) &=& \frac{1}{4 \, \Delta_x \, \Delta_y} \bigl[-q_{i+1,j+1}(\psi_{i,j+1}-\psi_{i+1,j})
+q_{i-1,j-1}(\psi_{i-1,j}-\psi_{i,j-1}) \nonumber \\
& & +q_{i-1,j+1}(\psi_{i,j+1}-\psi_{i-1,j})
-q_{i+1,j-1}(\psi_{i+1,j}-\psi_{i,j-1}) \bigr] .
\label{eq:j3}
\end{eqnarray} 
Note that $J_{1}$, which corresponds to the central second-order difference scheme,
is not sufficient for the conservation of energy, enstrophy, and skew-symmetry by the
numerical discretization.
\cite{arakawa1966computational} showed that the judicious combination of $J_{1}, J_{2}$, and $J_{3}$ in
Eq.~\eqref{eq:ja1} achieves the above discrete conservation properties.

\subsection{Time integration scheme}
\label{sec:time}
For the time discretization, we employ an optimal third-order total variation diminishing Runge-Kutta (TVDRK3) scheme \citep{gottlieb1998total}. For clarity of notation, we rewrite the governing equations in the following form:
\begin{equation}
\frac{dq_i}{dt} = R_i ,
\label{eq:rk}
\end{equation}
where subscript $i$ represents the layer index and $R_i$ denotes the discrete spatial derivative operator, including the nonlinear Jacobian of the convective term, the linear biharmonic diffusive term, the forcing term, and the subfilter-scale term. For each layer, the TVDRK3 scheme then becomes:
\begin{eqnarray}
q_{i}^{(1)} &=& q^{n} + \Delta t R_{i}^{(n)} , \nonumber \\
q_{i}^{(2)} &=& \frac{3}{4}  q_{i}^{n} + \frac{1}{4} q_{i}^{(1)} + \frac{1}{4}\Delta t R_{i}^{(1)} , \\
q_{i}^{n+1} &=& \frac{1}{3}  q_{i}^{n} + \frac{2}{3} q_{i}^{(2)} + \frac{2}{3}\Delta t R_{i}^{(2)} , \nonumber
\label{eq:TVDRK}
\end{eqnarray}
where $\Delta t$ is the adaptive time step size, which can be computed at the end of each time step by:
\begin{equation}
\Delta t = c \, \frac{\mbox{min}(\Delta_x,\Delta_y)}{\mbox{max}\left\{ |\frac{\partial \psi_i}{\partial x}|,|\frac{\partial \psi_i}{\partial y}| \right\} } ,
\label{eq:dt}
\end{equation}
where $c$ is known as the Courant-Friedrichs-Lewy (CFL) number.
To ensure the numerical stability of the time discretization scheme, we require that $c \leq 1$.

\subsection{Inversion subproblem}
\label{sec:subproblem}
Most of the demand on computing resources posed by QG models comes in the solution of the elliptic inversion subproblem \citep{miller2007numerical}. This is also true for our study.  However, we take advantage of the simple square shape of our domain and utilize one of the fastest available techniques \citep{moin2002fundamentals,san2012coarse}, which is the FFT based direct inversion to solve the subproblem:
\begin{eqnarray}
Q_1 &=& \mbox{Ro} \nabla^2\psi_1  + \frac{\mbox{Fr}}{\delta}(\psi_2-\psi_1) , \\
Q_2 &=& \mbox{Ro} \nabla^2\psi_2  + \frac{\mbox{Fr}}{1-\delta}(\psi_1-\psi_2) ,
\label{eq:sub-pr}
\end{eqnarray}
where $Q_1=q_1-y$ and $Q_2=q_2-y$. The impermeability boundary condition imposed as $\psi |_{\Omega} = 0$ suggests the use of a fast sine transform (an inverse transform) for each layer:
\begin{equation}
\hat{Q}_{1_{k,l}}=\frac{2}{N_x}\frac{2}{N_y} \sum_{i=1}^{N_x-1} \sum_{j=1}^{N_y-1} Q_{1_{i,j}} \sin \left(\frac{\pi k i}{N_x}\right) \sin \left(\frac{\pi l j}{N_y}\right) ,
\label{eq:ifft-1}
\end{equation}
\begin{equation}
\hat{Q}_{2_{k,l}}=\frac{2}{N_x}\frac{2}{N_y} \sum_{i=1}^{N_x-1} \sum_{j=1}^{N_y-1} Q_{2_{i,j}} \sin \left(\frac{\pi k i}{N_x}\right) \sin \left(\frac{\pi l j}{N_y}\right) ,
\label{eq:ifft-2}
\end{equation}
where $N_x$ and $N_y$ are the total number of grid points in $x$ and $y$ directions. Here the symbol hat is used to represent the corresponding Fourier coefficient of the physical grid data with a subscript pair $i,j$, where $i=0,1, ... N_x$ and $j=0,1, ... N_y$. As a second step, we directly solve the subproblem in Fourier space:
\begin{equation}
\hat{\psi}_{1_{k,l}}= \frac{\alpha_{k,l} \hat{Q}_{1_{k,l}} - \frac{\mbox{Fr}}{1-\delta}\hat{Q}_{1_{k,l}} - \frac{\mbox{Fr}}{\delta}\hat{Q}_{2_{k,l}}}{\alpha_{k,l} \left(\alpha_{k,l} -\frac{\mbox{Fr}}{\delta} - \frac{\mbox{Fr}}{1-\delta} \right)} ,
\label{eq:isub-1}
\end{equation}
\begin{equation}
\hat{\psi}_{2_{k,l}}= \frac{ \alpha_{k,l} \hat{Q}_{2_{k,l}} - \frac{\mbox{Fr}}{1-\delta}\hat{Q}_{1_{k,l}} - \frac{\mbox{Fr}}{\delta}\hat{Q}_{2_{k,l}}}{\alpha_{k,l} \left(\alpha_{k,l} -\frac{\mbox{Fr}}{\delta} - \frac{\mbox{Fr}}{1-\delta} \right)} ,
\label{eq:isub-2}
\end{equation}
where
\begin{equation}
\alpha_{k,l} = \frac{\mbox{Ro}}{\Delta_x^2}\left[2 \cos \left(\frac{\pi k }{N_x} \right) - 2  \right] + \frac{\mbox{Ro}}{\Delta_y^2}\left[2 \cos \left(\frac{\pi l }{N_y} \right)- 2  \right] .
\label{eq:ialpha}
\end{equation}
Finally, the streamfunction arrays for each layer are found by performing a forward sine transform:
\begin{equation}
\psi_{1_{i,j}}= \sum_{k=1}^{N_x-1} \sum_{l=1}^{N_y-1} \hat{\psi}_{1_{k,l}} \sin \left(\frac{\pi k i}{N_x}\right) \sin \left(\frac{\pi l j}{N_y}\right) ,
\label{eq:ffft-1}
\end{equation}
\begin{equation}
\psi_{2_{i,j}}= \sum_{k=1}^{N_x-1} \sum_{l=1}^{N_y-1} \hat{\psi}_{2_{k,l}} \sin \left(\frac{\pi k i}{N_x}\right) \sin \left(\frac{\pi l j}{N_y}\right) ,
\label{eq:ffft-2}
\end{equation}
The computational cost of this elliptic solver is $\displaystyle \mathcal{O}\left(N_x \, N_y \, \log(N_x) \, \log(N_y) \right)$. The FFT algorithm given by \cite{press1992numerical} is used for forward and inverse sine transforms.

\section{Results}
\label{s_results}

The main goal of this section is to test the new AD model \eqref{eq:adm-g1}-\eqref{eq:adm-s} in the numerical simulation of the two-layer QG model.
We also investigate the sensitivity of the AD model with respect to the model parameters.
It turns out that the most important modeling choice is the spatial filter employed in the AD procedure.
We consider two spatial filters in conjunction with the AD model:
the tridiagonal filter (Section \ref{ss_tf}) and the differential filter (Section \ref{ss_df}).
We denote the resulting models AD-TF and AD-DF, respectively.
To test the AD-TF and AD-DF models,
we utilize two different parameter sets, corresponding to two physical oceanic settings: (i) Experiment 1 represents a large ocean basin with the physical parameters used by \cite{tanaka2010alternating}, (ii) Experiment 2 represents a moderate ocean basin with the physical parameters used by \cite{özgökmen1998emergence}. In terms of the classification given by \cite{berloff1999large}, both sets of experiments lie under the chaotic regime. The physical parameters and corresponding dimensionless parameters are summarized in Table~\ref{tab:sets}.
All computations were carried out using a gfortran compiler on a Linux cluster system.
The rest of the section is organized as follows.
In Section \ref{ss_dns}, we present results from the {\it direct numerical simulation (DNS)} for the two settings, Experiment 1 and Experiment 2.
Section \ref{ss_adtf} presents results with the AD-TF model.
Finally, Section \ref{ss_addf} presents results with the AD-DF model.
\begin{table}[!t]
\centering
\caption{Physical parameter sets used in the numerical experiments. }
\label{tab:sets}       
\begin{tabular}{lll}
\hline\noalign{\smallskip}
Variable (unit) & Experiment 1 & Experiment 2 \\
\noalign{\smallskip}\hline\noalign{\smallskip}
  $L$ ($km$)   & 5000 & 2000 \\
  $\Delta x^{512^2}$ = $\Delta y^{512^2}$  ($km$)   & 9.765625 & 3.90625  \\
  $H_1$ ($km$)   & 0.6  & 1.0  \\
  $H_2$ ($km$)& 3.4    & 4.0  \\
  $f_0$ ($s^{-1}$) & $9.35\times10^{-5}$ & $9.35\times10^{-5}$ \\
  $\beta$ ($m^{-1}s^{-1}$) & $1.75\times10^{-11}$ & $1.75\times10^{-11}$ \\
  $\rho_1$ ($kgm^{-3}$)    & 1030    & 1030  \\
  $g'$ ($ms^{-2}$)    & 0.02    & 0.02  \\
  $\tau_0$ ($Nm^{-2}$)    & 0.1    & 0.1  \\
  $\gamma$ ($s^{-1}$)    & $4\times10^{-7}$    & $5\times10^{-8}$ \\
  $\nu$ ($m^{2}s^{-1}$)    & 100    & 50  \\
  $\delta_{M}$ ($km$)    & 17.88    & 14.19  \\
  $\delta_{S}$ ($km$)    & 22.86   & 2.86 \\
  $\delta_{I}$ ($km$)    & 25.77   & 31.56 \\
  $R_{d}$ ($km$)    & 31.16   & 42.79 \\
  $V$ ($ms^{-1}$)    & 0.0116   & 0.0174 \\
  $L/V$ ($year$)    & 13.64   & 3.64 \\
  $L/R_{d}$    & 160.5   & 46.74 \\
  Ro                     & $2.66\times10^{-5}$   & $2.49\times10^{-4}$ \\
  Fr                     & 0.073   & 0.087 \\
  $\sigma$              & $4.57\times10^{-3}$   & $1.43\times10^{-3}$ \\
  $A$          & $4.57\times10^{-8}$    & $3.57\times10^{-7}$  \\
  $\delta$            & 0.15   & 0.2 \\
  Re            & 580.97   & 697.16 \\
\noalign{\smallskip}\hline
\end{tabular}
\end{table}

\subsection{Direct numerical simulation}
\label{ss_dns}
\begin{figure}
\centering
\mbox{
\subfigure{\includegraphics[width=0.5\textwidth]{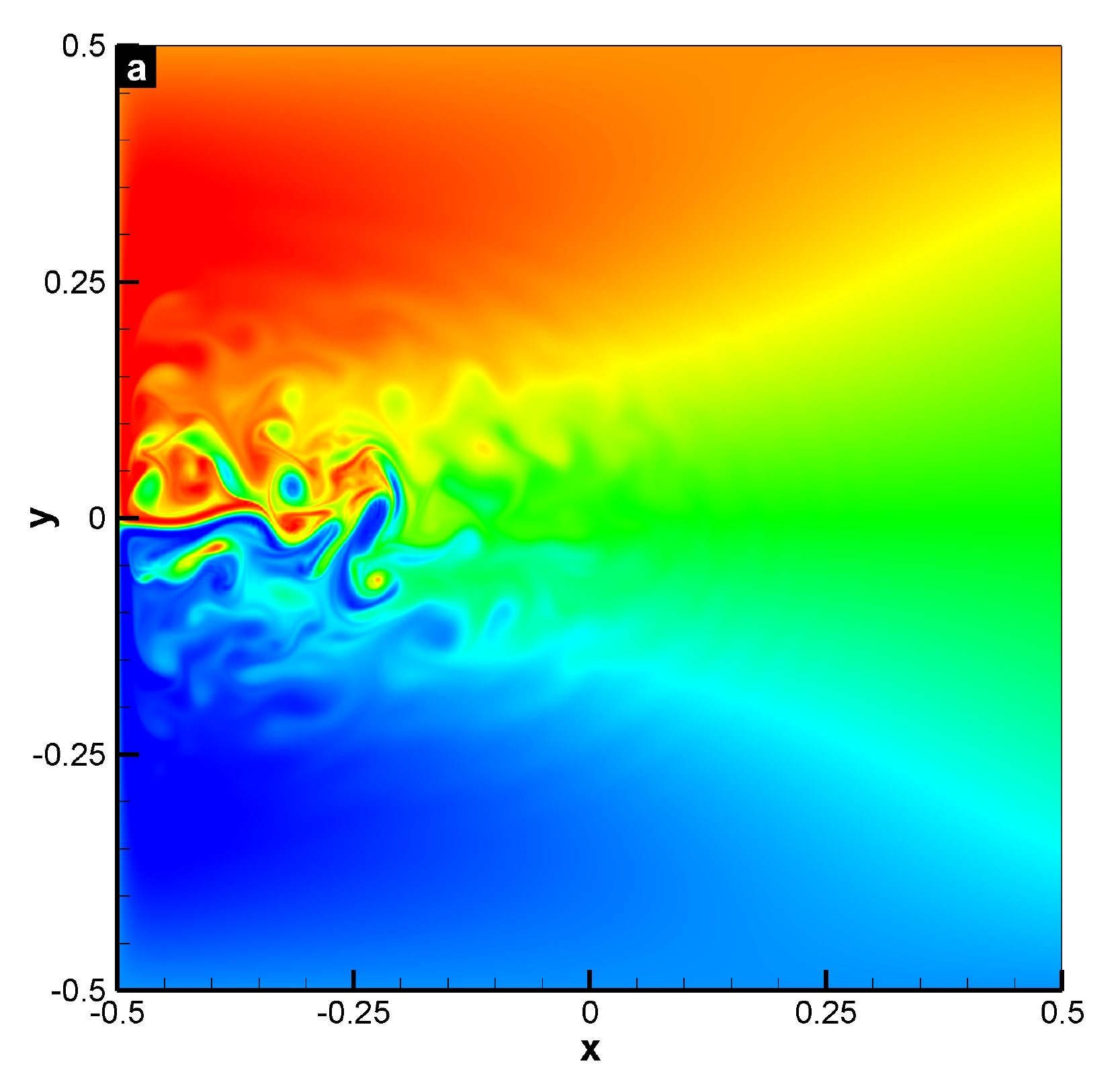}}
\subfigure{\includegraphics[width=0.5\textwidth]{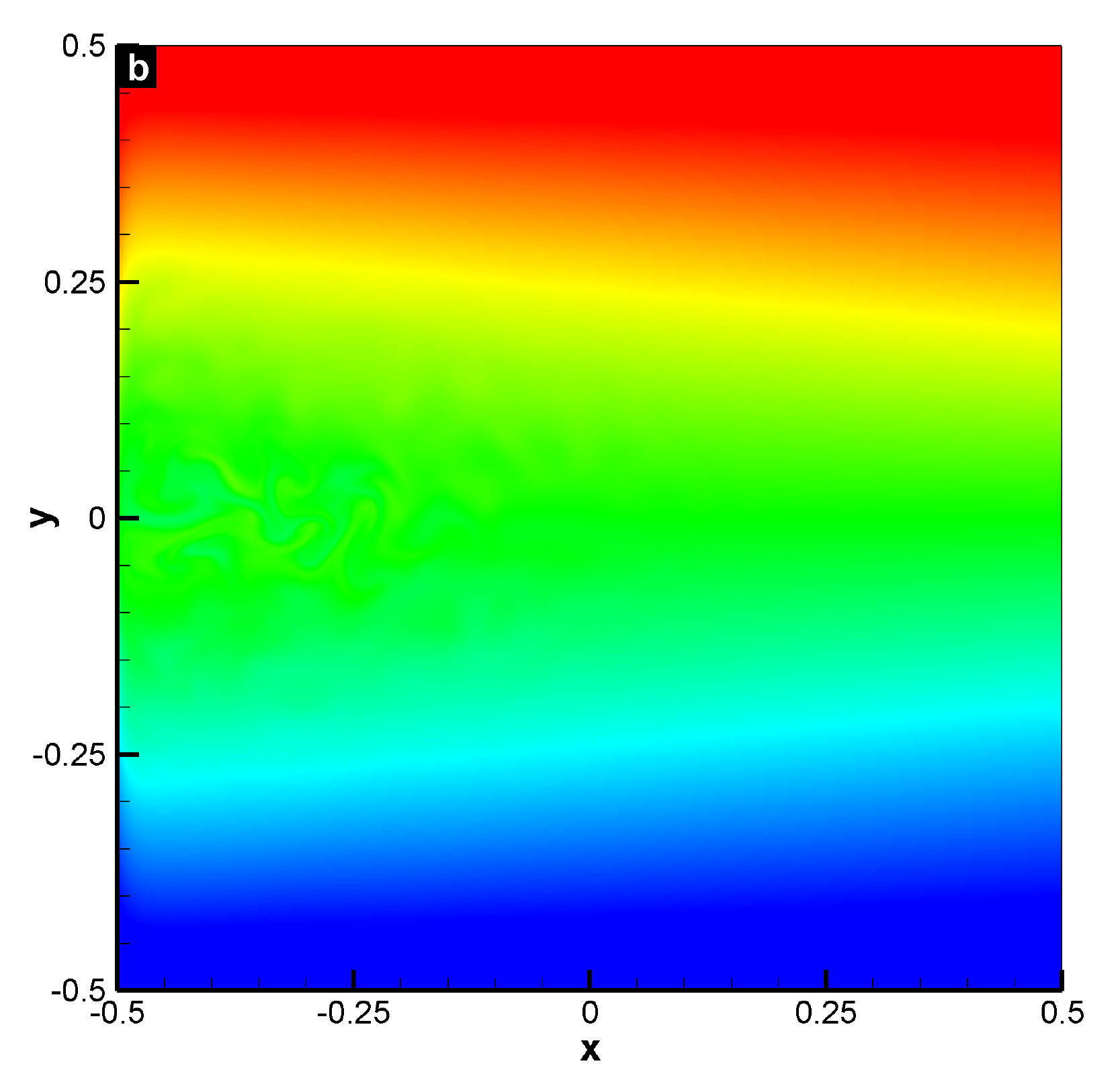}}
}
\caption{
Experiment 1: DNS results for
(a) instantaneous potential vorticity for the upper layer, and
(b) instantaneous potential vorticity for the lower layer.
}
\label{fig:ins-1}
\end{figure}

\begin{figure}
\centering
\mbox{
\subfigure{\includegraphics[width=0.5\textwidth]{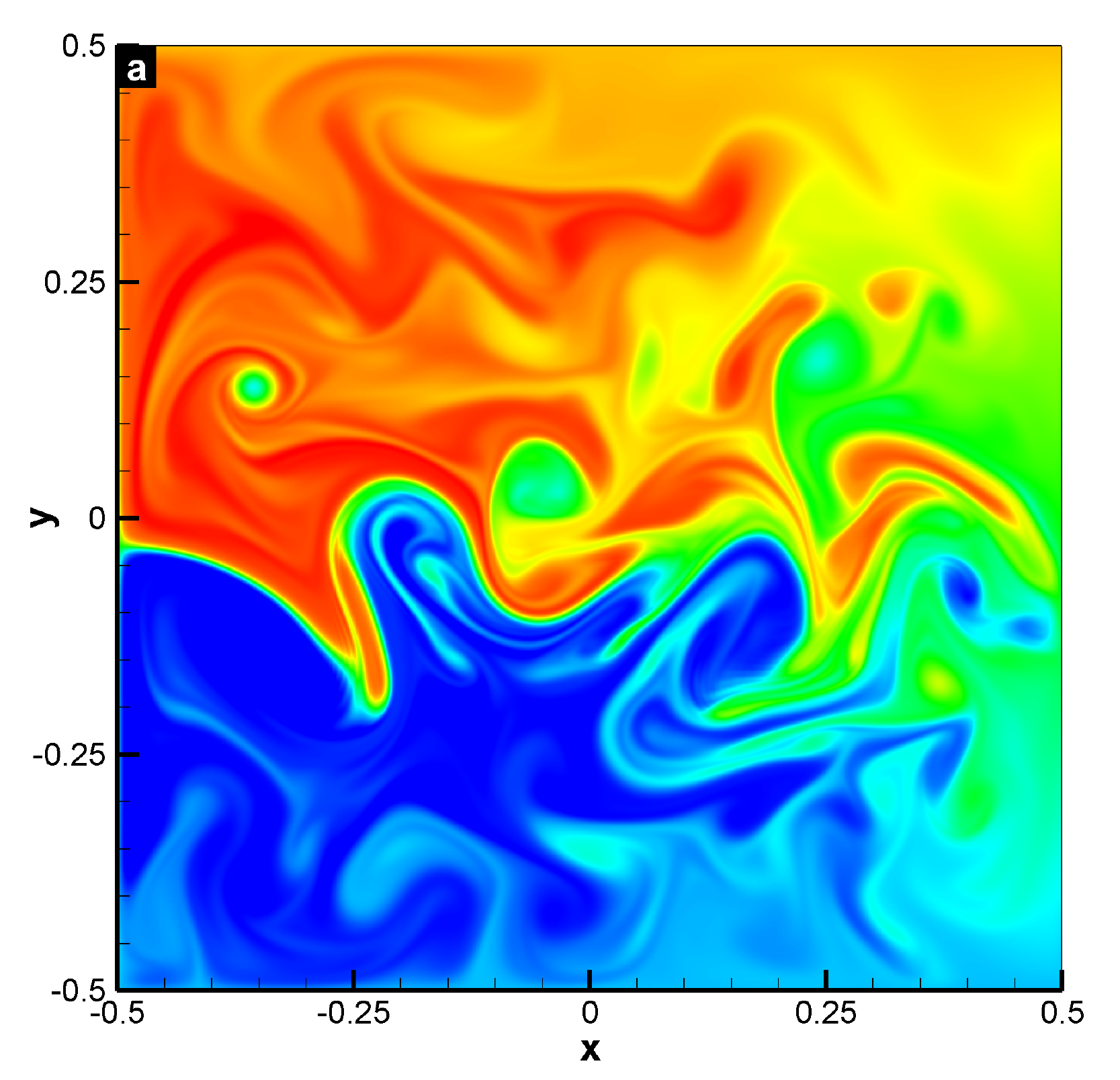}}
\subfigure{\includegraphics[width=0.5\textwidth]{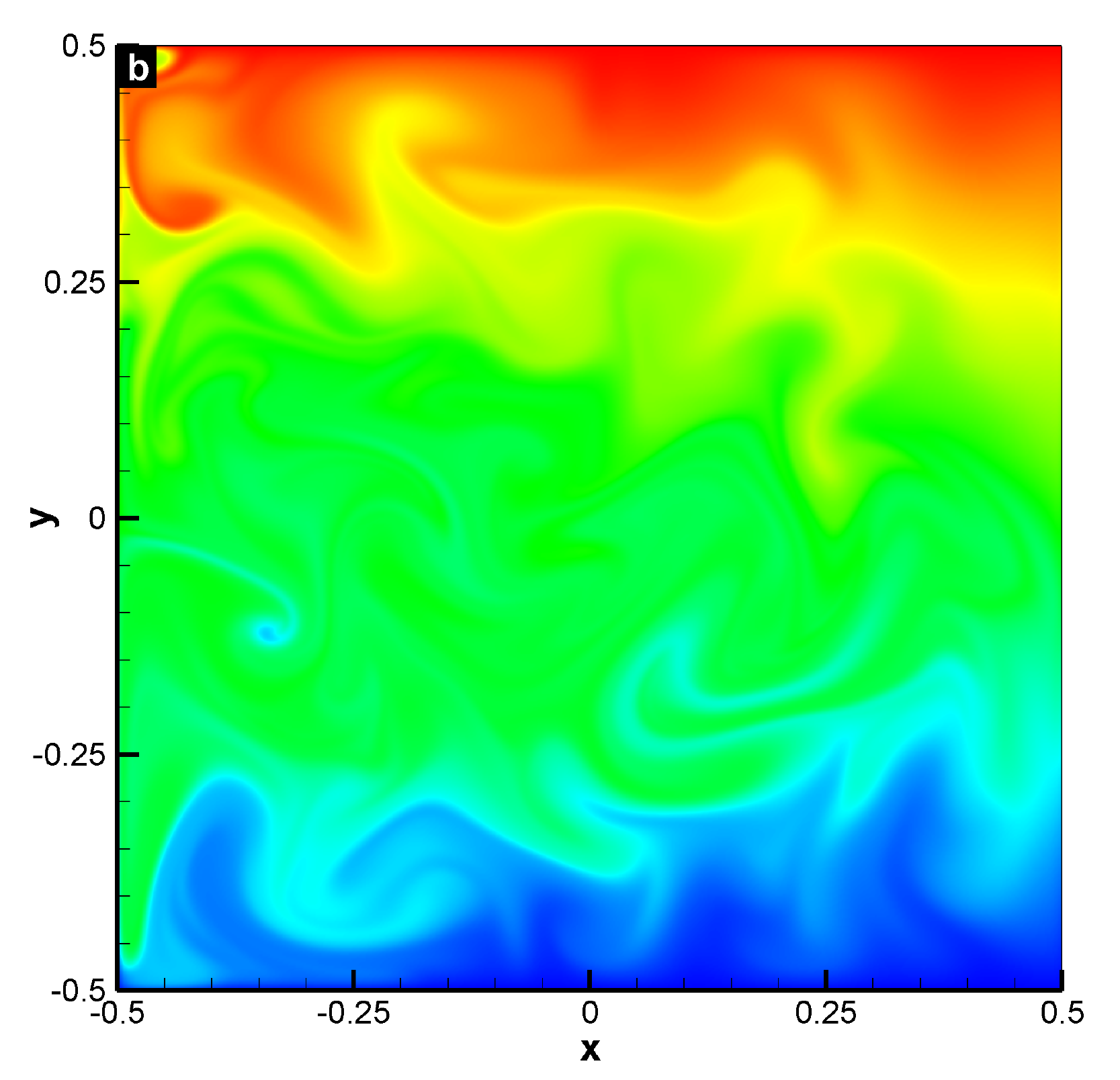}}
}
\caption{
Experiment 2: DNS results for
(a) instantaneous potential vorticity for the upper layer, and
(b) instantaneous potential vorticity for the lower layer.
}
\label{fig:ins-2}
\end{figure}

\begin{figure}
\centering
\mbox{
\subfigure{\includegraphics[width=0.5\textwidth]{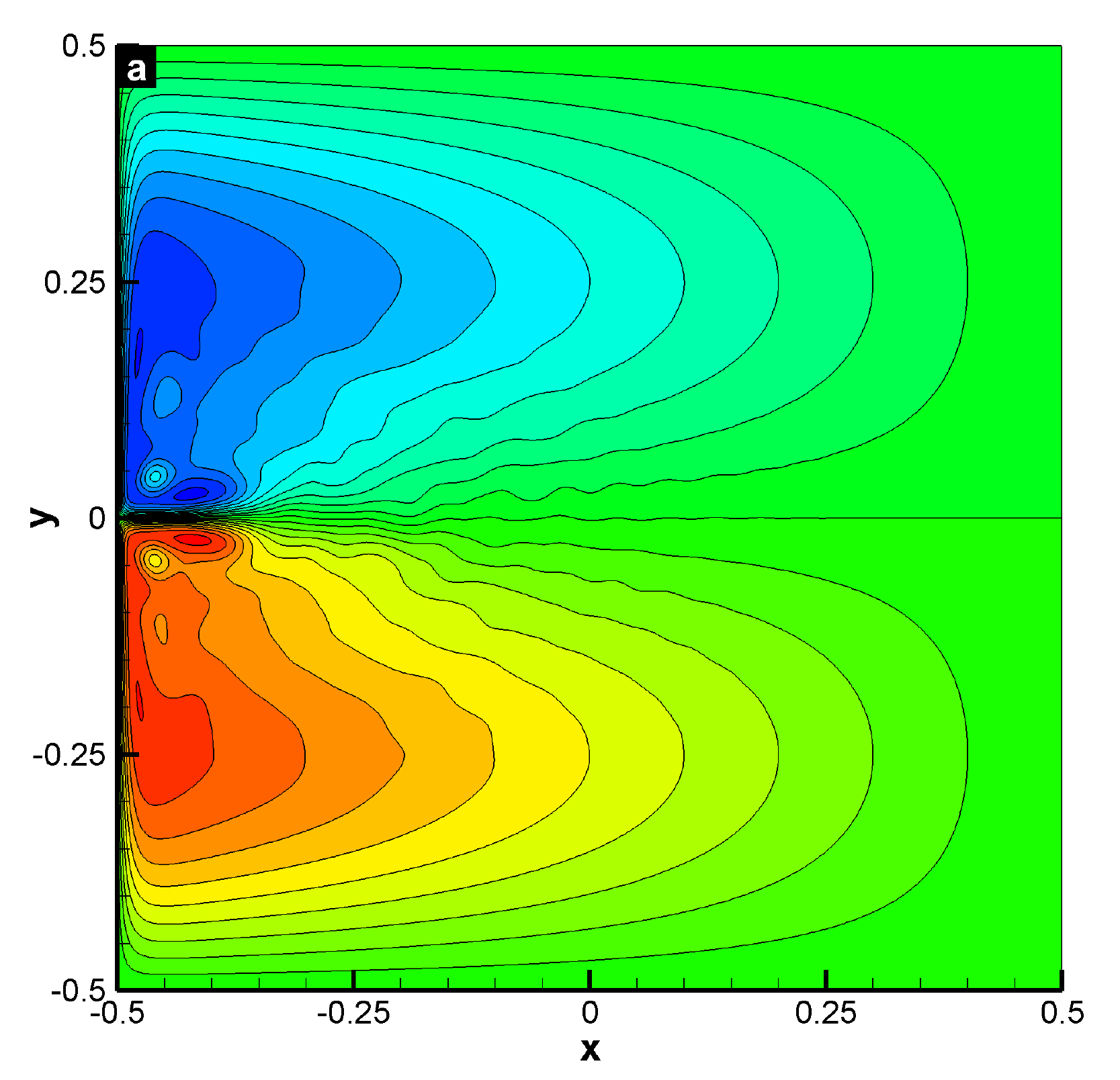}}
\subfigure{\includegraphics[width=0.5\textwidth]{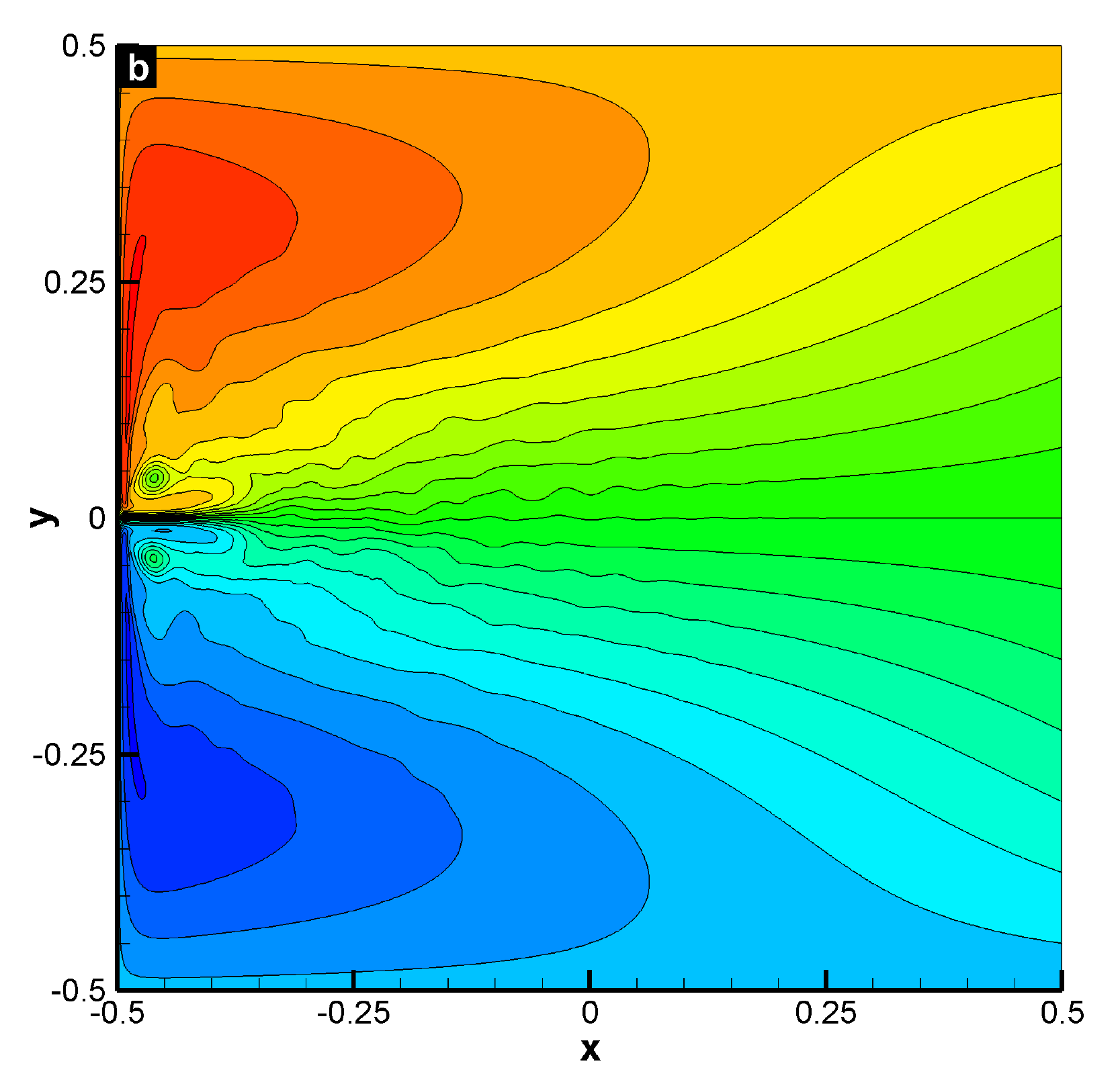}}
}
\mbox{
\subfigure{\includegraphics[width=0.5\textwidth]{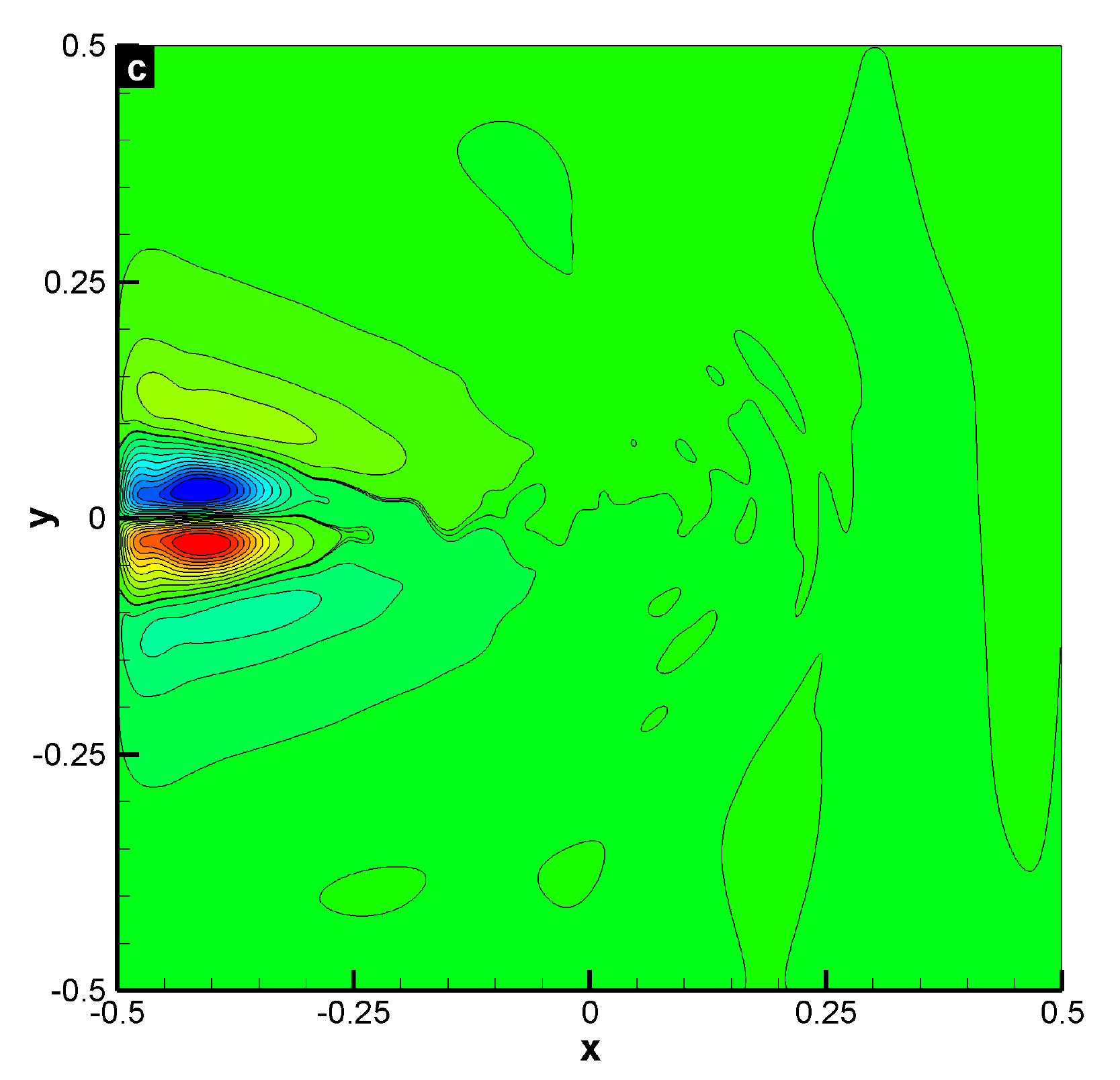}}
\subfigure{\includegraphics[width=0.5\textwidth]{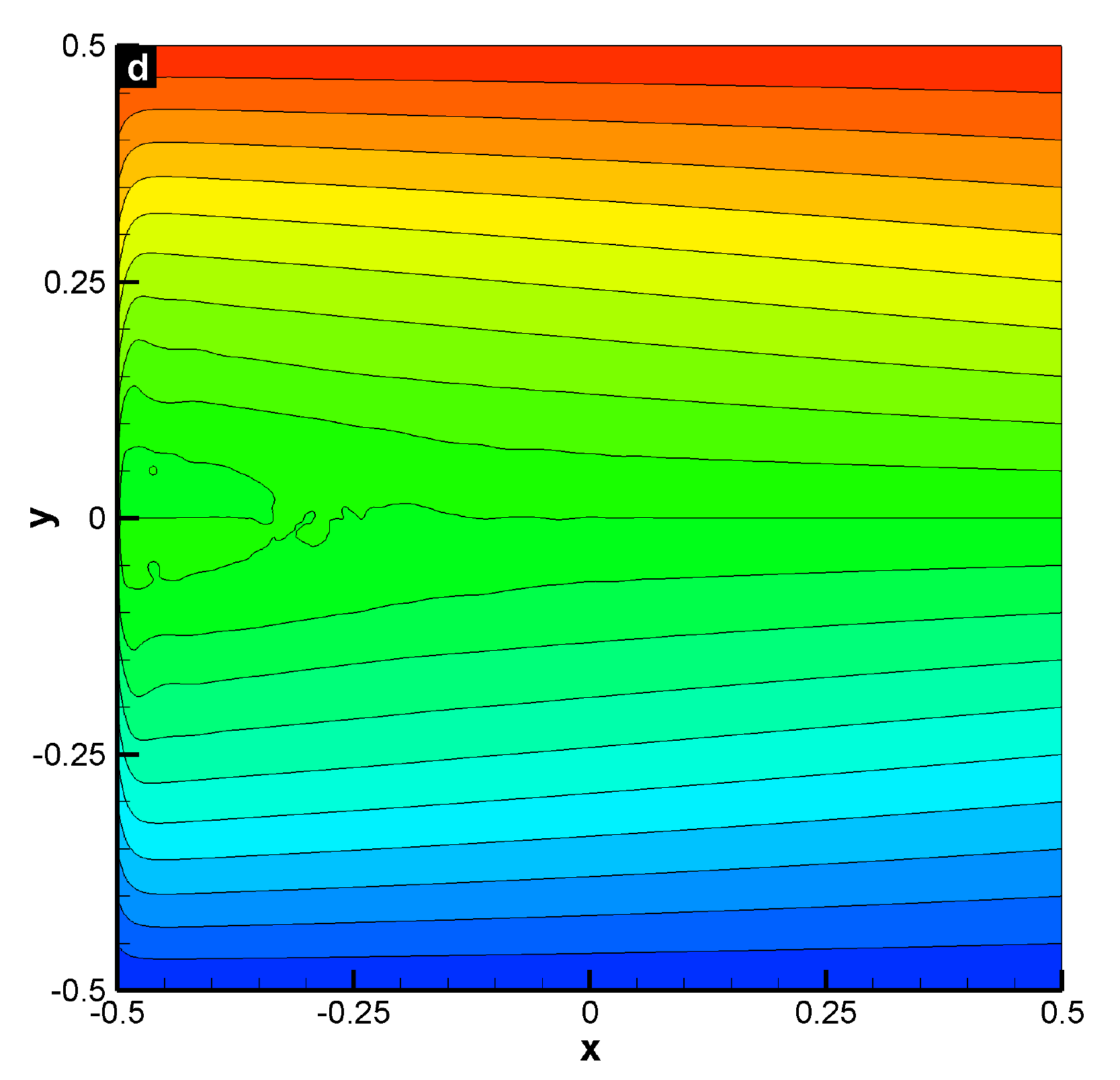}}
}
\caption{
Experiment 1: DNS results for
(a) mean streamfunction contours for the upper layer,
(b) mean potential vorticity contours for the upper layer,
(c) mean streamfunction contours for the lower layer, and
(d) mean potential vorticity contours for the lower layer.
}
\label{fig:mean-1}
\end{figure}

\begin{figure}
\centering
\mbox{
\subfigure{\includegraphics[width=0.5\textwidth]{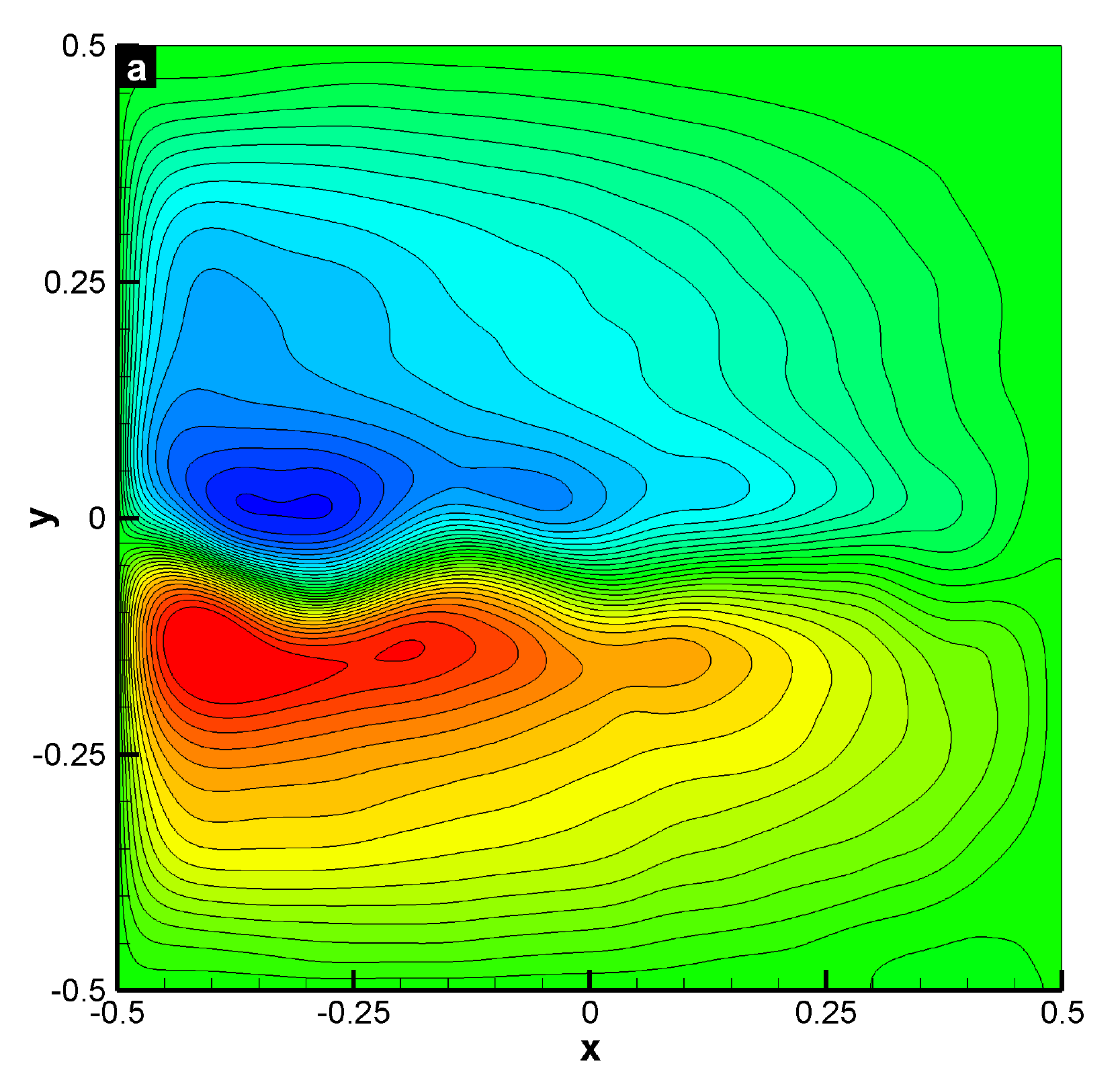}}
\subfigure{\includegraphics[width=0.5\textwidth]{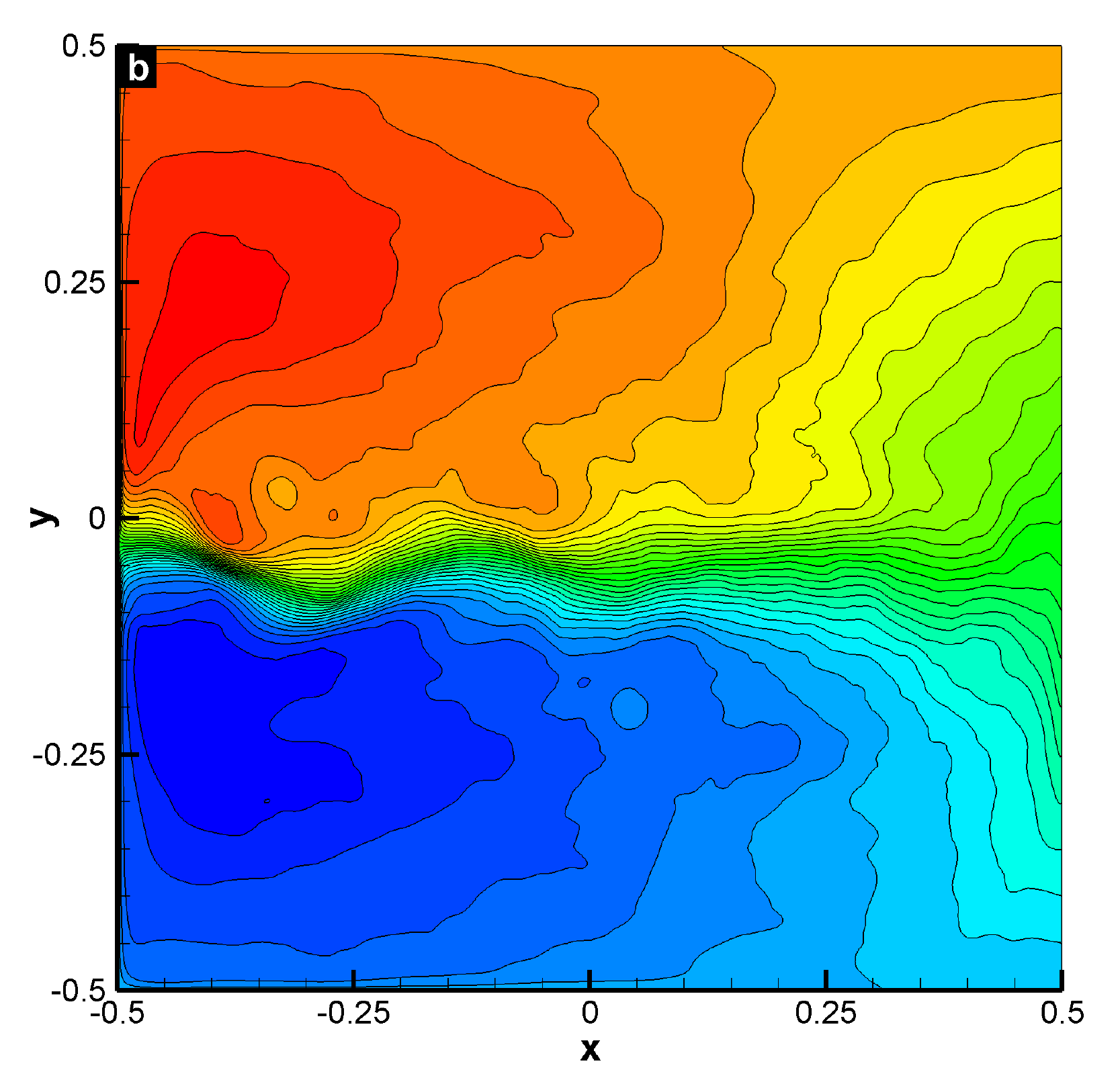}}
}
\mbox{
\subfigure{\includegraphics[width=0.5\textwidth]{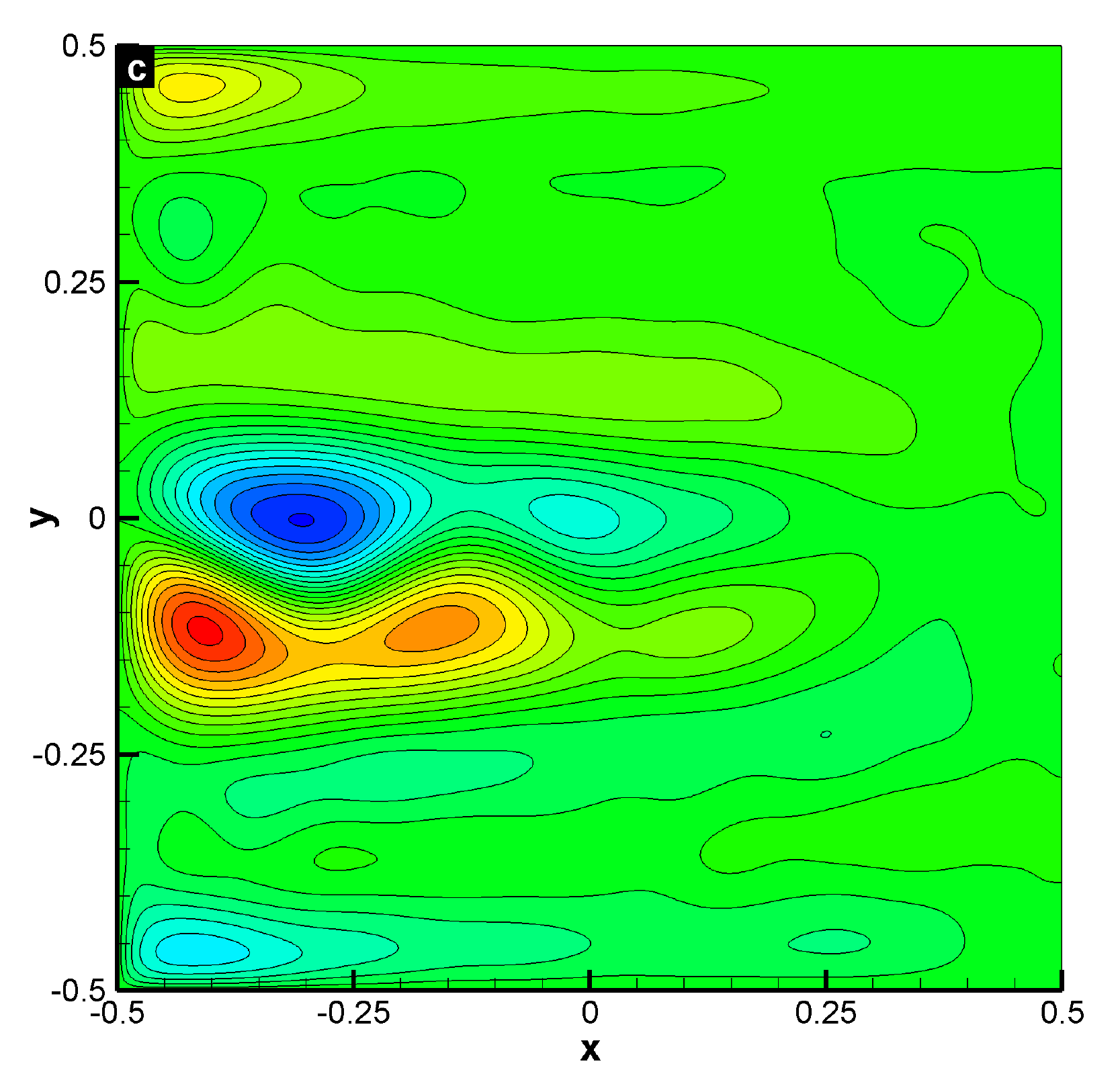}}
\subfigure{\includegraphics[width=0.5\textwidth]{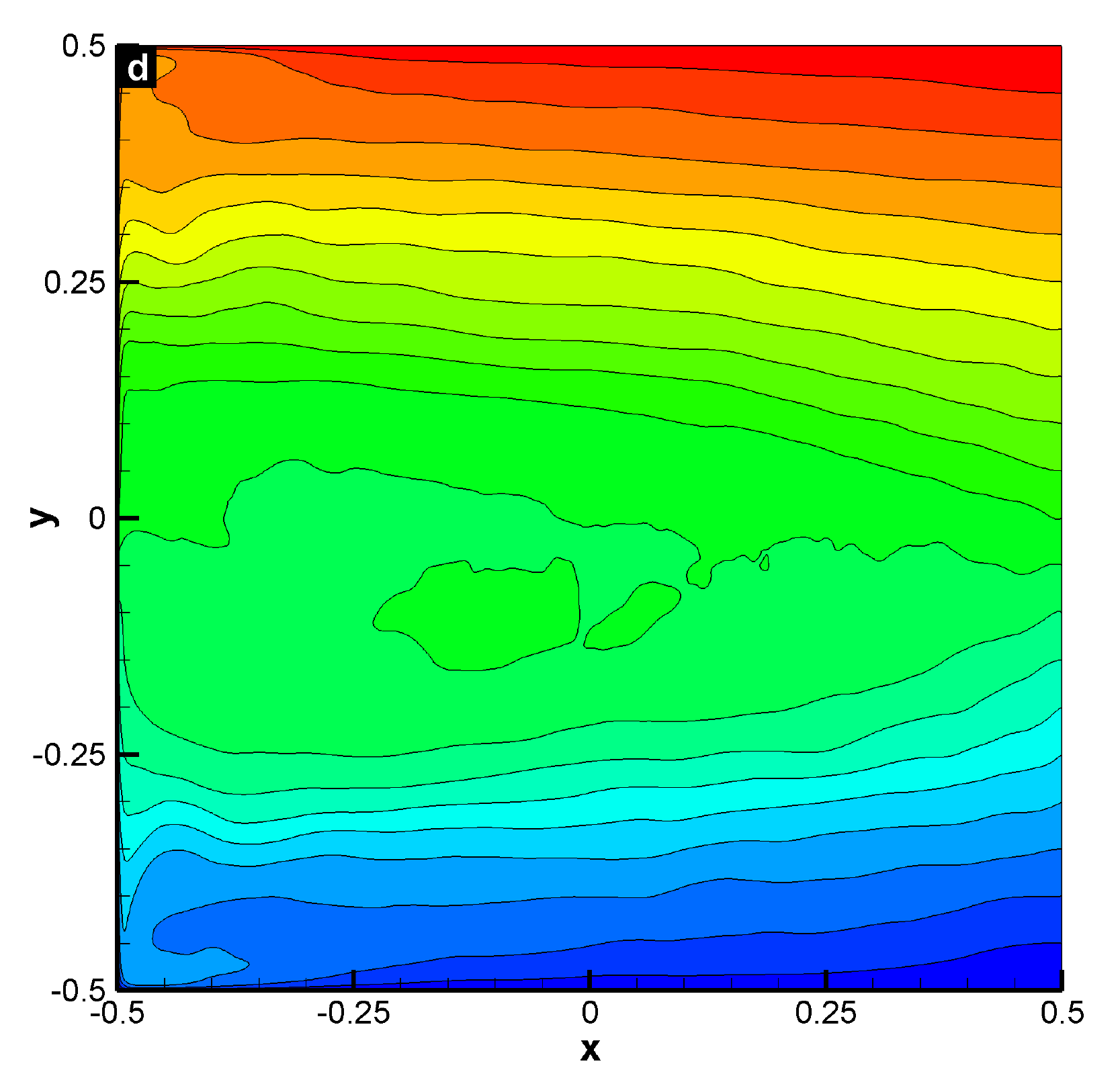}}
}
\caption{
Experiment 2: DNS results for
(a) mean streamfunction contours for the upper layer,
(b) mean potential vorticity contours for the upper layer,
(c) mean streamfunction contours for the lower layer, and
(d) mean potential vorticity contours for the lower layer.
}
\label{fig:mean-2}
\end{figure}

We start by performing a DNS on a fine mesh of $512^2$ spatial resolution.
We emphasize that the term DNS in this study is not meant to indicate that a fully detailed solution is being computed on the molecular viscosity scale, but instead refers to resolving the simulation down to the Munk scale via the specified lateral eddy viscosity parameterization. We also emphasize that the DNS results are given by the numerical solution of Eqs.~\eqref{eq:adm-g1} and \eqref{eq:adm-g} with $S^{*}_{1} = S^{*}_{2} = 0$.
A statistically steady state solution is obtained after an initial transient spin-up process. Instantaneous contour plots for the potential vorticities in the upper and lower layers are shown in Fig.~\ref{fig:ins-1} and Fig.~\ref{fig:ins-2} for Experiment 1 and Experiment 2, respectively. The length scales in these two experiments are quite different. For example, the ratio of the basin length scale $L$ to the Rossby deformation radius $R_d$ is $L/R_d=160.5$ for Experiment 1 and $L/R_d=46.74$ for Experiment 2. Therefore, the structure of the eastward jet formation on the western boundary for Experiment 1 is different from that of Experiment 2. This difference becomes more obvious in the mean flow field. The results for time-averaged mean field data obtained from 2000 snapshots in the statistically steady state are given in Fig.~\ref{fig:mean-1} and Fig.~\ref{fig:mean-2}. The results show strong western boundary currents with cyclonic (counter-clockwise rotating) subpolar gyres and anticyclonic (clockwise rotating) subtropical gyres producing a strong eastward jet in both experiments. However, the produced eastward jet formation in Experiment 2 shows swirling structure and almost reaches the eastern boundary of the basin.
Compared to Experiment 1, the bottom layer is more active in Experiment 2.
Since in Experiment 2 we used the same parameters and boundary conditions as in \cite{özgökmen1998emergence}, the plot in Fig.~\ref{fig:mean-2} is similar to Fig. 2 in  \cite{özgökmen1998emergence}.
Although in Experiment 1 we have used the same parameters as those used in \cite{tanaka2010alternating}, the boundary conditions we used are different from their boundary conditions: we used the slip boundary conditions, whereas they used the no-slip boundary conditions.
Thus, the plot in Fig.~\ref{fig:mean-1} is different from the corresponding one in \cite{tanaka2010alternating}.

\begin{figure}
\centering
\mbox{
\subfigure[]{\includegraphics[width=0.5\textwidth]{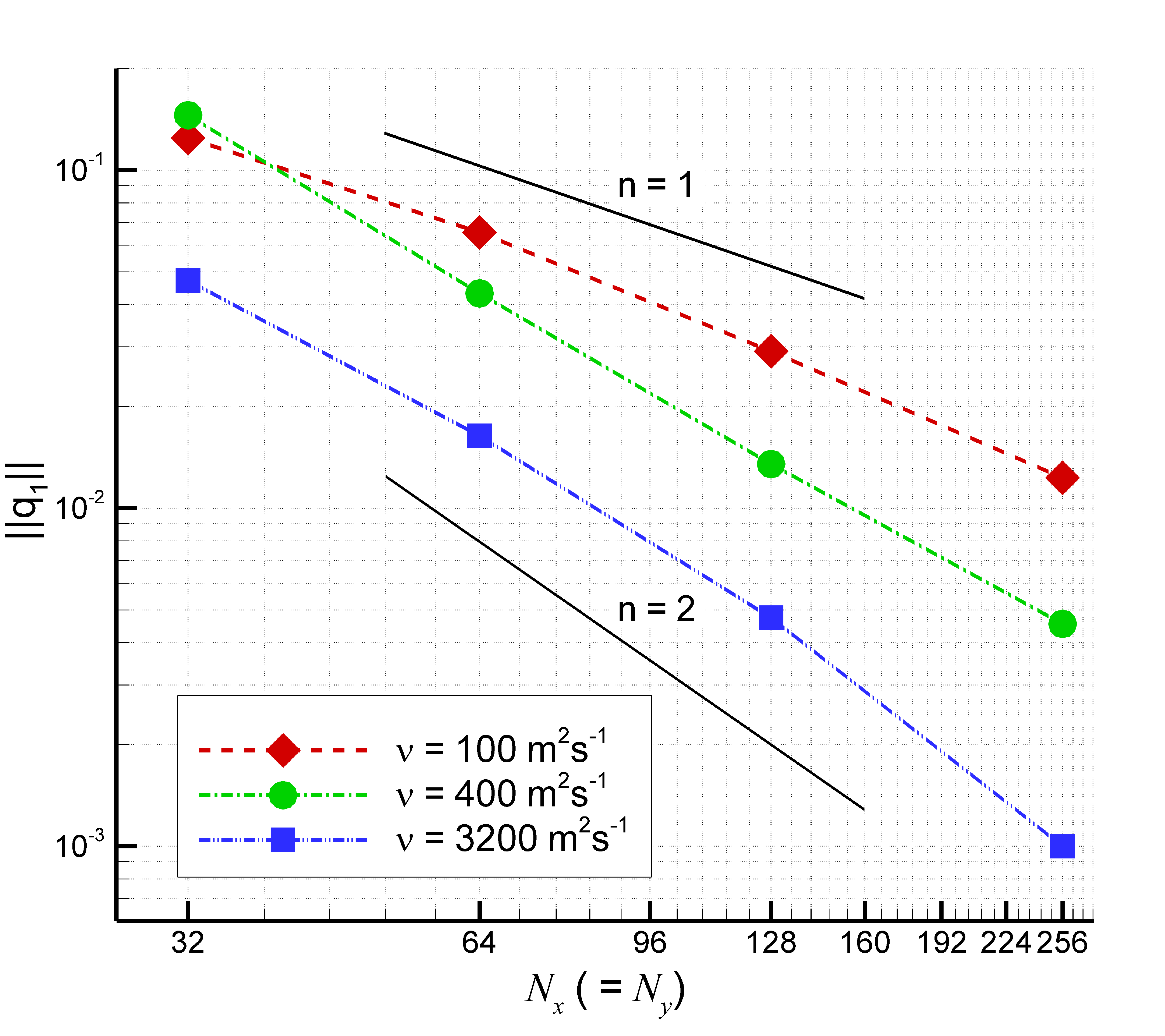}}
\subfigure[]{\includegraphics[width=0.5\textwidth]{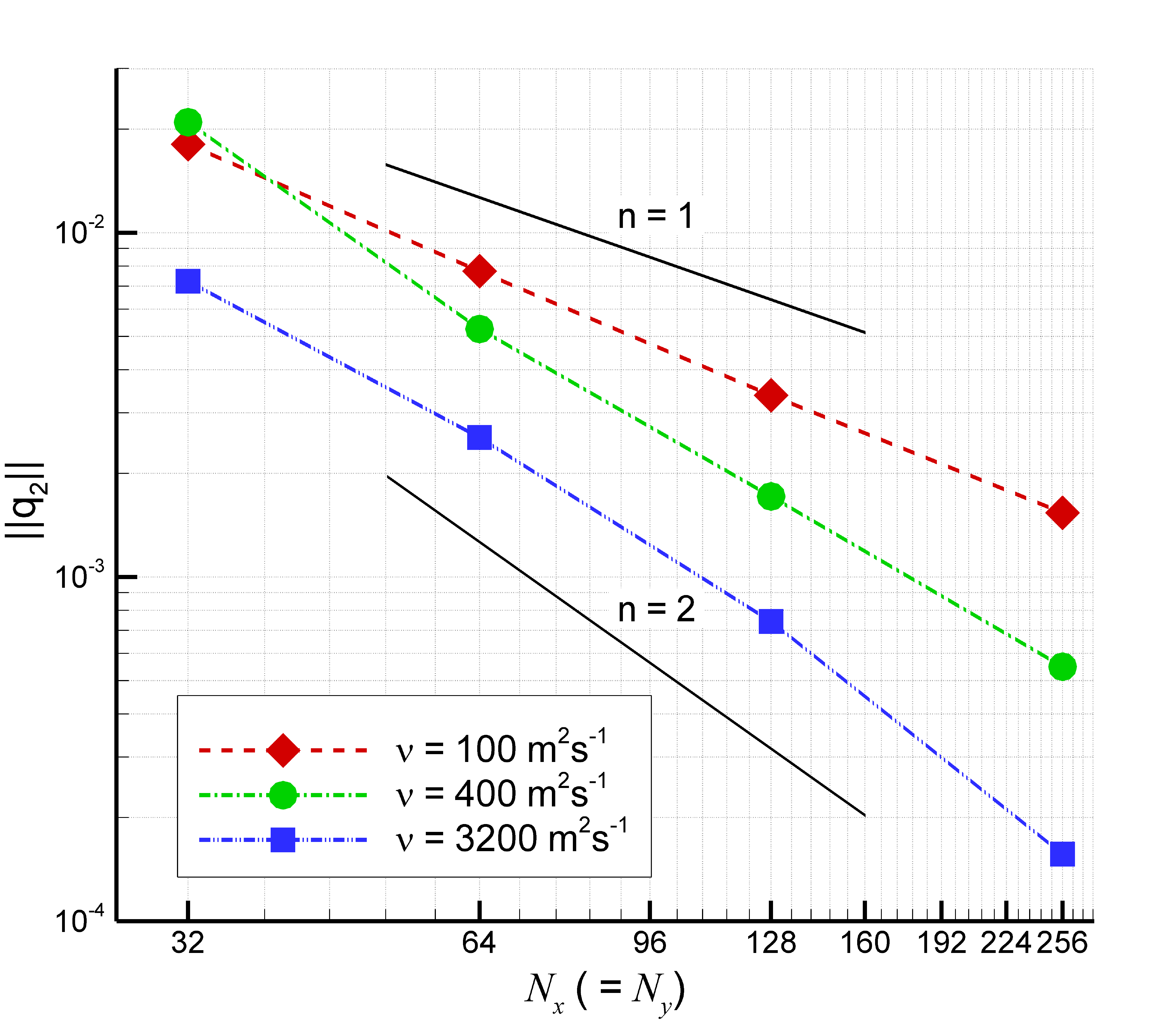}}
}
\caption{Experiment 1:  Log-log plot of the time-averaged $L^2$ norm of the error of the potential vorticity in the two layers, $\| q_1 \|$ and $\| q_2 \|$, for varying eddy viscosity coefficients $\nu$.
The reference solution used in the computation of the error is the numerical approximation obtained at a grid resolution of $512^2$.}
\label{fig:norms}
\end{figure}

To quantify the effect of the numerical discretization on the numerical results, we vary the grid resolution ($N_{x} \times N_{y}$), the time step ($\Delta t$), and the eddy viscosity coefficient ($\nu$) in the QG2 model.
The following quantities are monitored.
The first quantity is the time-averaged $L^2$ norm of the error of the potential vorticity, denoted as $\| q_i \|$, where the subscript $i$ represents the layer index.
The reference solution used in the computation of the error is the numerical approximation obtained at a grid resolution of $512^2$.
The second quantity is the time-averaged basin-integrated kinetic energy, $E_i$, which is defined as
\begin{equation}
E_i = \frac{1}{T_2 - T_1}\int_{t=T_1}^{t=T_2} E_i(t) \, dt ,
\label{eq:definition_E}
\end{equation}
where, again, the subscript $i$ represents the layer index and $T_1=6$ and $T_2=8$ are the temporal bounds for the averaging window.
The integrand $E_i(t)$ in \eqref{eq:definition_E} is the instantaneous basin integrated kinetic energy in each layer and is defined as
\begin{equation}
E_{i}(t)
= \frac{1}{2} \iint \left(\frac{\partial \psi_i}{\partial x}\right)^2
+ \left(\frac{\partial \psi_i}{\partial y} \right)^2 dx \, dy ,
\label{eq:26}
\end{equation}
%
%
%
First, we investigate the effect of the grid resolution on the numerical results.
To this end, we fix the time step $\Delta t= 2\times 10^{-5}$ and vary the grid resolution, $N_{x} \times N_{y}$, and the eddy viscosity coefficient in the QG2 model, $\nu$.
Table~\ref{tab:mean_energy} presents the time-averaged basin-integrated kinetic energy of the upper layer, $E_1$, defined in \eqref{eq:definition_E}.
This table shows that, for most grid resolutions, accurate results are obtained for the high values of the eddy viscosity coefficient, $\nu$.
For the lowest values of $\nu$, however, the results are inaccurate at the lower grid resolutions, and relatively accurate at the higher grid resolutions.
This behavior is natural, since when the grid spacing is larger than the Munk scale, the smallest scale are not resolved, thereby producing grid-scale variability in the solution, which degrades the accuracy.
Table~\ref{tab:L2-DNS} presents the time-averaged $L^2$ norm of the error of the potential vorticity in the two layers, $\| q_1 \|$ and $\| q_2 \|$.
This table shows that, as expected, the error decreases as the grid resolution increases.
We note that this decrease in the error is faster for the high values of $\nu$.
This behavior is similar to that observed in Table~\ref{tab:mean_energy}.
Finally, the results in Table~\ref{tab:L2-DNS} and are also plotted in Fig.~\ref{fig:norms}.
This figure clearly shows that a second-order spatial accuracy is obtained for the high values of $\nu$, and a first-order spatial accuracy is obtained for the lowest values of $\nu$.
However, one important thing to note in Fig.~\ref{fig:norms} is that the convergence approaches second-order when $N_x=N_y$ aproaches 256. This is because the mininum number of grid points at which we can start to expect convergence is when the Munk scale is resolved, i.e., when $N_x=N_y=280$ for Experiment 1. Therefore, we emphasize that the use of $512^2$ resolution should suffice for a DNS, although just barely.

\begin{figure}[!t]
\centering
\includegraphics[width=0.5\textwidth]{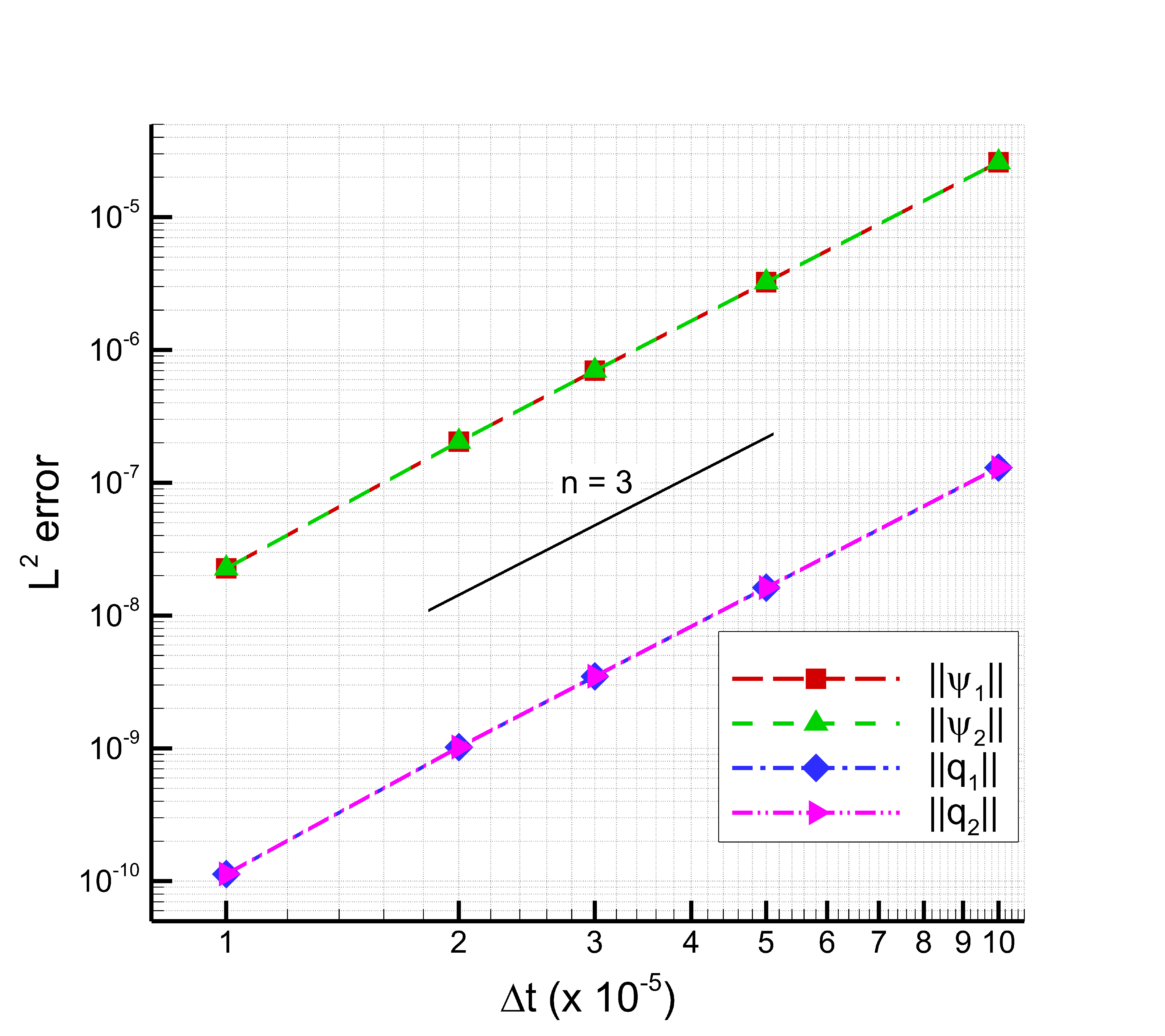}
\caption{Experiment 1:  Log-log plot of the $L^2$ norm of the error of the streamfunction, $\| \psi_1 \|$ and $\| \psi_2 \|$, and potential vorticity, $\| q_1 \|$ and $\| q_2 \|$, in the two layers at $t = 0.0075$, using the eddy viscosity coefficient $\nu = 100$ $m^{2}s^{-1}$ and a resolution of $512^2$.
The reference solution used in the computation of the error is the numerical approximation obtained by using the time step $\Delta t = 5 \times 10^{-6}$.}
\label{fig:time-norm}
\end{figure}

Next, we investigate the effect of the time step on the numerical results.
To this end, we fix the the eddy viscosity coefficient, $\nu = 100$ $m^{2}s^{-1}$, and vary the grid resolution, $N_{x} \times N_{y}$, and the time step, $\Delta t$.
Table~\ref{tab:level_time} presents the time-averaged basin-integrated kinetic energy of the two layers, $E_1$ and $E_2$.
This table shows that, for a fixed spatial resolution, varying the time step does not yield a significant change in the numerical results.
To perform the time-accuracy analysis for the  third-order Runge-Kutta time integration scheme, we fix the grid resolution at $512^2$ and plot in Fig.~\ref{fig:time-norm} the $L^2$ norm of the error at $t = 0.0075$ for different time step sizes.
The data obtained with $\Delta t = 5 \times10^{-6}$ is used as reference solution to compute the error norms.
We present results at an early integration time to prevent the numerical instability that could appear later on as a result of the violation of the CFL criterion.
The log-log plot in Fig.~\ref{fig:time-norm} clearly shows that the time integration scheme achieves the expected third-order temporal accuracy for both the upper and lower layers in Experiment 1.
To investigate the effects of the adaptive time discretization described in Section \ref{sec:time}, we performed the same numerical experiments as those in Table~\ref{tab:level_time}, this time, however, using the adaptive time-stepping scheme with a fixed CFL number $c=0.95$.
This approach yielded the same qualitative results as those in Table~\ref{tab:level_time}.

The above numerical studies quantify the effects of the numerical discretization described in Section~\ref{sec:numerical_methods}.
The following general conclusions can be drawn.
The spatial discretization is optimal (second-order) for high values of the eddy viscosity coefficient, and is suboptimal (first-order) for the low values that we use in this study.
The time discretization error appears to be dominated by the spatial discretization error.
Indeed, for a fixed grid resolution, changing the time step had only negligible effects on the numerical results.
Although it is hard to decouple the numerical effects from the LES modeling effects, the above numerical studies will serve as a guide in the subsequent interpretation of the LES results.
Furthermore, a more detailed presentation of error estimates for the spatial and temporal schemes utilized here can be found in a recent study on two-dimensional decaying turbulence conducted by \cite{san2012high}.
%
%
%


\begin{table}[!t]
\centering
\caption{Experiment 1: Time-averaged basin-integrated kinetic energy of the upper layer, $E_1$, for varying grid resolutions, $N_x \times N_y$, varying eddy viscosity coefficients, $\nu$, and fixed time step $\Delta t= 2\times10^{-5}$.}
\label{tab:mean_energy}       
\begin{tabular}{p{0.13\textwidth}p{0.13\textwidth}p{0.12\textwidth}p{0.12\textwidth}p{0.12\textwidth}p{0.12\textwidth}p{0.11\textwidth}}
\hline\noalign{\smallskip}
{$N_x \times N_y$} & $\nu=100$ & $\nu=200$ & $\nu=400$ & $\nu=800$ & $\nu=1600$ & $\nu=3200$ \\
\noalign{\smallskip}\hline\noalign{\smallskip}
$32^2$  & 195.028 & 200.188 & 151.178 & 89.139 & 57.016 & 36.500 \\
$64^2$  & 103.787 & 77.749  & 59.083 & 43.305 & 33.567 & 27.878 \\
$128^2$ & 77.617  & 63.618  & 51.364 & 42.545 & 34.003 & 27.661 \\
$256^2$ & 79.478  & 65.560  & 52.646 & 42.208 & 34.764 & 27.851 \\
$512^2$ & 81.609  & 66.084  & 52.787 & 42.051 & 35.096 & 27.921 \\
\noalign{\smallskip}\hline
\end{tabular}
\end{table}

\begin{table}[!t]
\centering
\caption{Experiment 1:  Time-averaged $L^2$ norm of the error of the potential vorticity in the two layers, $\| q_1 \|$ and $\| q_2 \|$, for varying grid resolutions, $N_x \times N_y$, varying eddy viscosity coefficients, $\nu$, and fixed time step $\Delta t = 2\times10^{-5}$.
The reference solution used in the computation of the error is the numerical approximation obtained at a grid resolution of $512^2$.}
\label{tab:L2-DNS}
\begin{tabular}{p{0.13\textwidth}p{0.12\textwidth}p{0.12\textwidth}p{0.12\textwidth}p{0.12\textwidth}p{0.12\textwidth}p{0.12\textwidth}}
\hline\noalign{\smallskip}
\multirow{2}{*}{$N_x \times N_y$}  &
\multicolumn{2}{l}{\underline{$\nu=100$ \quad \quad \quad \quad \quad \quad}} &
\multicolumn{2}{l}{\underline{$\nu=400$ \quad \quad \quad \quad \quad \quad}} &
\multicolumn{2}{l}{\underline{$\nu=3200$ \quad \quad \quad \quad \quad \quad}} \\
 & $\parallel q_1 \parallel$  & $\parallel q_2 \parallel$ & $ \parallel q_1 \parallel$ & $\parallel q_2 \parallel$ & $\parallel q_1 \parallel$ & $\parallel q_2 \parallel$ \\
\noalign{\smallskip}\hline\noalign{\smallskip}
$32^2$   & 1.2446E-1 & 1.8075E-2 & 1.4552E-1  & 2.0959E-2  & 4.7177E-2 & 7.2268E-3    \\
$64^2$   & 6.5465E-2 & 7.7261E-3 & 4.3220E-2  & 5.2517E-3  & 1.6356E-2 & 2.5420E-3 \\
$128^2$  & 2.9121E-2 & 3.3675E-3 & 1.3513E-2  & 1.7138E-3  & 4.7441E-3 & 7.4199E-4  \\
$256^2$  & 1.2296E-2 & 1.5355E-3 & 4.5496E-3  & 5.4868E-4  & 1.0002E-3 & 1.5671E-4  \\
\noalign{\smallskip}\hline
\end{tabular}
\end{table}

\begin{table}[!t]
\centering
\caption{Experiment 1: Time-averaged basin-integrated kinetic energy of the two layers, $E_1$ and $E_2$, for varying grid resolutions, $N_x \times N_y$, and fixed eddy viscosity coefficient, $\nu = 100$ $m^{2}s^{-1}$.}
\label{tab:level_time}
\begin{tabular}{p{0.13\textwidth}p{0.12\textwidth}p{0.12\textwidth}p{0.12\textwidth}p{0.12\textwidth}p{0.12\textwidth}p{0.12\textwidth}}
\hline\noalign{\smallskip}
\multirow{2}{*}{$N_x \times N_y$}  &
\multicolumn{2}{l}{\underline{$\Delta t=1 \times 10^{-5}$ \quad \quad \quad}} &
\multicolumn{2}{l}{\underline{$\Delta t=2 \times 10^{-5}$ \quad \quad \quad}} &
\multicolumn{2}{l}{\underline{$\Delta t=4 \times 10^{-5}$ \quad \quad \quad}} \\
 & $E_1$  & $E_2$ & $ E_1$ & $E_2$ & $E_1$ & $E_2$ \\
\noalign{\smallskip}\hline\noalign{\smallskip}
$32^2$   & 198.862 & 1.124 & 195.028 & 1.086  & 196.293  & 1.095 \\
$64^2$   & 104.332 & 0.874 & 103.787 & 0.876  & 104.143  & 0.875 \\
$128^2$  & 78.210  & 1.195 & 77.617  & 1.961  & 77.768   & 1.952 \\
$256^2$  & 79.194  & 2.532 & 79.478  & 2.523  & 79.416   & 2.538 \\
$512^2$  & 81.277  & 2.592 & 81.609  & 2.594  & 80.996   & 2.601 \\
\noalign{\smallskip}\hline
\end{tabular}
\end{table}

\subsection{Approximate deconvolution model with the tridiagonal filter (AD-TF)}
\label{ss_adtf}
To test the new AD-TF model \eqref{eq:adm-g}-\eqref{eq:adm-s},
we employ the standard LES methodology:
We first run a DNS on a fine mesh (of $512^2$ spatial resolution).
We then run on a much coarser mesh (of $32^2$ spatial resolution) an under-resolved numerical simulation (denoted in what follows as QG2$_{c}$).
We emphasize that QG2$_{c}$ does not employ any subfilter-scale model.
Finally, we employ the new AD-TF model on the same coarse mesh utilized in QG2$_{c}$ (of $32^2$ spatial resolution).
The criterion used in assessing the success of the new AD-TF model is its ability to produce more accurate (i.e., closer to the DNS data) results than those for QG2$_{c}$, without a significant increase in computational time.
Following \cite{san2011}, in the AD-TF model, we use the tridiagonal filtering procedure with $N=5$ and $\alpha=0.25$.
To compare the DNS, the QG2$_{c}$, and the AD-TF model, we utilize data that is time-averaged between $t=6$ and $t=8$ by using 2000 snapshots of the field. Note that that this averaging period corresponds to $27.28$ years for Experiment 1.

\begin{figure}
\centering
\mbox{
\subfigure{\includegraphics[width=0.33\textwidth]{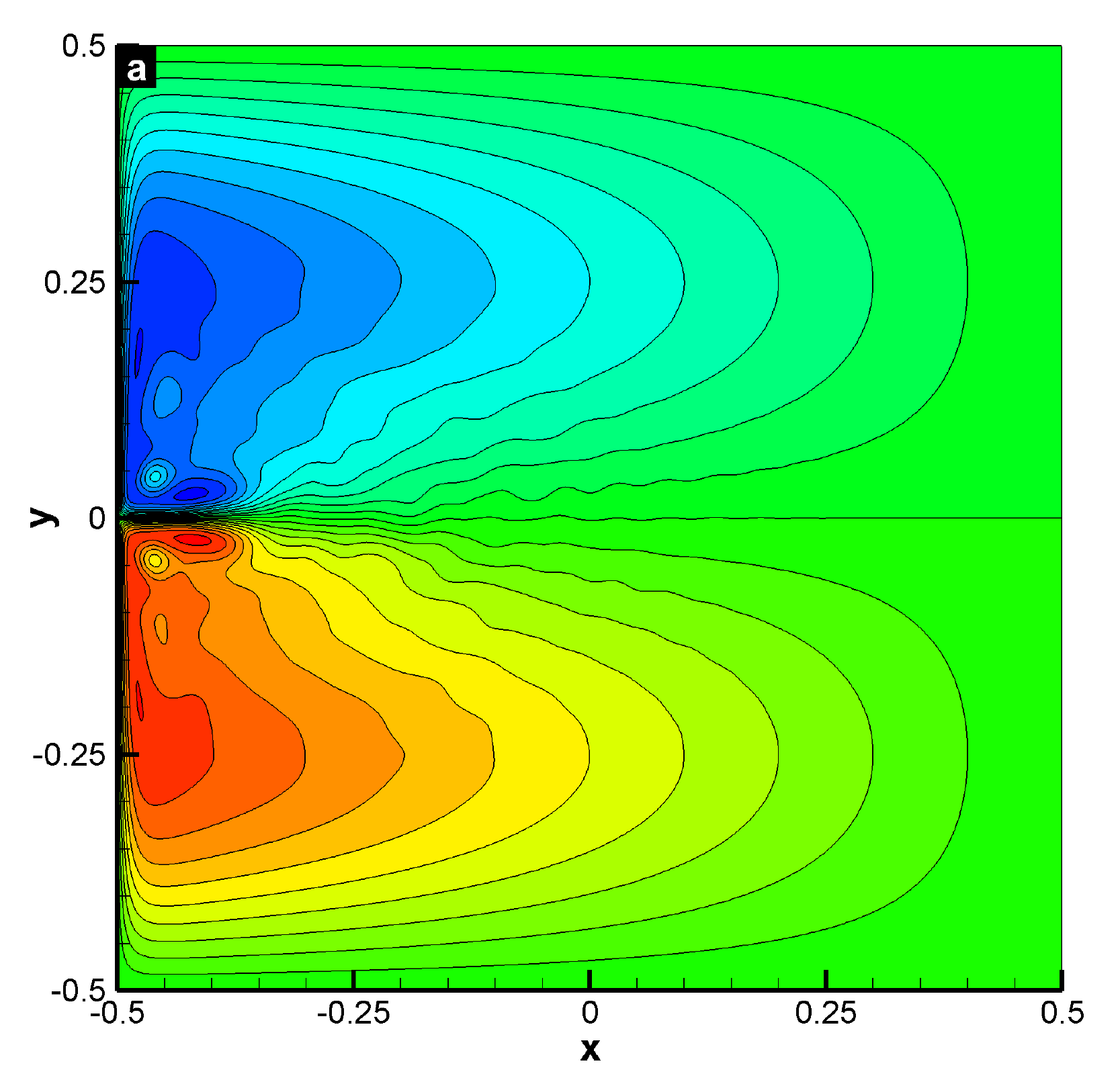}}
\subfigure{\includegraphics[width=0.33\textwidth]{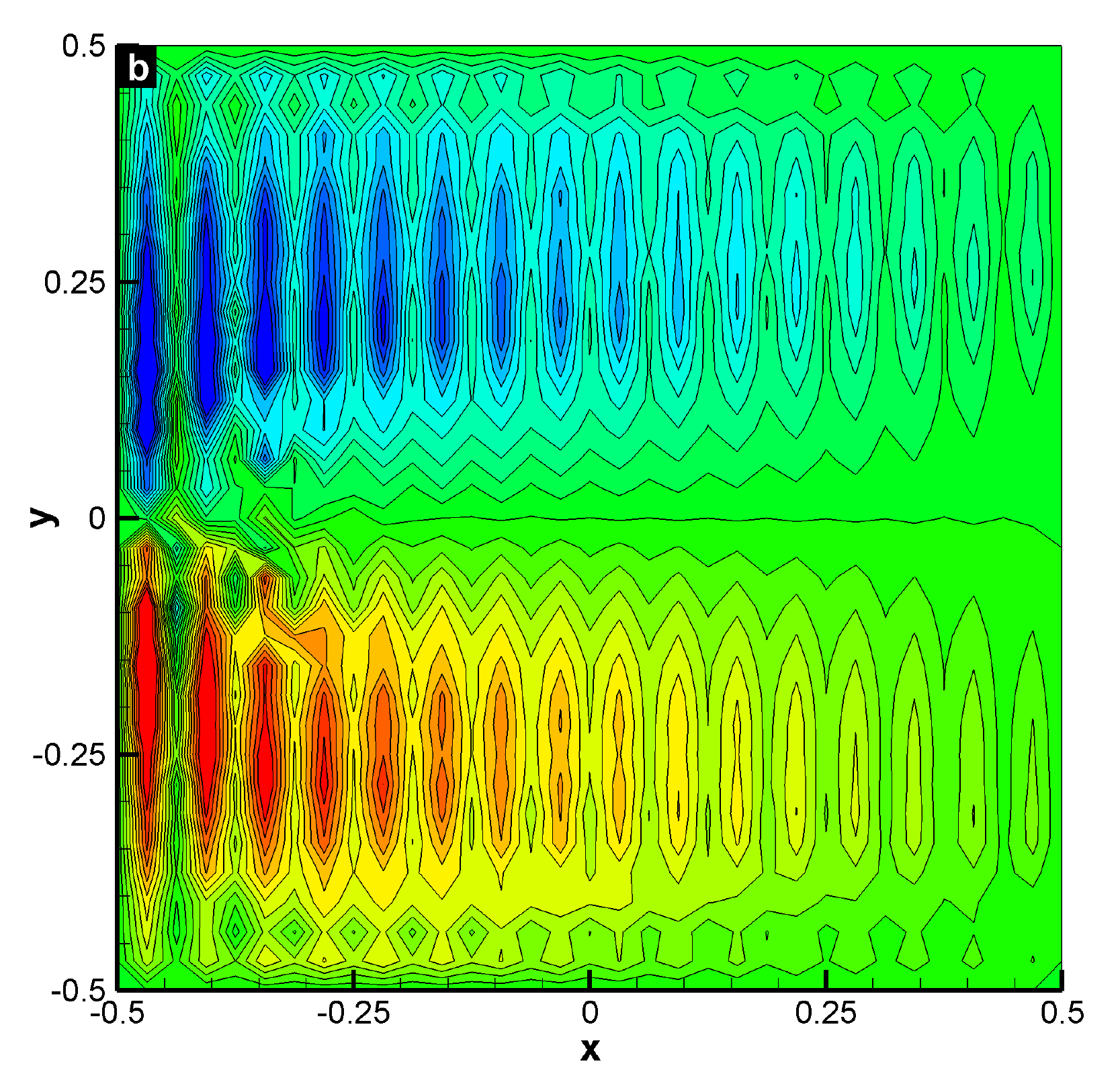}}
\subfigure{\includegraphics[width=0.33\textwidth]{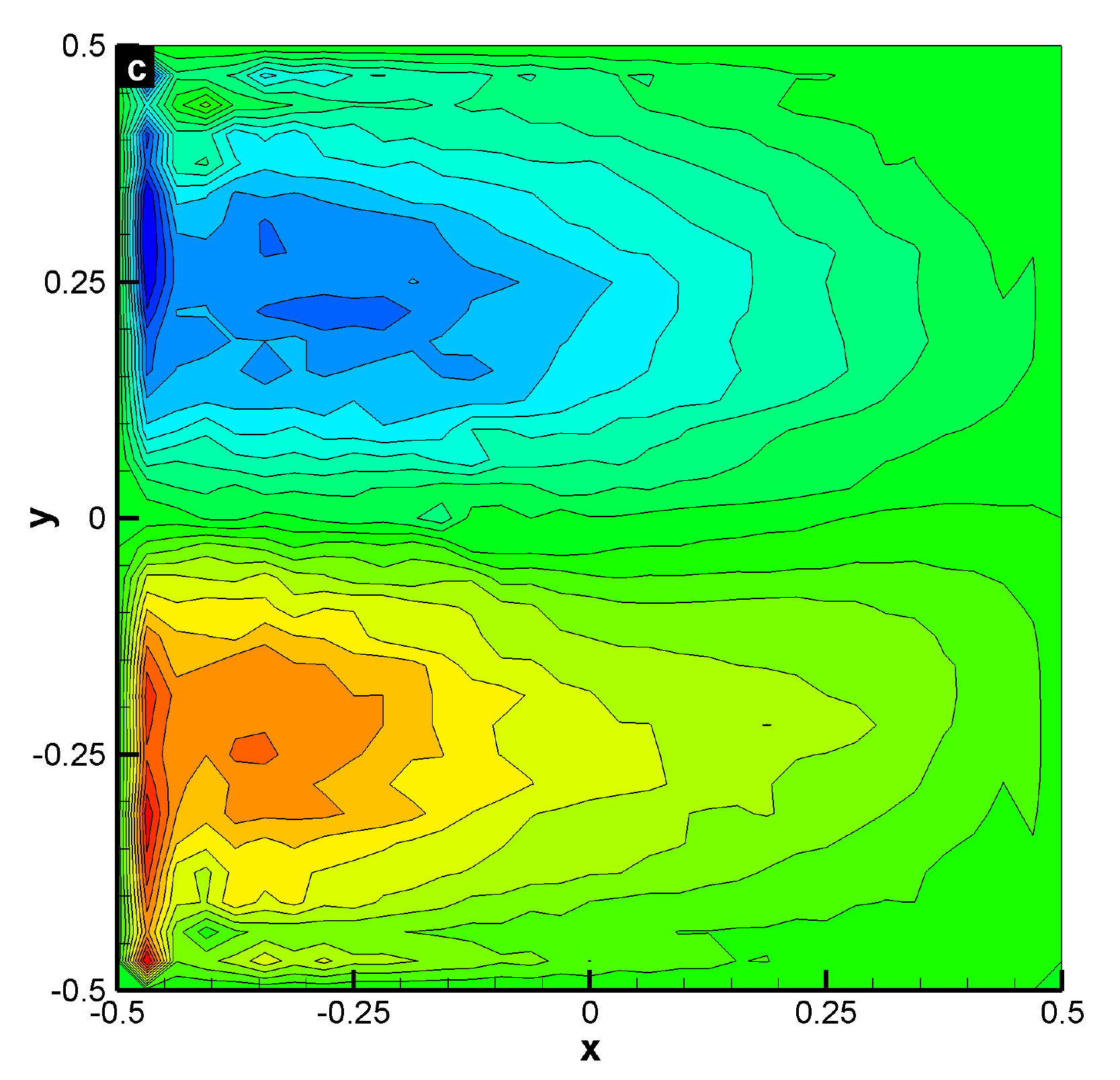}} }
\caption{
Experiment 1: Time-averaged streamfunction contours for the upper layer:
(a) DNS results at a resolution of $512^2$;
(b) QG2$_{c}$ (under-resolved numerical simulation without any subfilter-scale model) results at a resolution of $32^2$; and
(c) AD-TF results at a resolution of  $32^2$.
The contour layouts are identical.
Note that the AD-TF results are significantly better than the QG2$_{c}$ results.
}
\label{fig:E1-s}
\end{figure}

\begin{figure}
\centering
\mbox{
\subfigure{\includegraphics[width=0.33\textwidth]{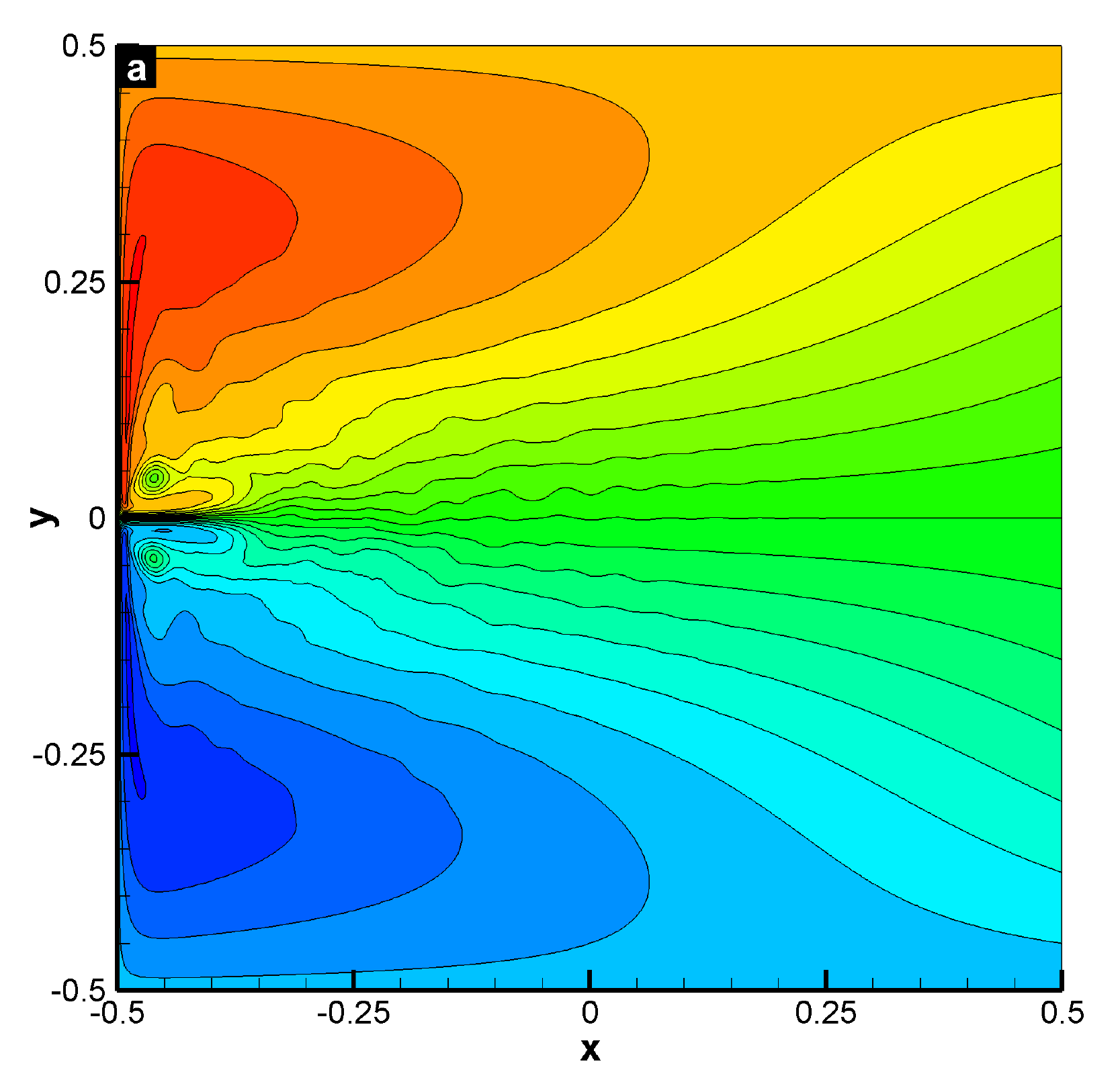}}
\subfigure{\includegraphics[width=0.33\textwidth]{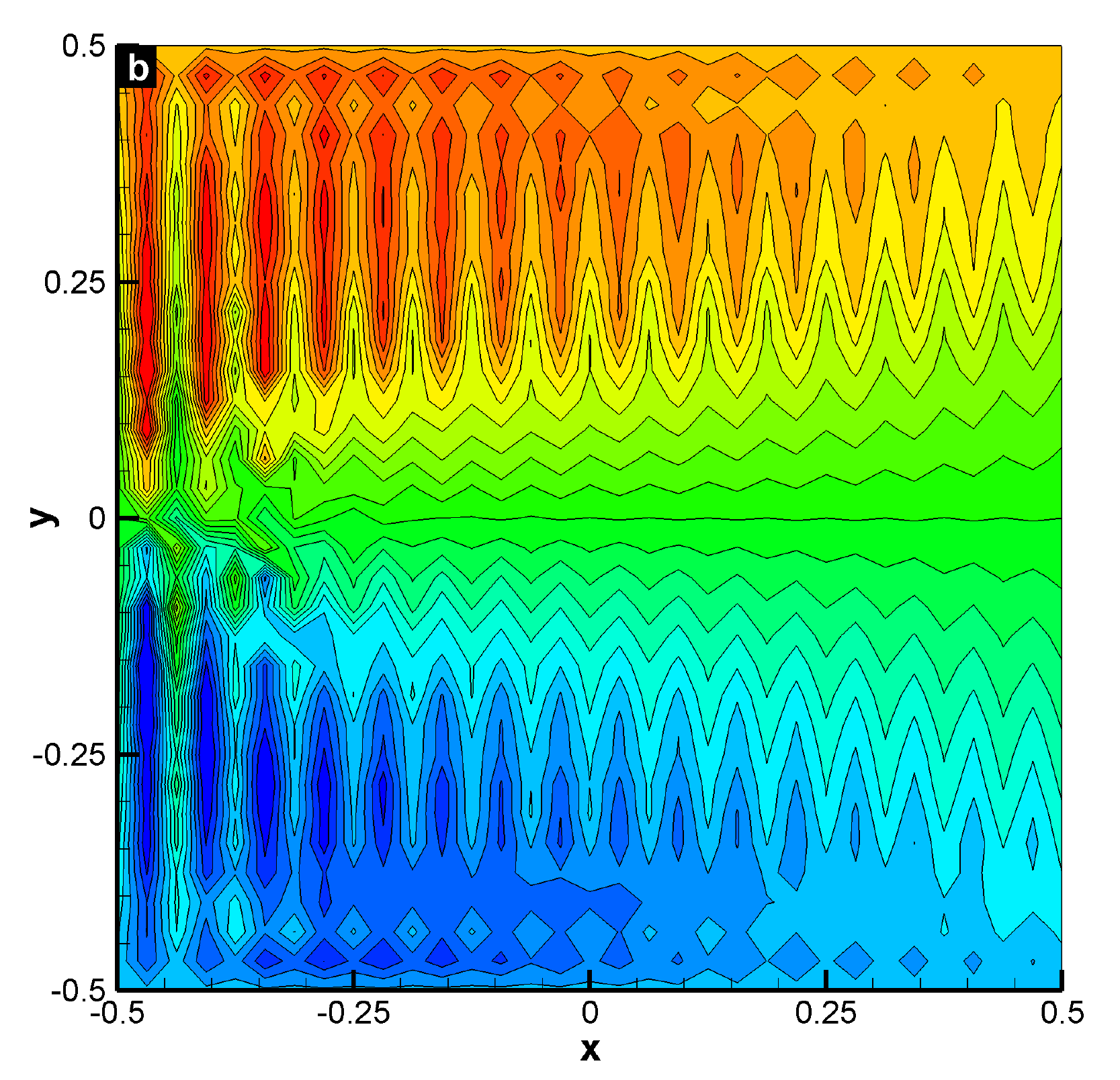}}
\subfigure{\includegraphics[width=0.33\textwidth]{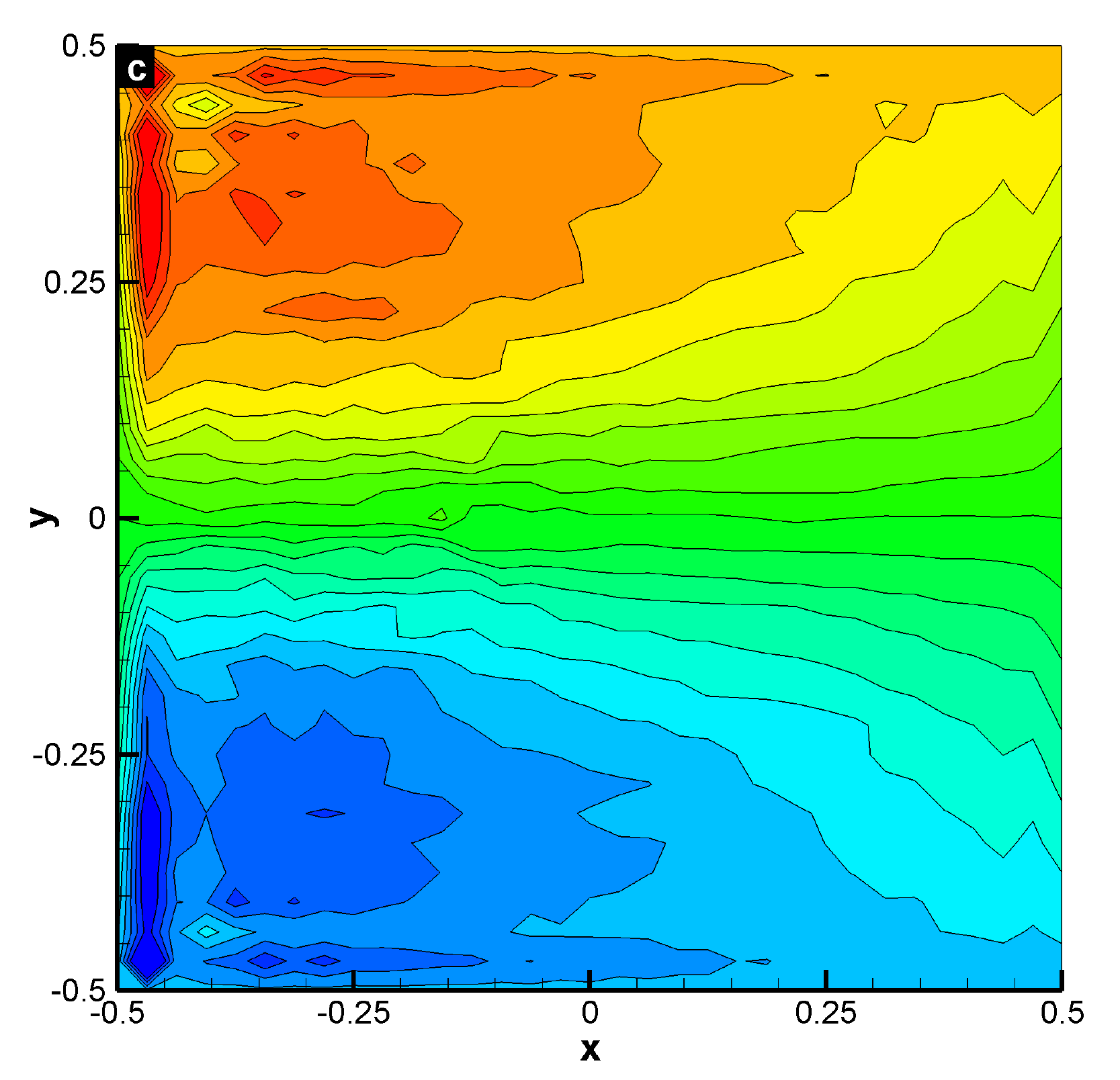}} }
\caption{
Experiment 1: Time-averaged potential vorticity contours for the upper layer:
(a) DNS results at a resolution of $512^2$;
(b) QG2$_{c}$ (under-resolved numerical simulation without any subfilter-scale model) results at a resolution of $32^2$; and
(c) AD-TF results at a resolution of  $32^2$.
The contour layouts are identical.
Note that the AD-TF results are significantly better than the QG2$_{c}$ results.
}
\label{fig:E1-q}
\end{figure}

\begin{figure}
\centering
\mbox{
\subfigure{\includegraphics[width=0.33\textwidth]{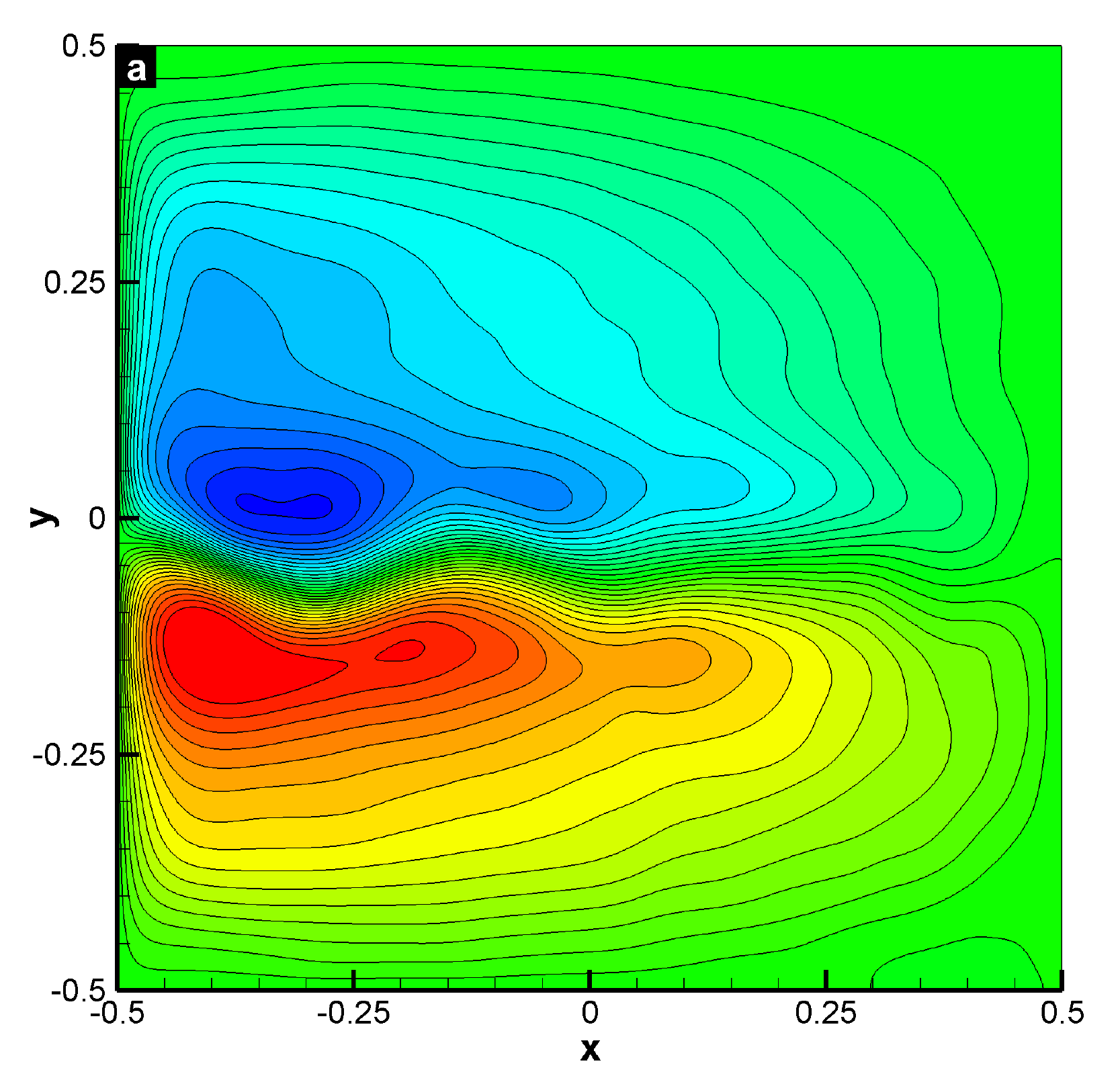}}
\subfigure{\includegraphics[width=0.33\textwidth]{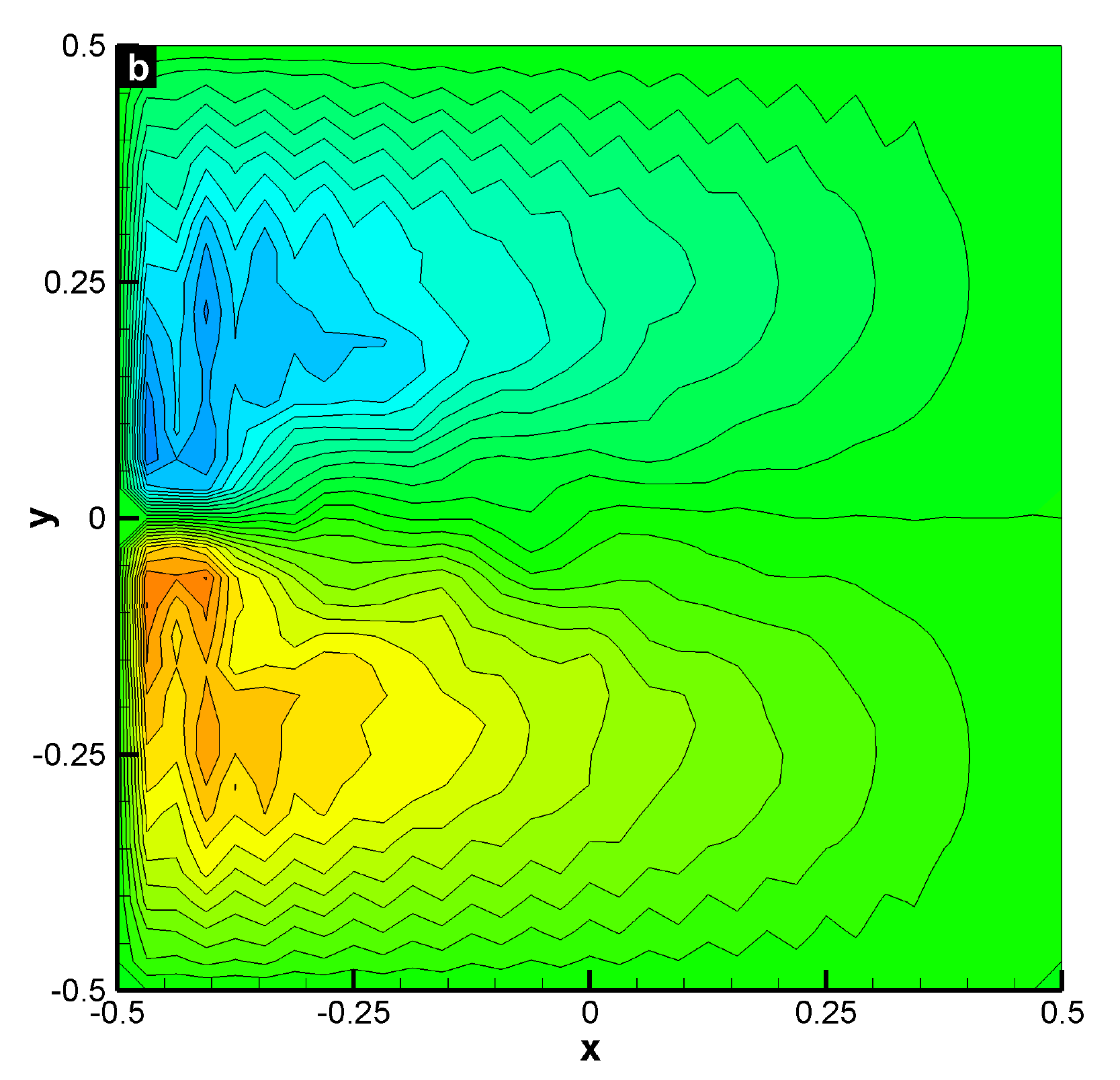}}
\subfigure{\includegraphics[width=0.33\textwidth]{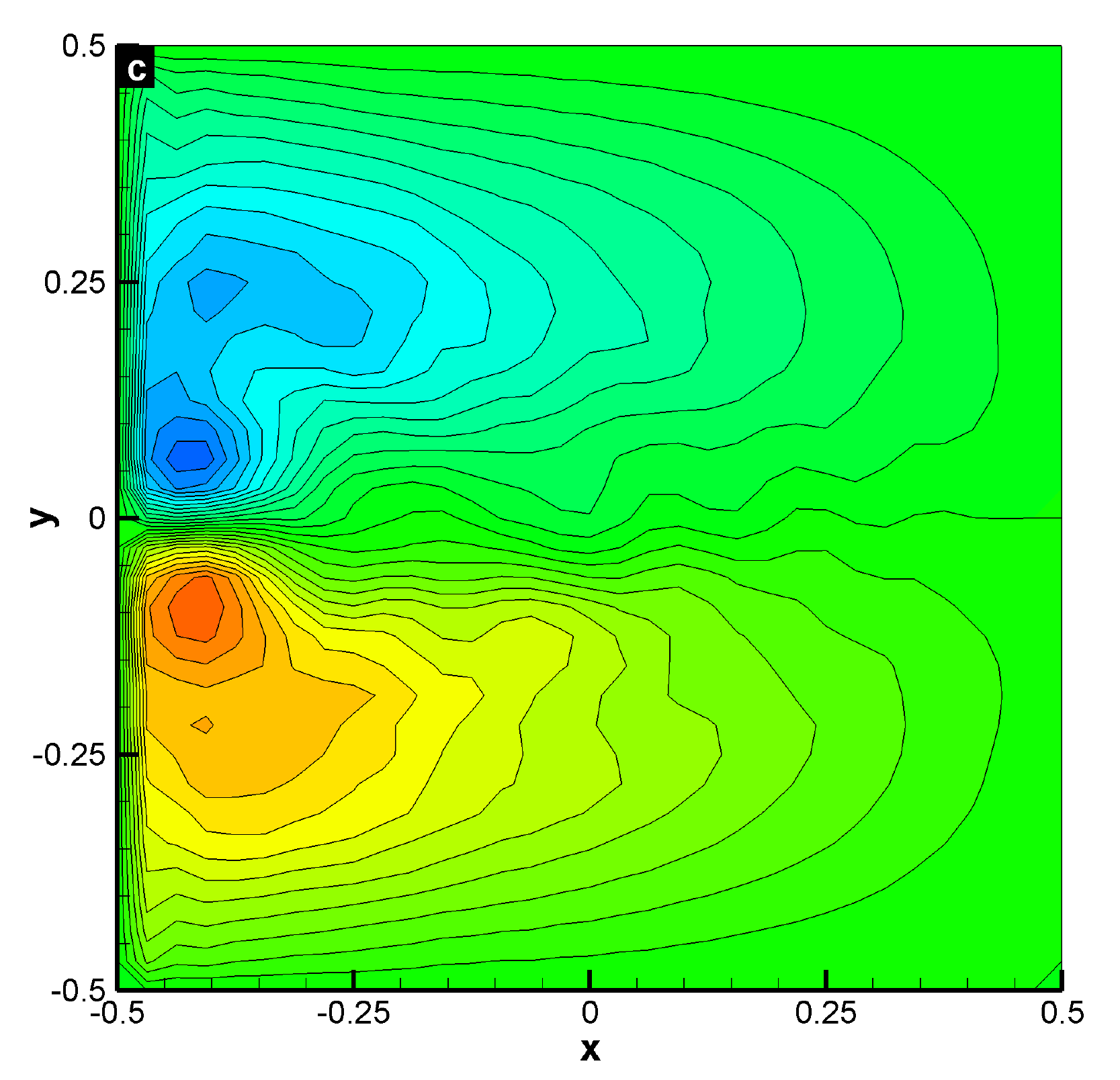}} }
\caption{
Experiment 2: Time-averaged streamfunction contours for the upper layer:
(a) DNS results at a resolution of $512^2$;
(b) QG2$_{c}$ (under-resolved numerical simulation without any subfilter-scale model) results at a resolution of $32^2$; and
(c) AD-TF results at a resolution of  $32^2$.
The contour layouts are identical.
Note that the AD-TF results are significantly better than the QG2$_{c}$ results.
}
\label{fig:E2-s}
\end{figure}

\begin{figure}
\centering
\mbox{
\subfigure{\includegraphics[width=0.33\textwidth]{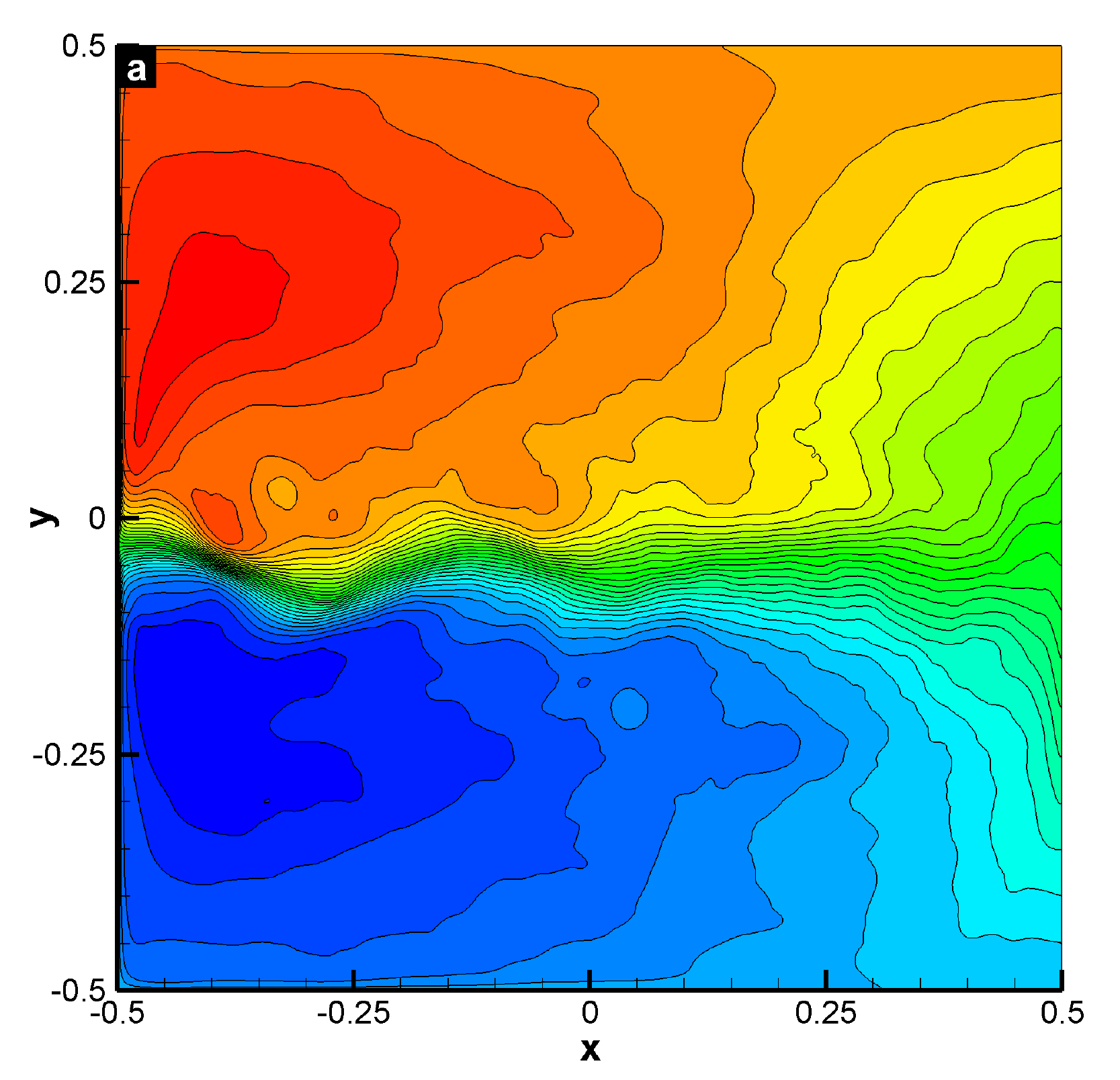}}
\subfigure{\includegraphics[width=0.33\textwidth]{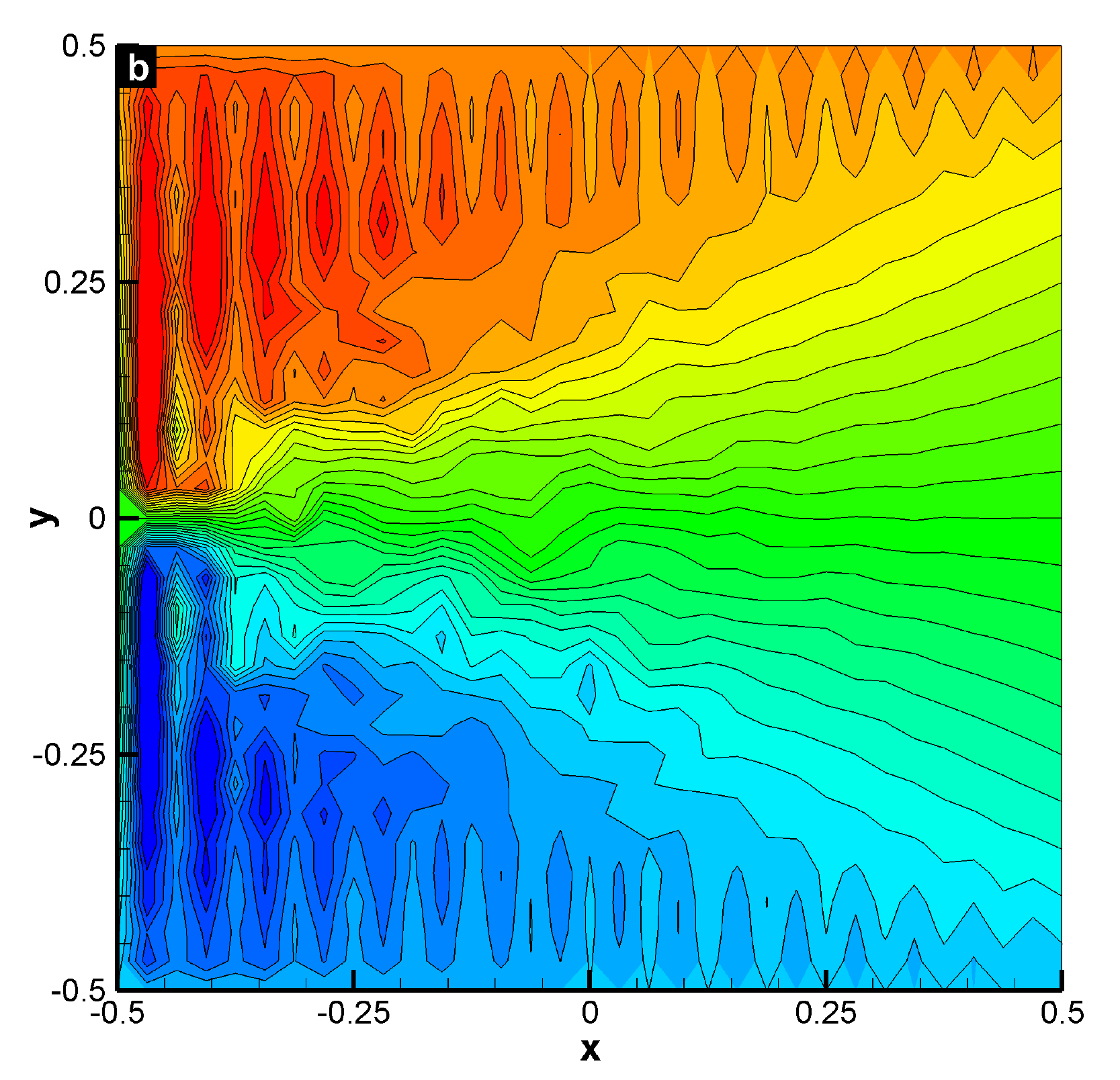}}
\subfigure{\includegraphics[width=0.33\textwidth]{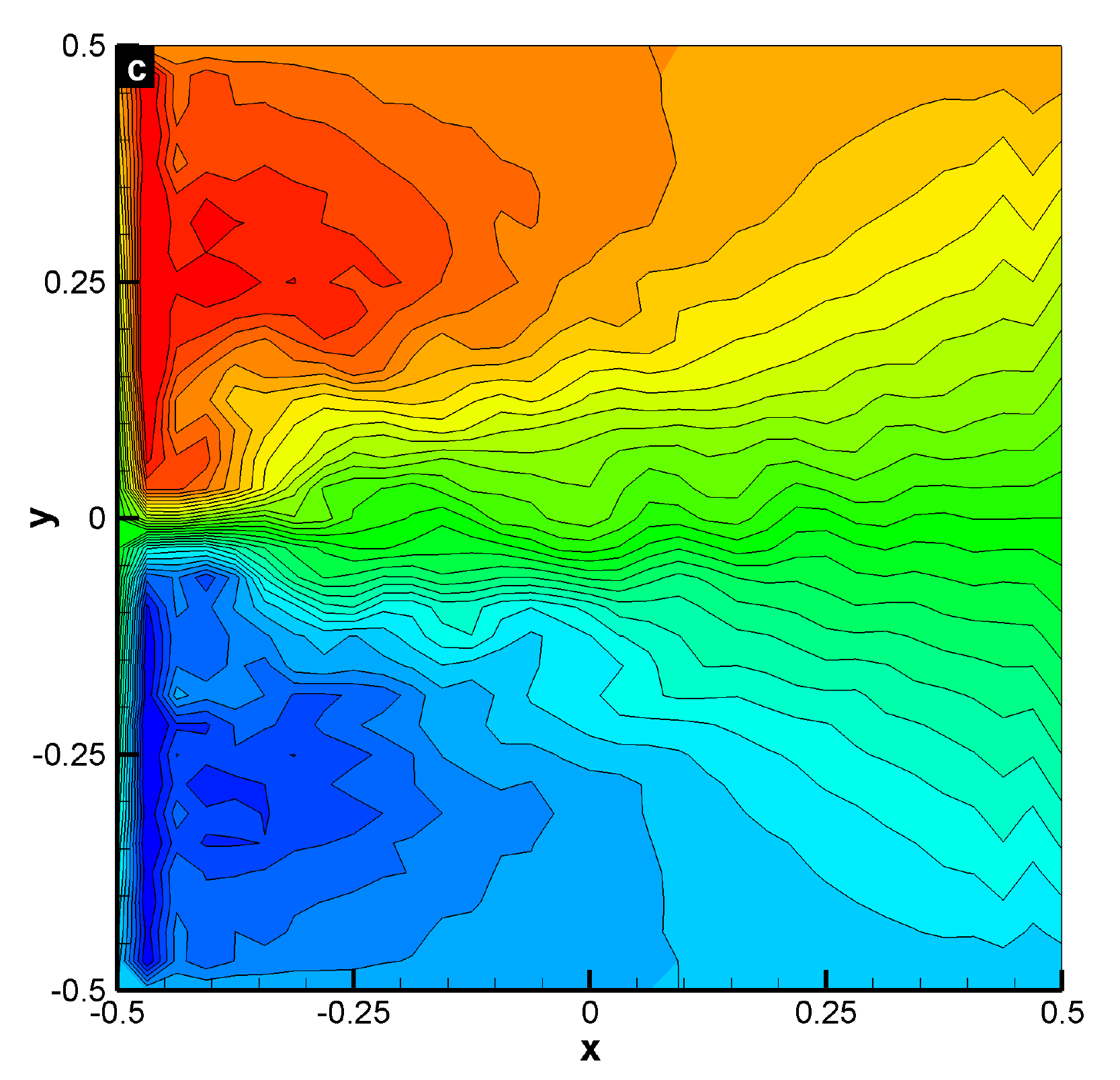}} }
\caption{
Experiment 2: Time-averaged potential vorticity contours for the upper layer:
(a) DNS results at a resolution of $512^2$;
(b) QG2$_{c}$ (under-resolved numerical simulation without any subfilter-scale model) results at a resolution of $32^2$; and
(c) AD-TF results at a resolution of  $32^2$.
The contour layouts are identical.
Note that the AD-TF results are significantly better than the QG2$_{c}$ results.
}
\label{fig:E2-q}
\end{figure}

For Experiment 1, we plot the mean streamfunction and potential vorticity contours in Figs.~\ref{fig:E1-s} and~\ref{fig:E1-q}, respectively. The new AD-TF model yields results that are significantly better than those corresponding to the under-resolved QG2$_{c}$ run. Similarly, we plot the mean streamfunction and potential vorticity contours in Figs.~\ref{fig:E2-s} and~\ref{fig:E2-q} for Experiment 2. We note that the proposed AD-TF model yields again improved results by smoothing out the numerical oscillations present in the under-resolved QG2$_{c}$ simulations.
We also note that the computational cost of the new AD-TF model is significantly lower than that of the DNS, and is comparable to the computational cost of the QG2$_{c}$.
Indeed, the CPU time is $119.6$ hrs. for the DNS, $141.8$ secs. for QG2$_{c}$, and $174.7$ secs. for the AD-TF model.
The numerical results for both experiments clearly suggest that the
the AD-TF model can provide relatively accurate results for under-resolved geophysical flows at a low computational cost.


\begin{figure}[!t]
\centering
\mbox{
\subfigure{\includegraphics[width=0.33\textwidth]{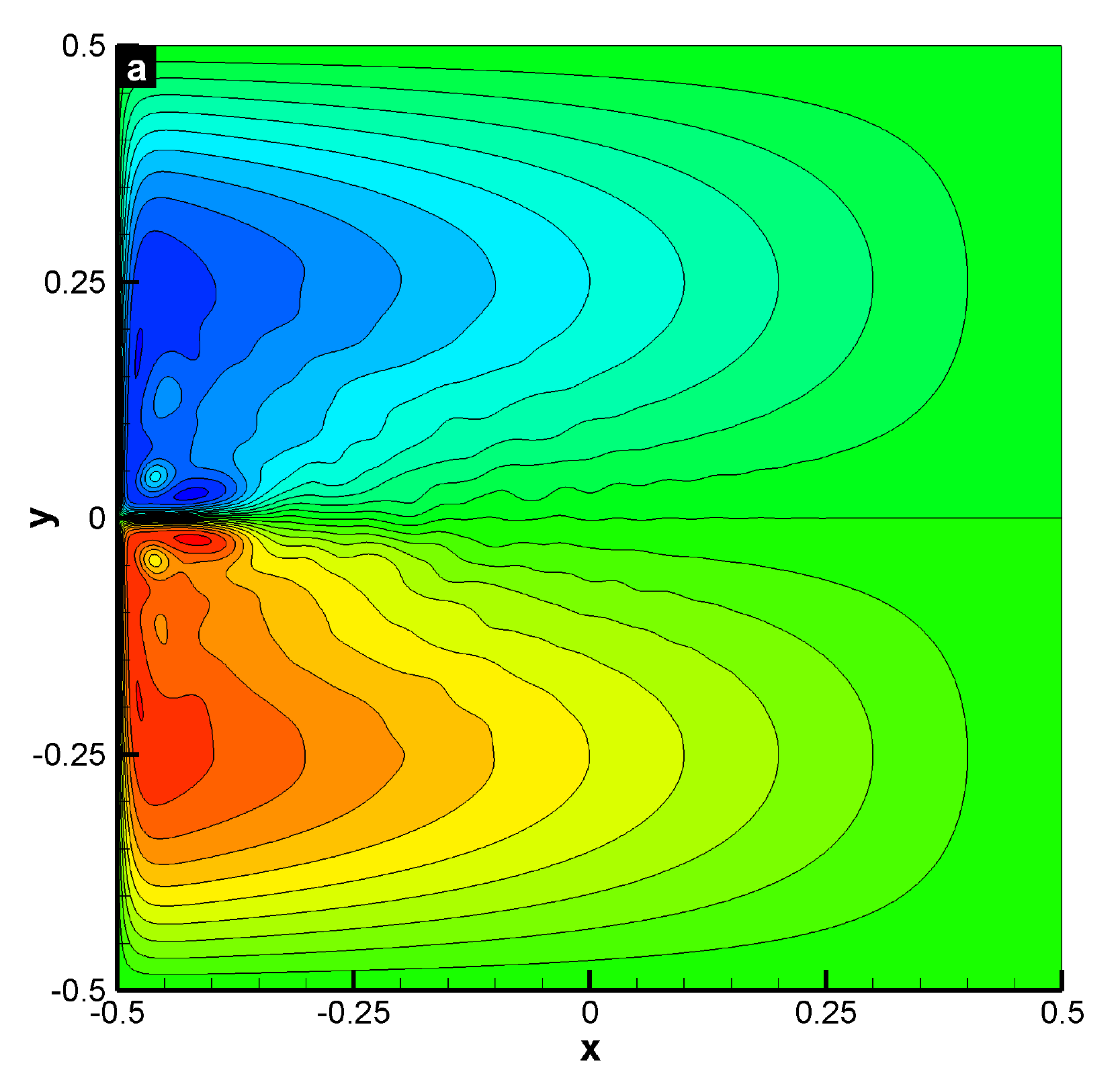}}
\subfigure{\includegraphics[width=0.33\textwidth]{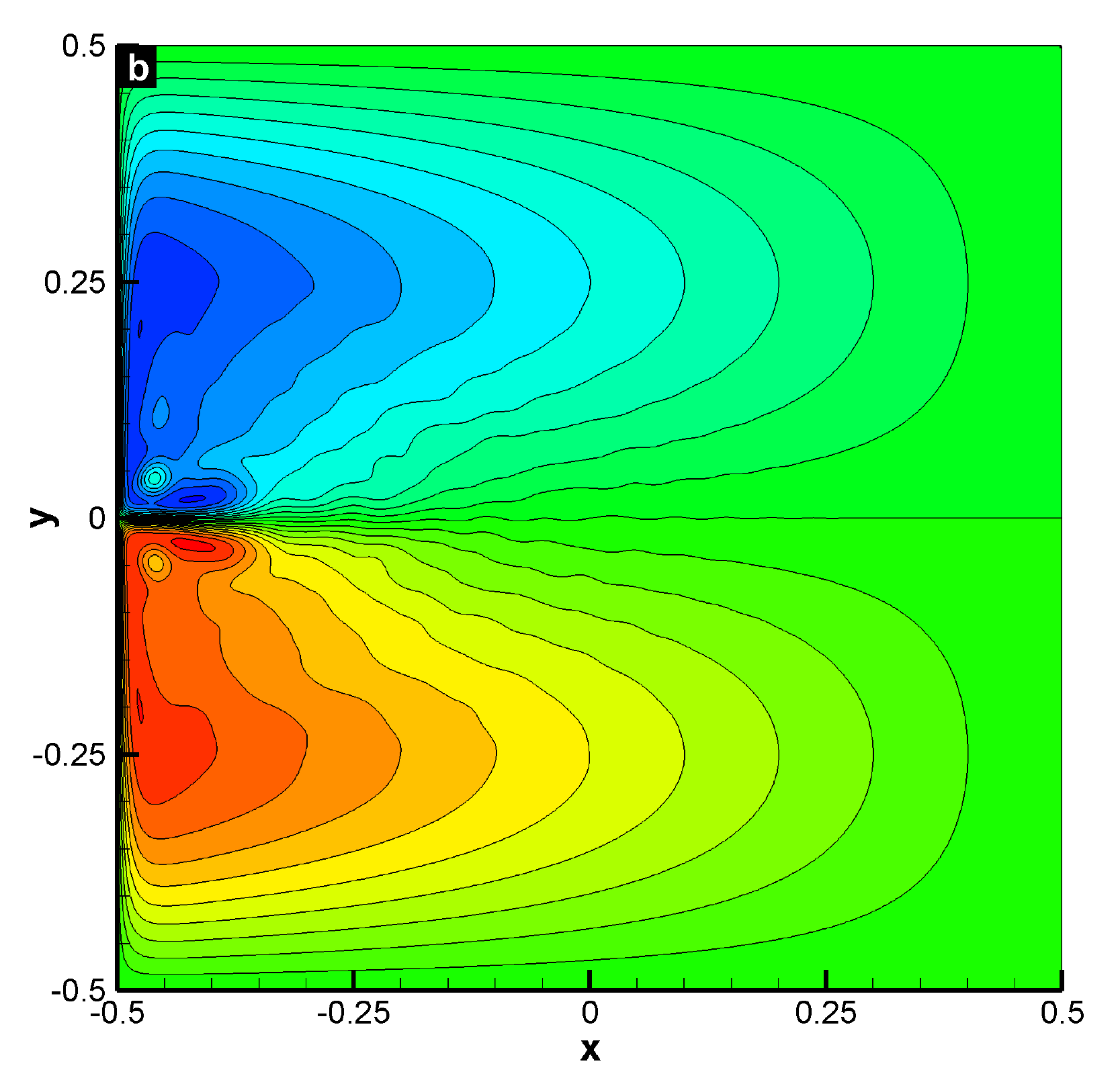}}
\subfigure{\includegraphics[width=0.33\textwidth]{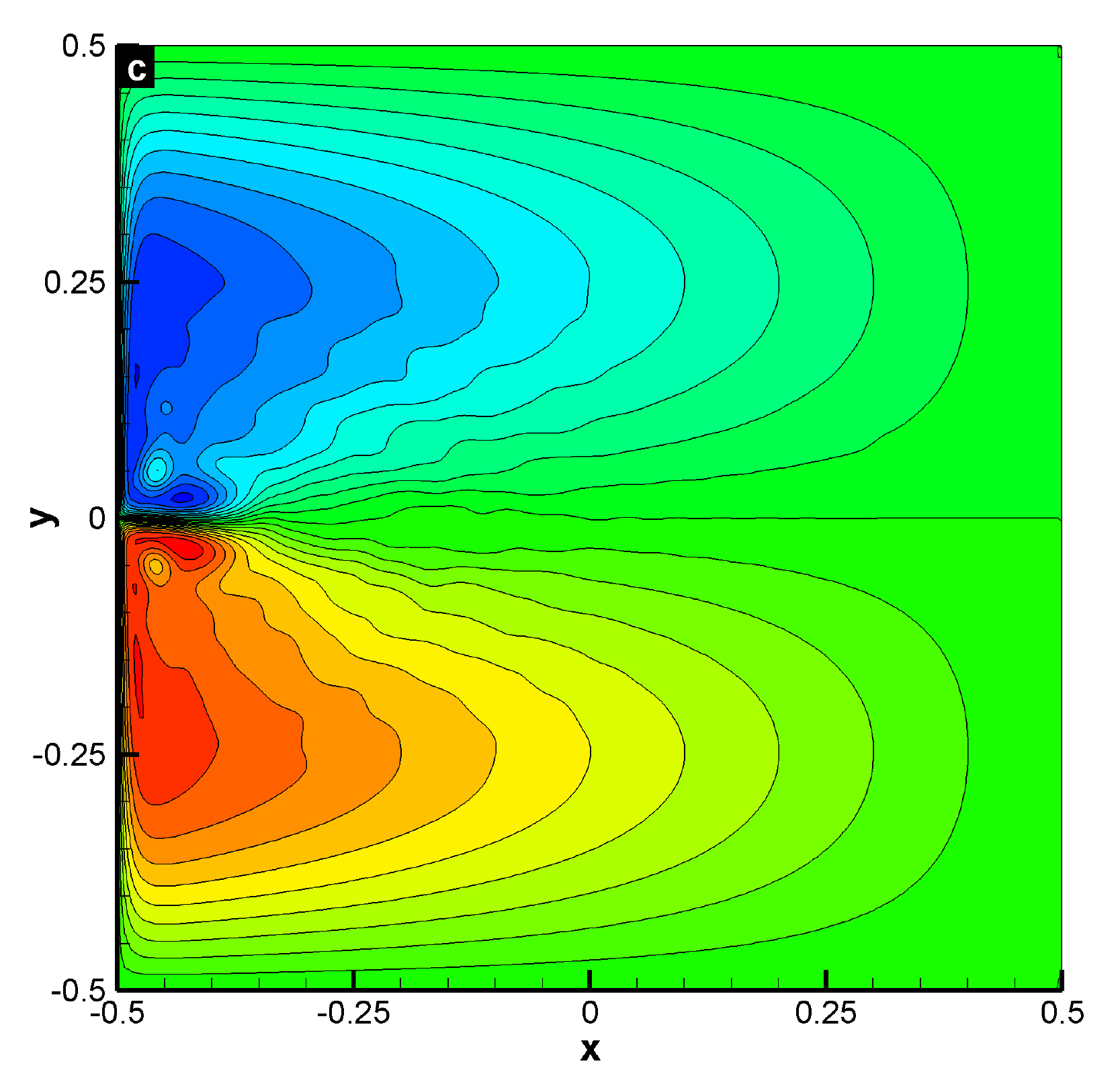}} }
\mbox{
\subfigure{\includegraphics[width=0.33\textwidth]{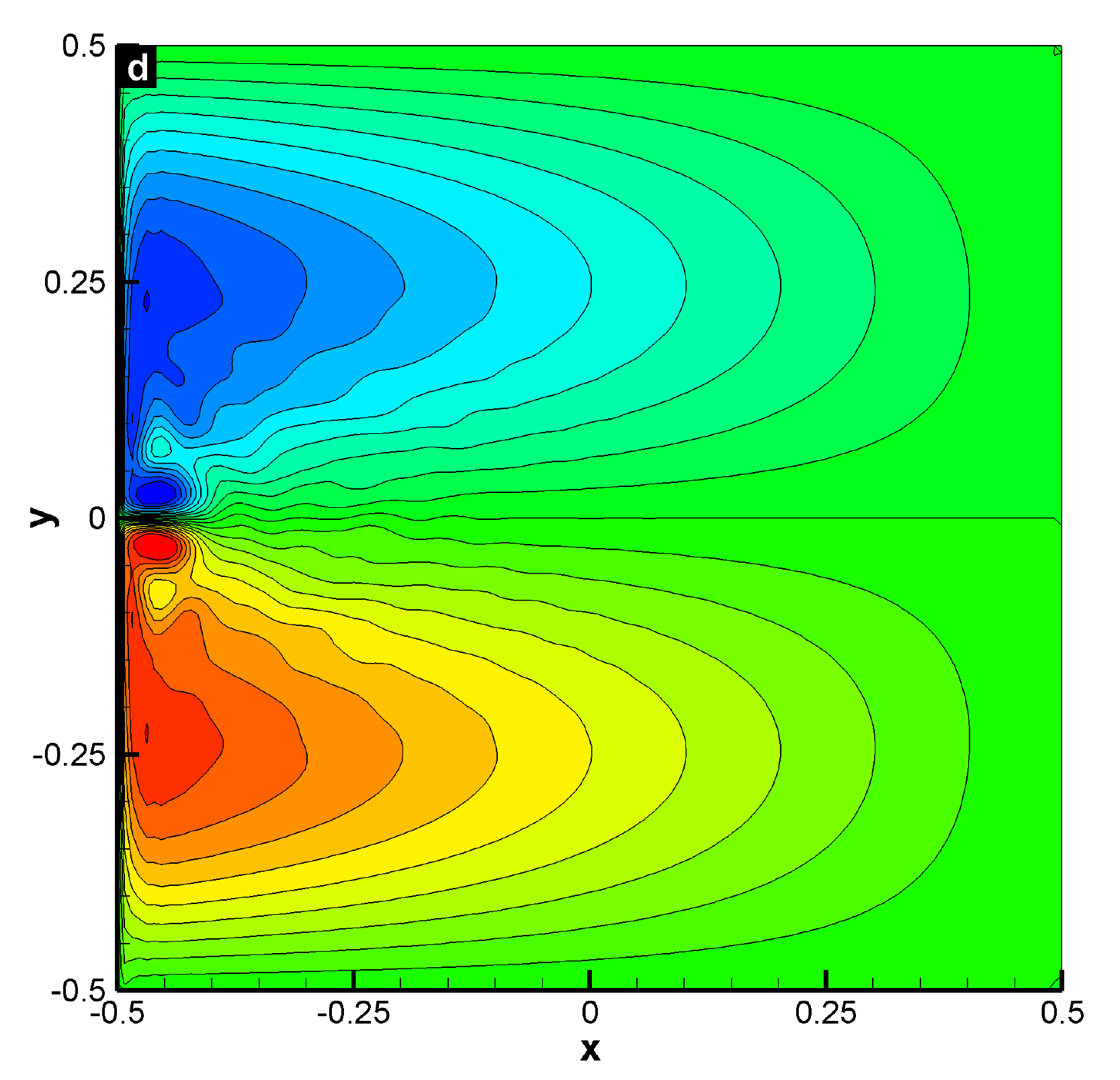}}
\subfigure{\includegraphics[width=0.33\textwidth]{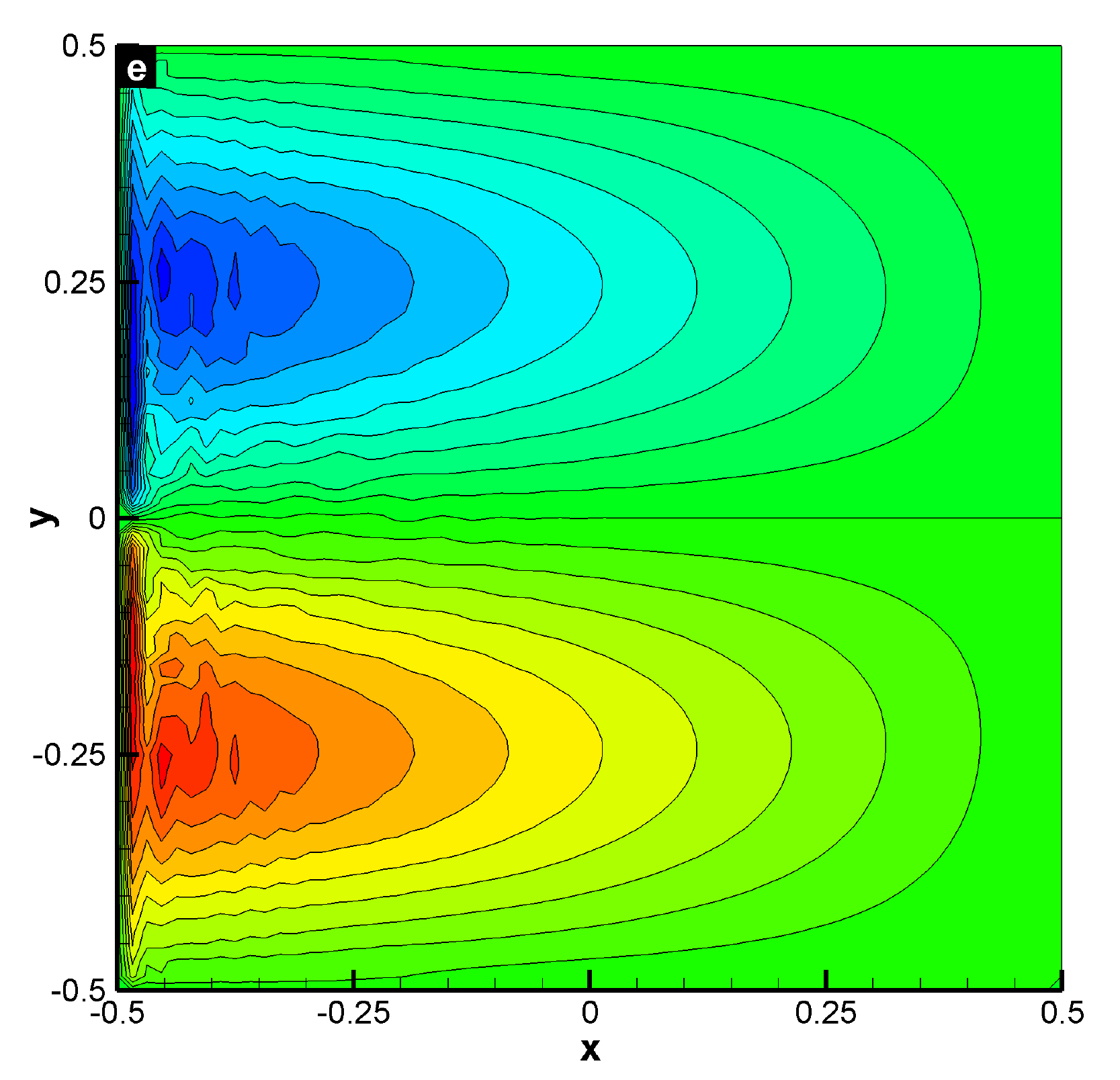}}
\subfigure{\includegraphics[width=0.33\textwidth]{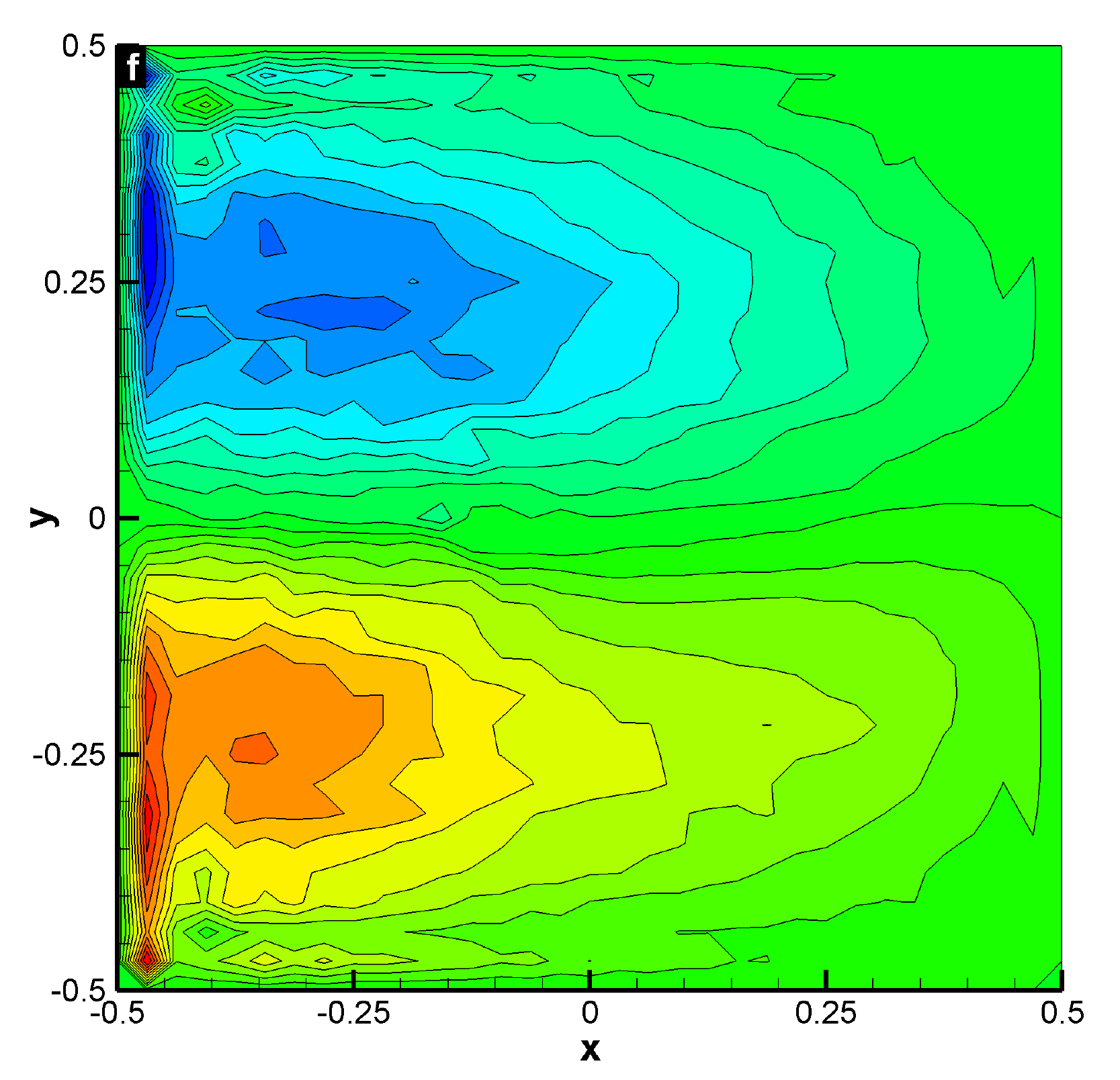}} }
\caption{
Experiment 1: Time-averaged streamfunction for the upper layer:
(a) DNS results at a resolution of $512^2$;
(b) AD-TF results at a resolution of  $512^2$;
(c) AD-TF results at a resolution of  $256^2$;
(d) AD-TF results at a resolution of  $128^2$;
(e) AD-TF results at a resolution of  $64^2$; and
(f) AD-TF results at a resolution of  $32^2$.
The contour layouts are identical.
Note:
(i) the accuracy of the AD-TF results at the $128^2$ resolution; and
(ii) the consistency of the AD-TF results with respect to the mesh size.
}
\label{fig:E1-s-g}
\end{figure}

\begin{figure}[!t]
\centering
\mbox{
\subfigure{\includegraphics[width=0.33\textwidth]{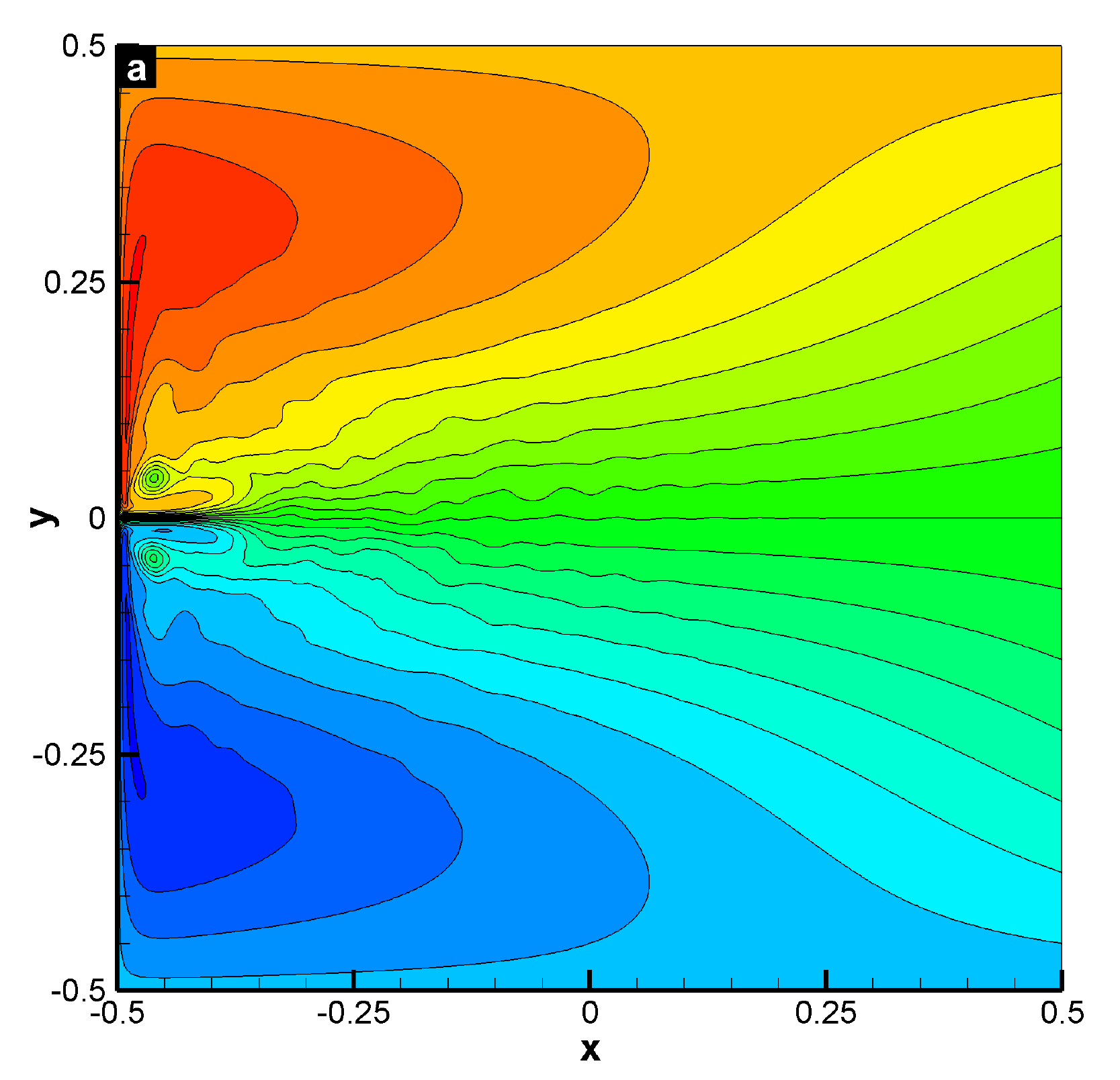}}
\subfigure{\includegraphics[width=0.33\textwidth]{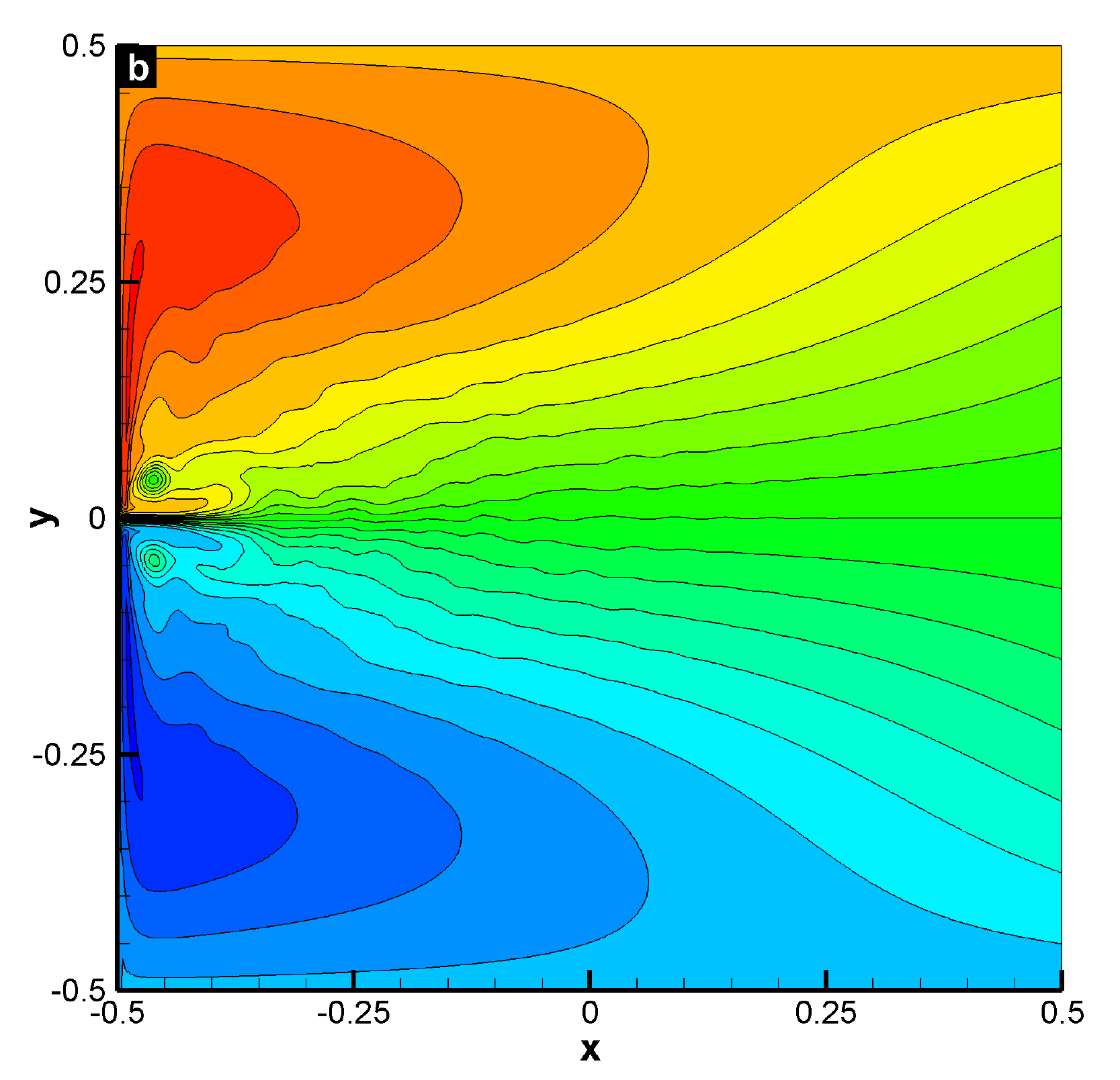}}
\subfigure{\includegraphics[width=0.33\textwidth]{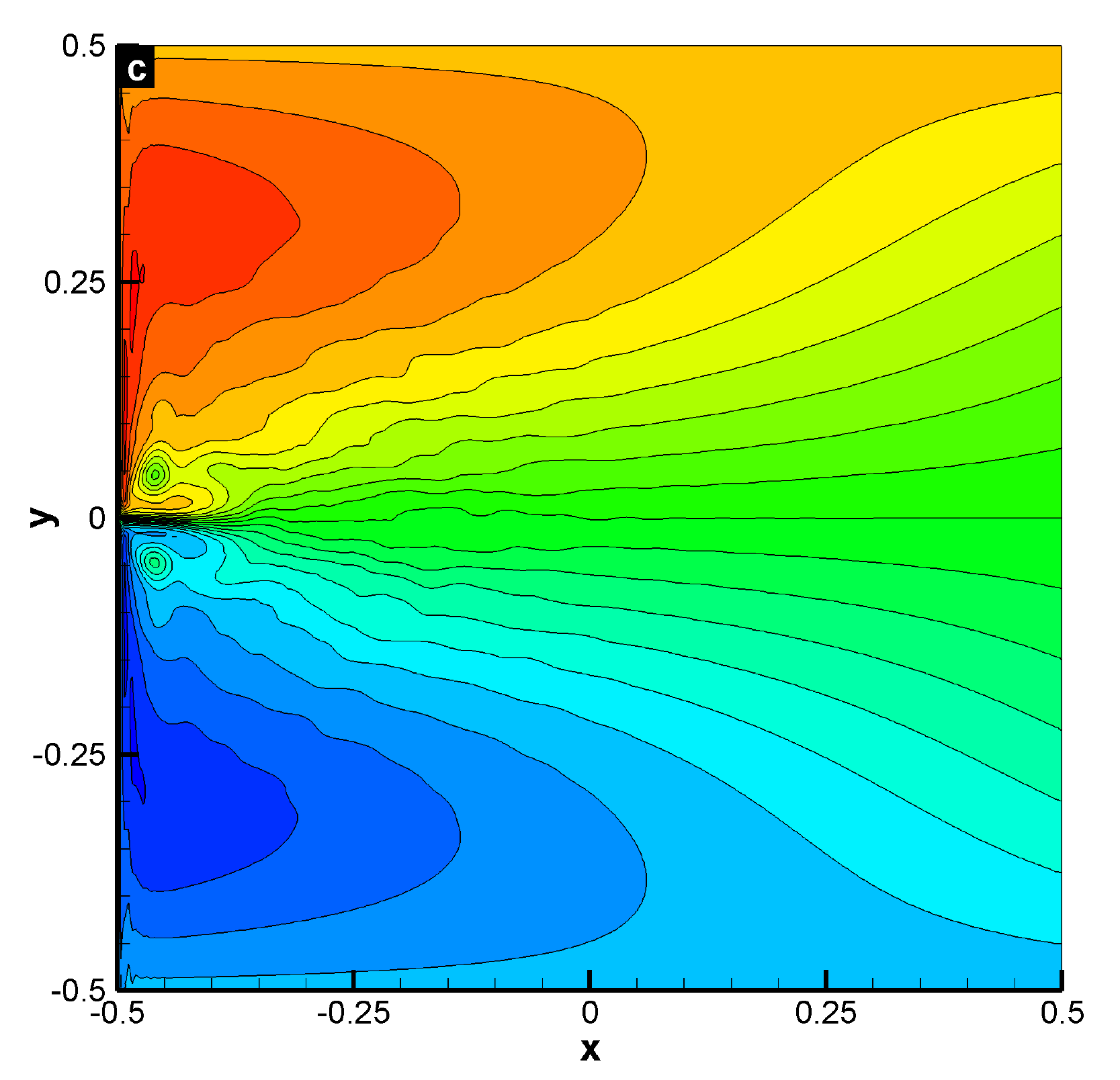}} }
\mbox{
\subfigure{\includegraphics[width=0.33\textwidth]{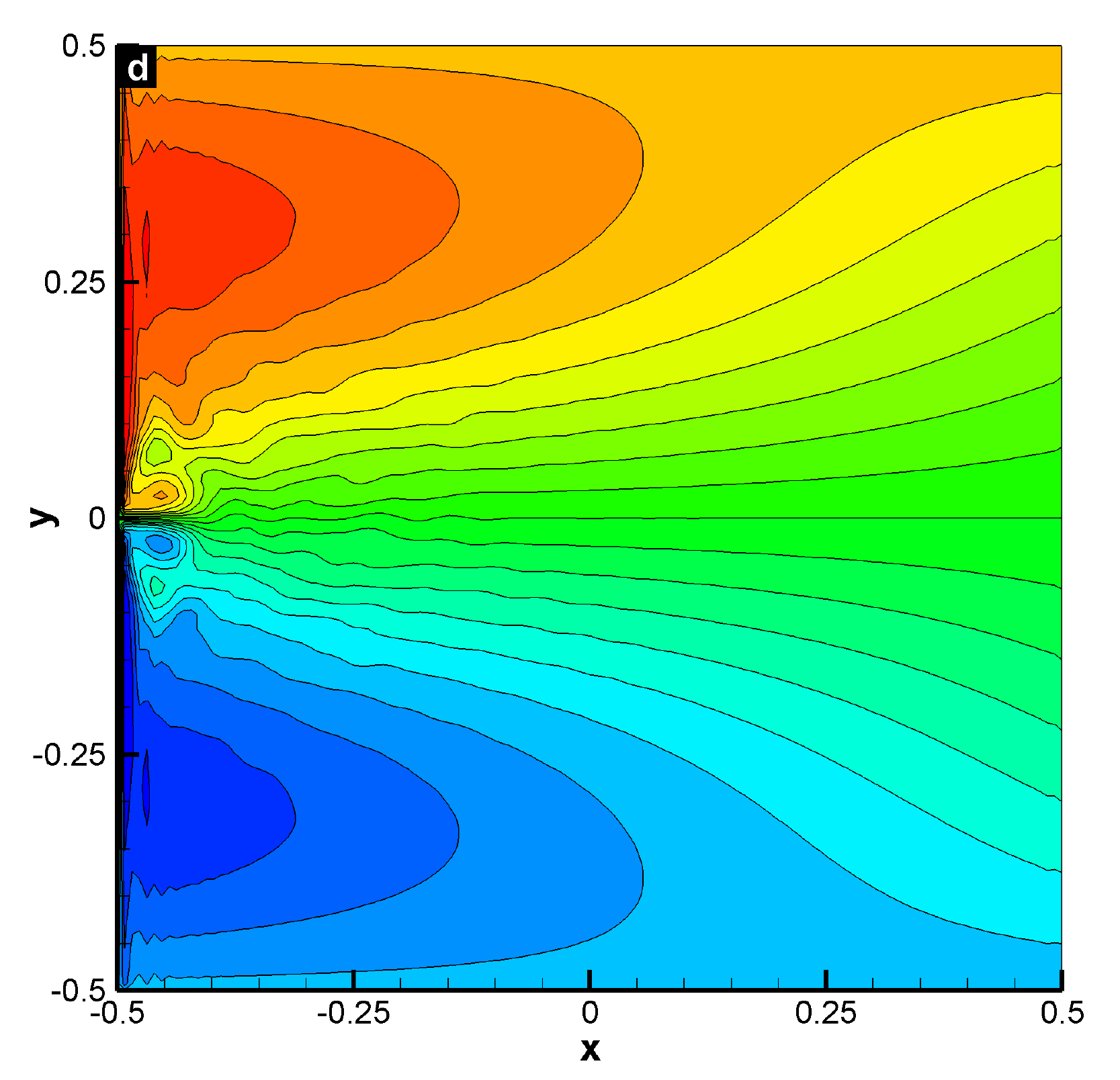}}
\subfigure{\includegraphics[width=0.33\textwidth]{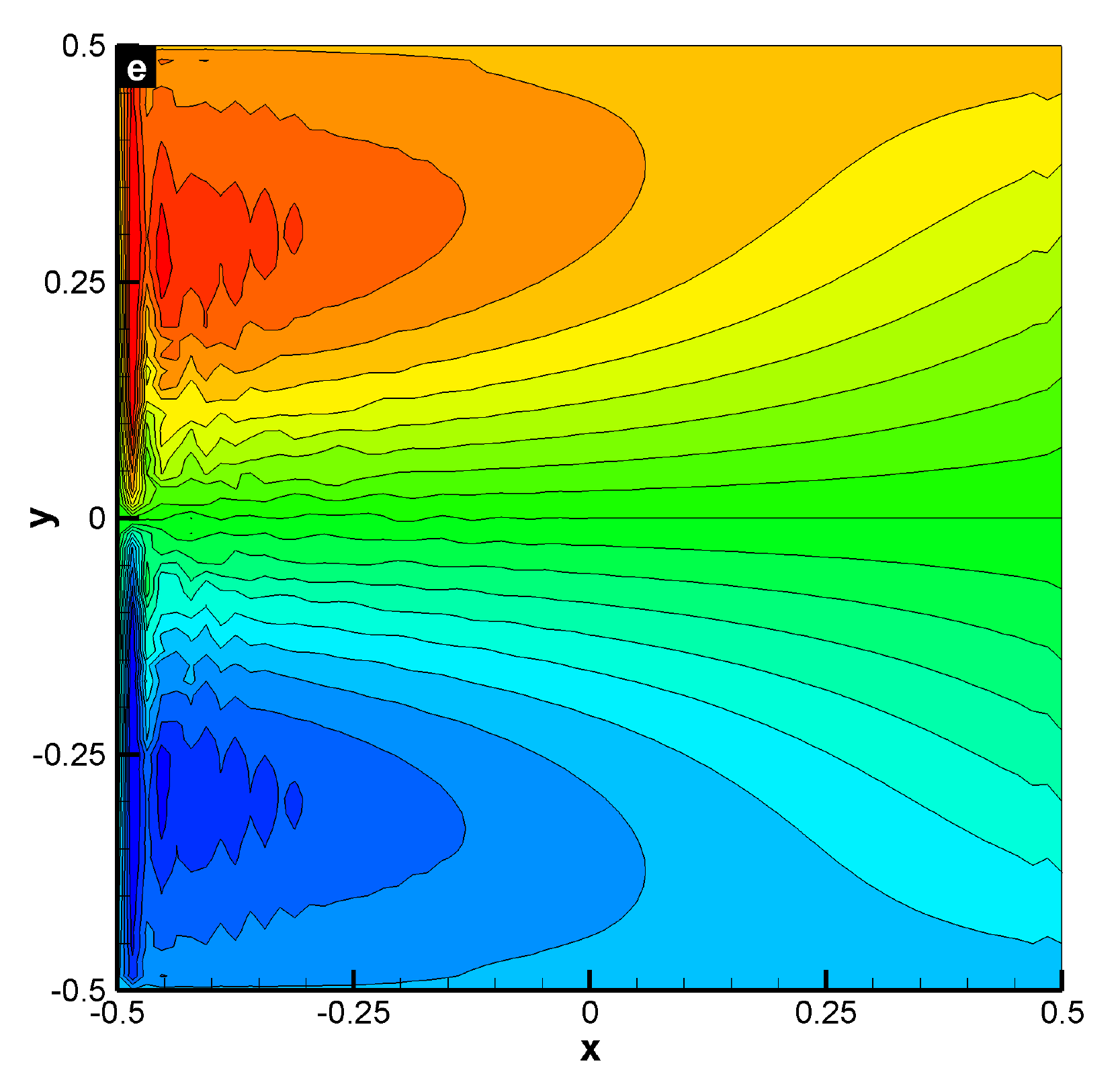}}
\subfigure{\includegraphics[width=0.33\textwidth]{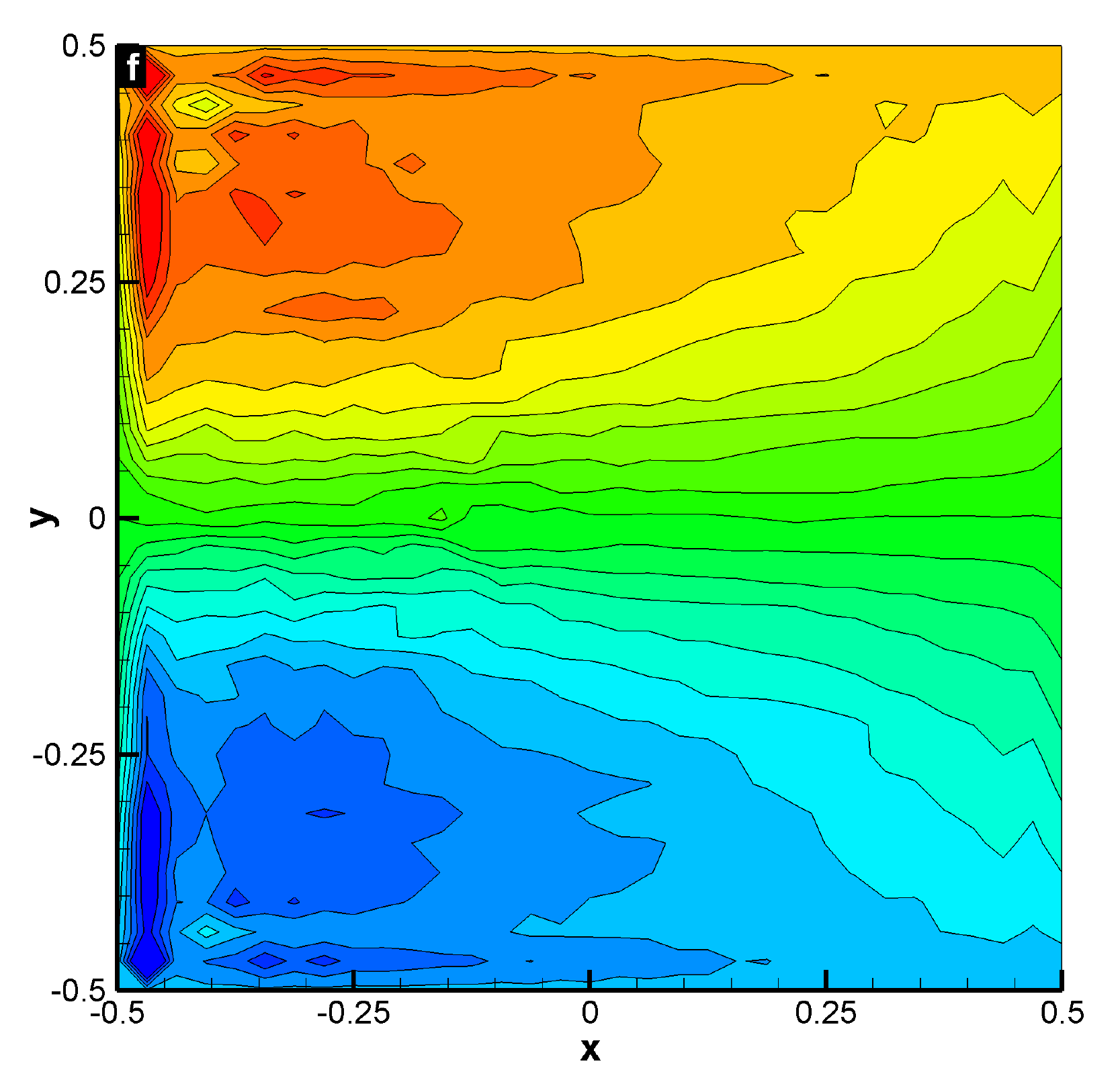}} }
\caption{
Experiment 1: Time-averaged potential vorticity for the upper layer:
(a) DNS results at a resolution of $512^2$;
(b) AD-TF results at a resolution of  $512^2$;
(c) AD-TF results at a resolution of  $256^2$;
(d) AD-TF results at a resolution of  $128^2$;
(e) AD-TF results at a resolution of  $64^2$; and
(f) AD-TF results at a resolution of  $32^2$.
The contour layouts are identical.
Note:
(i) the accuracy of the AD-TF results at the $128^2$ resolution; and
(ii) the consistency of the AD-TF results with respect to the mesh size.
}
\label{fig:E1-q-g}
\end{figure}

\begin{figure}
\centering
\mbox{
\subfigure[]{\includegraphics[width=0.5\textwidth]{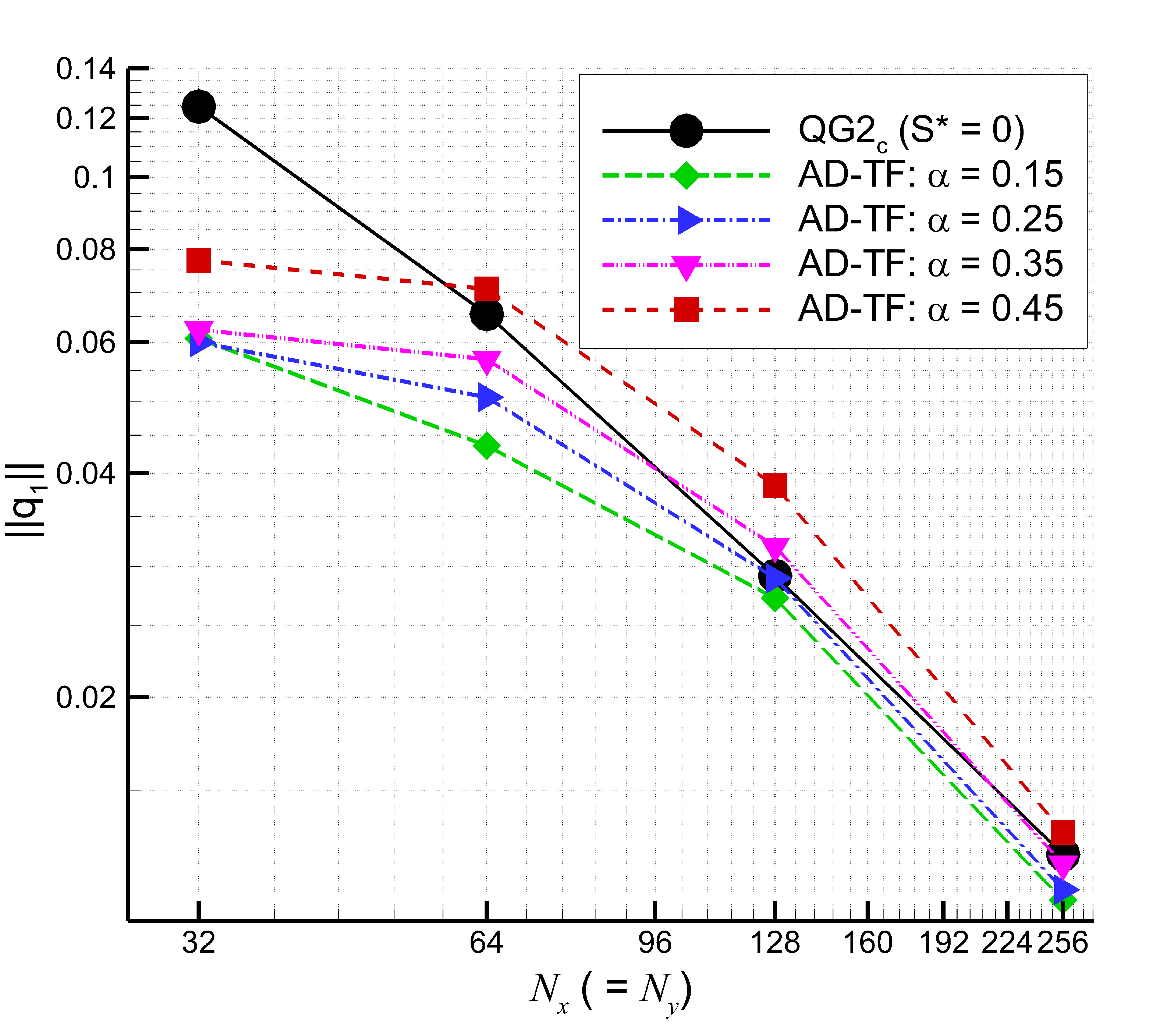}}
\subfigure[]{\includegraphics[width=0.5\textwidth]{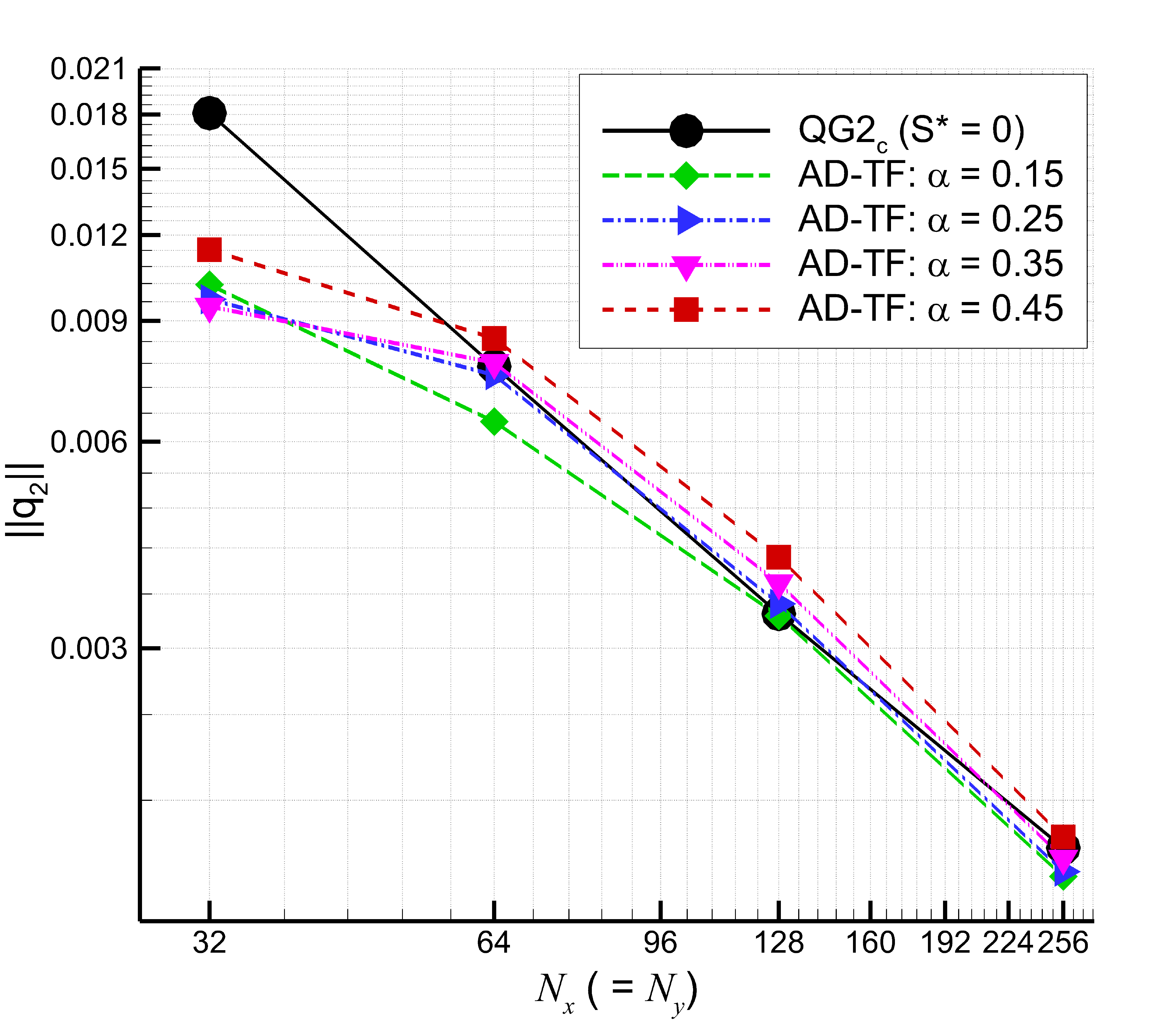}}
}
\caption{Experiment 1:  Log-log plot of the time-averaged $L^2$ norm of the error of the potential vorticity in the two layers, (a) $\| q_1 \|$ and (b) $\| q_2 \|$, for varying parameter $\alpha$ in the AD-TF model.
The results obtained with QG2$_c$ (the under-resolved numerical simulation without any subfilter-scale model) are also included for comparison purposes.
The reference solution used in the computation of the error is the numerical approximation obtained at a grid resolution of $512^2$.}
\label{fig:AD-TF-norms}
\end{figure}

Although the AD-TF model performs well given the coarse mesh utilized,
a natural question is whether we can increase its accuracy by using a finer mesh.
Thus, we investigate the behavior of the AD-TF model for various resolutions: $512^2$, $256^2$, $128^2$, $64^2$, and $32^2$. Since similar conclusions hold for both experiments, we only discuss the results for Experiment 1.
The time-averaged streamfunction and potential vorticity contour plots for the upper layers are shown in Figs.~\ref{fig:E1-s-g} and~\ref{fig:E1-q-g}, respectively.
The qualitative results displayed in these figures are reinforced by the quantitative results in Table~\ref{tab:L2-TF-alpha}, which presents the time-averaged $L^2$ norm of the error of the potential vorticity in the two layers, $\| q_1 \|$ and $\| q_2 \|$, for fixed truncation order, $N=5$, varying grid resolutions, $N_x \times N_y$, and varying free parameter $\alpha$.
These results are also compared graphically in Fig.~\ref{fig:AD-TF-norms}.
The CPU time is 296 hrs for the DNS results, 48.5 hrs for the $256^2$ resolution, 4.1 hrs for the $128^2$ resolution, 0.34 hrs for the $64^2$ resolution, and 2.9 mins for the $32^2$ resolution.
The main conclusion that can be drawn from the plots in Figs.~\ref{fig:E1-s-g},~\ref{fig:E1-q-g},~\ref{fig:AD-TF-norms}, Table~\ref{tab:L2-TF-alpha}, and the computational efficiency study is that at the lowest resolutions the AD-TF model achieves a very high speed-up factor and an acceptable order of accuracy with respect to the DNS results (significantly higher than the accuracy of QG2$_{c}$, i.e., the under-resolved numerical simulation without any subfilter-scale model).
We also conclude that the AD-TF model is consistent with the original set of equations, since the AD-TF results converge to the DNS results when the mesh size approaches zero.

Finally, we perform a sensitivity study of the free smoothing parameter $\alpha$ and the order $N$ in the AD-TF model.
For comparison purposes, we also include results for QG2$_c$ (the under-resolved numerical simulation without any subfilter-scale model).
In order to quantify the results of the AD-TF model, we compute the error norms with respect to the DNS results with a resolution of $512^2$.
In both DNS and QG2$_c$ computations, the subfilter-scale term is set to zero: $S^{*}_{1} = S^{*}_{2} = 0$.



We start by investigating the sensitivity of the AD-TF model with respect to the parameter $\alpha$.
Table~\ref{tab:L2-TF-alpha} and Fig.~\ref{fig:AD-TF-norms} show that the sensitivity of the results to the free parameter $\alpha$ decreases with increasing mesh refinement.
Indeed, at the coarsest resolution (i.e., $32^2$), the values $\alpha = 0.15$, $\alpha = 0.25$, and $\alpha = 0.35$ yield practically indistinguishable results.
The value $\alpha = 0.45$ yields the most inaccurate results.
At the $64^2$ resolution, the value $\alpha = 0.15$ yields the best results, whereas the value $\alpha = 0.45$ yields again the most inaccurate ones.
At the $128^2$ and $256^2$ resolutions, the results are similar for all the values of $\alpha$.
In conclusion, the value $\alpha = 0.25$ appears to be optimal, since it yields the best results at the $32^2$ resolution in Table~\ref{tab:L2-TF-alpha} and Fig.~\ref{fig:AD-TF-norms}.
We note, however, that the values $\alpha = 0.15$ and $\alpha = 0.35$ yield similar results.
We also note that, for low values of $\alpha$, the AD-TF model performs better than QG2$_c$ at all resolutions.
For higher values of $\alpha$, the AD-TF model performs better than QG2$_c$ at the lowest resolution, but its accuracy starts to degrade at higher resolutions.

\begin{figure}
\centering
\includegraphics[width=0.5\textwidth]{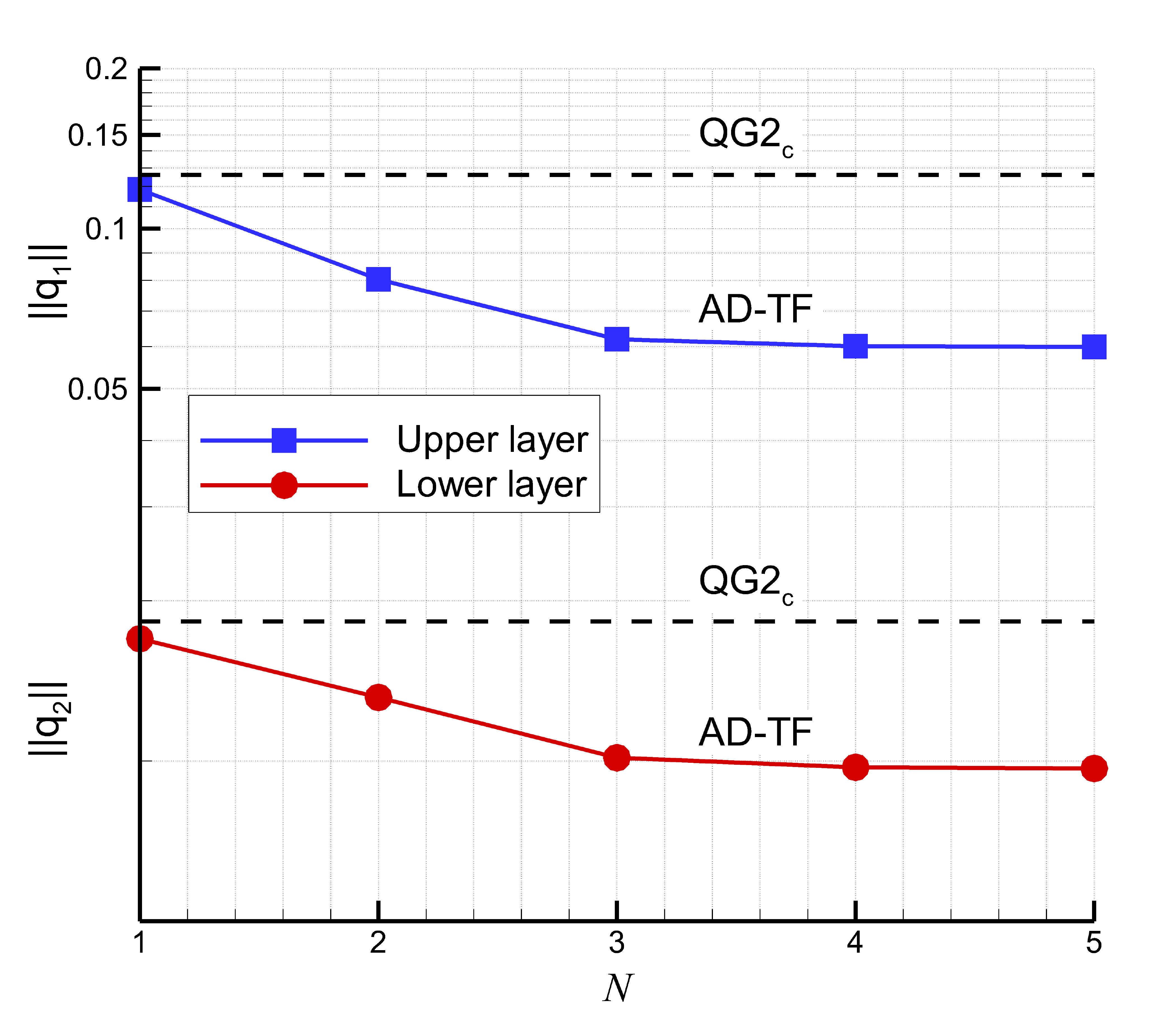}
\caption{
Experiment 1: The time-averaged $L^2$ norm of the error of the potential vorticities in the two layers, $\| q_1 \|$ and $\| q_2 \|$, for varying orders $N$ in the AD-TF model, at a grid resolution of $32^2$, and for a fixed parameter $\alpha = 0.25$.
The results obtained with QG2$_c$ (the under-resolved numerical simulation without any subfilter-scale model) are also included (dashed lines) for comparison purposes.
The reference solution used in the computation of the error is the numerical approximation obtained at a grid resolution of $512^2$.}
\label{fig:AD-TF-N}
\end{figure}

Next, we investigate the sensitivity of the AD-TF model with respect to the order $N$.
To this end, in Table~\ref{tab:L2-TF-N}, we fix the parameter $\alpha = 0.25$ and the grid resolution $32^2$, and present the time-averaged $L^2$ norm of the error of the streamfunction, $\| \psi_1 \|$ and $\| \psi_2 \|$, and potential vorticity in the two layers, $\| q_1 \|$ and $\| q_2 \|$, for varying orders $N$ in the AD-TF model.
These results are also compared graphically in Fig.~\ref{fig:AD-TF-N}.
We note that, at this coarse resolution, the AD-TF model performs better than QG2$_c$ for all values of $N$.
Based on the results in Table~\ref{tab:L2-TF-N} and Fig.~\ref{fig:AD-TF-N}, we conclude that the truncation order $N=3$ is the optimal value for the AD-TF model.
Indeed, increasing the value of $N$ from $1$ to $3$, results in a significant decrease in the error.
For higher values of $N$, however, the decrease in the error is negligible.
Since increasing the value of $N$ implies more filtering operations in the computation of the subfilter-scale term and, thus, a higher computational time, the value $N=3$ yields the best results in terms of combined accuracy and efficiency.

\begin{table}[!t]
\centering
\caption{Experiment 1:  Time-averaged $L^2$ norm of the error of the potential vorticity in the two layers, $\| q_1 \|$ and $\| q_2 \|$, for varying grid resolutions, $N_x \times N_y$, and varying the parameter $\alpha$ of the AD-TF model.
The results obtained with QG2$_c$ (the under-resolved numerical simulation without any subfilter-scale model) are also included for comparison purposes.
The reference solution used in the computation of the error is the numerical approximation obtained at a grid resolution of $512^2$.}
\label{tab:L2-TF-alpha}
\begin{tabular}{lllllllllll}
\hline\noalign{\smallskip}
\multirow{2}{*}{$N_x \times N_y$}  &
\multicolumn{2}{l}{\underline{$\alpha=0.15$  \quad \quad \quad}} &
\multicolumn{2}{l}{\underline{$\alpha=0.25$  \quad \quad \quad}} &
\multicolumn{2}{l}{\underline{$\alpha=0.35$  \quad \quad \quad}} &
\multicolumn{2}{l}{\underline{$\alpha=0.45$  \quad \quad \quad}} &
\multicolumn{2}{l}{\underline{QG2$_c$ ($S^{*}_{i}=0$) \quad}} \\
 & $\parallel q_1 \parallel$  & $\parallel q_2 \parallel$ & $ \parallel q_1 \parallel$ & $\parallel q_2 \parallel$ & $\parallel q_1 \parallel$ & $\parallel q_2 \parallel$ & $\parallel q_1 \parallel$ & $\parallel q_2 \parallel$ & $\parallel q_1 \parallel$ & $\parallel q_2 \parallel$ \\
\noalign{\smallskip}\hline\noalign{\smallskip}
$32^2$   & 6.07E-2 & 1.02E-2 & 6.00E-2  & 9.68E-3  & 6.24E-2 & 9.45E-3 & 7.73E-2 & 1.14E-2  & 1.24E-1 & 1.81E-2 \\
$64^2$   & 4.36E-2 & 6.42E-3 & 5.06E-2  & 7.50E-3  & 5.69E-2 & 7.82E-3 & 7.08E-2 & 8.48E-3  & 6.55E-2 & 7.73E-3 \\
$128^2$  & 2.72E-2 & 3.34E-3 & 2.90E-2  & 3.48E-3  & 3.19E-2 & 3.73E-3 & 3.85E-2 & 4.07E-3  & 2.91E-2 & 3.37E-3 \\
$256^2$  & 1.07E-2 & 1.39E-3 & 1.10E-2  & 1.42E-3  & 1.19E-2 & 1.48E-3 & 1.32E-2 & 1.59E-3  & 1.23E-2 & 1.54E-3 \\
\noalign{\smallskip}\hline
\end{tabular}
\end{table}

\begin{table}
\centering
\caption{Experiment 1:  Time-averaged $L^2$ norm of the error of the streamfunction, $\| \psi_1 \|$ and $\| \psi_2 \|$, and potential vorticity in the two layers, $\| q_1 \|$ and $\| q_2 \|$, for a fixed parameter $\alpha = 0.25$, and varying orders $N$ in the AD-TF model.
The reference solution used in the computation of the error is the numerical approximation obtained at a grid resolution of $512^2$.}
\label{tab:L2-TF-N}
\begin{tabular}{p{0.22\textwidth}p{0.16\textwidth}p{0.16\textwidth}p{0.16\textwidth}p{0.16\textwidth}}
\hline\noalign{\smallskip}
Method ($N_x \times N_y$) & $\parallel \psi_1 \parallel$  & $\parallel \psi_2 \parallel$ & $ \parallel q_1 \parallel$ & $\parallel q_2 \parallel$ \\
\noalign{\smallskip}\hline\noalign{\smallskip}
QG2$_c$ ($32^2$)        & 2.2090E-1 & 2.6845E-2 & 1.2446E-1  & 1.8075E-2    \\
AD-TF; $N=1$ ($32^2$)   & 2.0484E-1 & 1.9669E-2 & 1.1848E-1  & 1.6965E-2  \\
AD-TF; $N=2$ ($32^2$)   & 1.4563E-1 & 3.0302E-2 & 8.0293E-2  & 1.3159E-2  \\
AD-TF; $N=3$ ($32^2$)   & 1.1990E-1 & 2.5521E-2 & 6.1966E-2  & 1.0139E-2  \\
AD-TF; $N=4$ ($32^2$)   & 1.1642E-1 & 2.4431E-2 & 6.0171E-2  & 9.7303E-3  \\
AD-TF; $N=5$ ($32^2$)   & 1.1701E-1 & 2.3525E-2 & 5.9979E-2  & 9.6823E-3  \\
\noalign{\smallskip}\hline
\end{tabular}
\end{table}


\subsection{Approximate deconvolution model with the differential filter (AD-DF)}
\label{ss_addf}

\begin{table}[!t]
\centering
\caption{Experiment 1:  Time-averaged $L^2$ norm of the error of the potential vorticity in the two layers, $\| q_1 \|$ and $\| q_2 \|$, for varying grid resolutions, $N_x \times N_y$, and varying the parameter $\lambda$ of the AD-DF model.
The results obtained with QG2$_c$ (the under-resolved numerical simulation without any subfilter-scale model) are also included for comparison purposes.
The reference solution used in the computation of the error is the numerical approximation obtained at a grid resolution of $512^2$.}
\label{tab:L2-DF-lambda}
\begin{tabular}{lllllllllll}
\hline\noalign{\smallskip}
\multirow{2}{*}{$N_x \times N_y$}  &
\multicolumn{2}{l}{\underline{$\lambda=0.4 h$  \quad \quad \quad}} &
\multicolumn{2}{l}{\underline{$\lambda=0.6 h$  \quad \quad \quad}} &
\multicolumn{2}{l}{\underline{$\lambda=0.8 h$  \quad \quad \quad}} &
\multicolumn{2}{l}{\underline{$\lambda=1.0 h$  \quad \quad \quad}} &
\multicolumn{2}{l}{\underline{QG2$_c$ ($S^{*}_{i}=0$) \quad}} \\
 & $\parallel q_1 \parallel$  & $\parallel q_2 \parallel$ & $ \parallel q_1 \parallel$ & $\parallel q_2 \parallel$ & $\parallel q_1 \parallel$ & $\parallel q_2 \parallel$ & $\parallel q_1 \parallel$ & $\parallel q_2 \parallel$ & $\parallel q_1 \parallel$ & $\parallel q_2 \parallel$ \\
\noalign{\smallskip}\hline\noalign{\smallskip}
$32^2$   & 8.72E-2 & 1.29E-2 & 7.03E-2  & 1.07E-2  & 7.09E-2 & 1.23E-2 & 7.28E-2 & 1.16E-2  & 1.24E-1 & 1.81E-2 \\
$64^2$   & 5.08E-2 & 7.00E-3 & 5.01E-2  & 7.36E-3  & 5.33E-2 & 8.00E-3 & 5.68E-2 & 8.62E-3  & 6.55E-2 & 7.73E-3 \\
$128^2$  & 2.74E-2 & 3.48E-3 & 3.07E-2  & 3.76E-3  & 3.53E-2 & 4.38E-3 & 3.93E-2 & 4.93E-3  & 2.91E-2 & 3.37E-3 \\
$256^2$  & 1.24E-2 & 1.49E-3 & 1.38E-2  & 1.66E-3  & 1.75E-2 & 1.89E-3 & 2.27E-2 & 2.23E-3  & 1.23E-2 & 1.54E-3 \\
\noalign{\smallskip}\hline
\end{tabular}
\end{table}

\begin{table}[!t]
\centering
\caption{Experiment 1:  Time-averaged $L^2$ norm of the error of the streamfunction, $\| \psi_1 \|$ and $\| \psi_2 \|$, and potential vorticity in the two layers, $\| q_1 \|$ and $\| q_2 \|$, for a fixed parameter $\lambda = 0.6 h$, and varying orders $N$ in the AD-DF model.
The reference solution used in the computation of the error is the numerical approximation obtained at a grid resolution of $512^2$.}
\label{tab:L2-DF-N}
\begin{tabular}{p{0.22\textwidth}p{0.16\textwidth}p{0.16\textwidth}p{0.16\textwidth}p{0.16\textwidth}}
\hline\noalign{\smallskip}
Method ($N_x \times N_y$) & $\parallel \psi_1 \parallel$  & $\parallel \psi_2 \parallel$ & $ \parallel q_1 \parallel$ & $\parallel q_2 \parallel$ \\
\noalign{\smallskip}\hline\noalign{\smallskip}
QG2$_c$ ($32^2$)        & 2.2090E-1 & 2.6845E-2 & 1.2446E-1  & 1.8075E-2    \\
AD-DF; $N=1$ ($32^2$)   & 3.3117E-1 & 2.0303E-2 & 1.9157E-1  & 2.7892E-2  \\
AD-DF; $N=2$ ($32^2$)   & 1.7354E-1 & 2.0192E-2 & 9.7860E-2  & 1.4304E-2  \\
AD-DF; $N=3$ ($32^2$)   & 1.4690E-1 & 2.0361E-2 & 8.0382E-2  & 1.1957E-2  \\
AD-DF; $N=4$ ($32^2$)   & 1.3297E-1 & 2.0021E-2 & 7.0474E-2  & 1.0715E-2  \\
AD-DF; $N=5$ ($32^2$)   & 1.3343E-1 & 1.9745E-2 & 7.0305E-2  & 1.0723E-2  \\
\noalign{\smallskip}\hline
\end{tabular}
\end{table}

\begin{figure}
\centering
\mbox{
\subfigure[]{\includegraphics[width=0.5\textwidth]{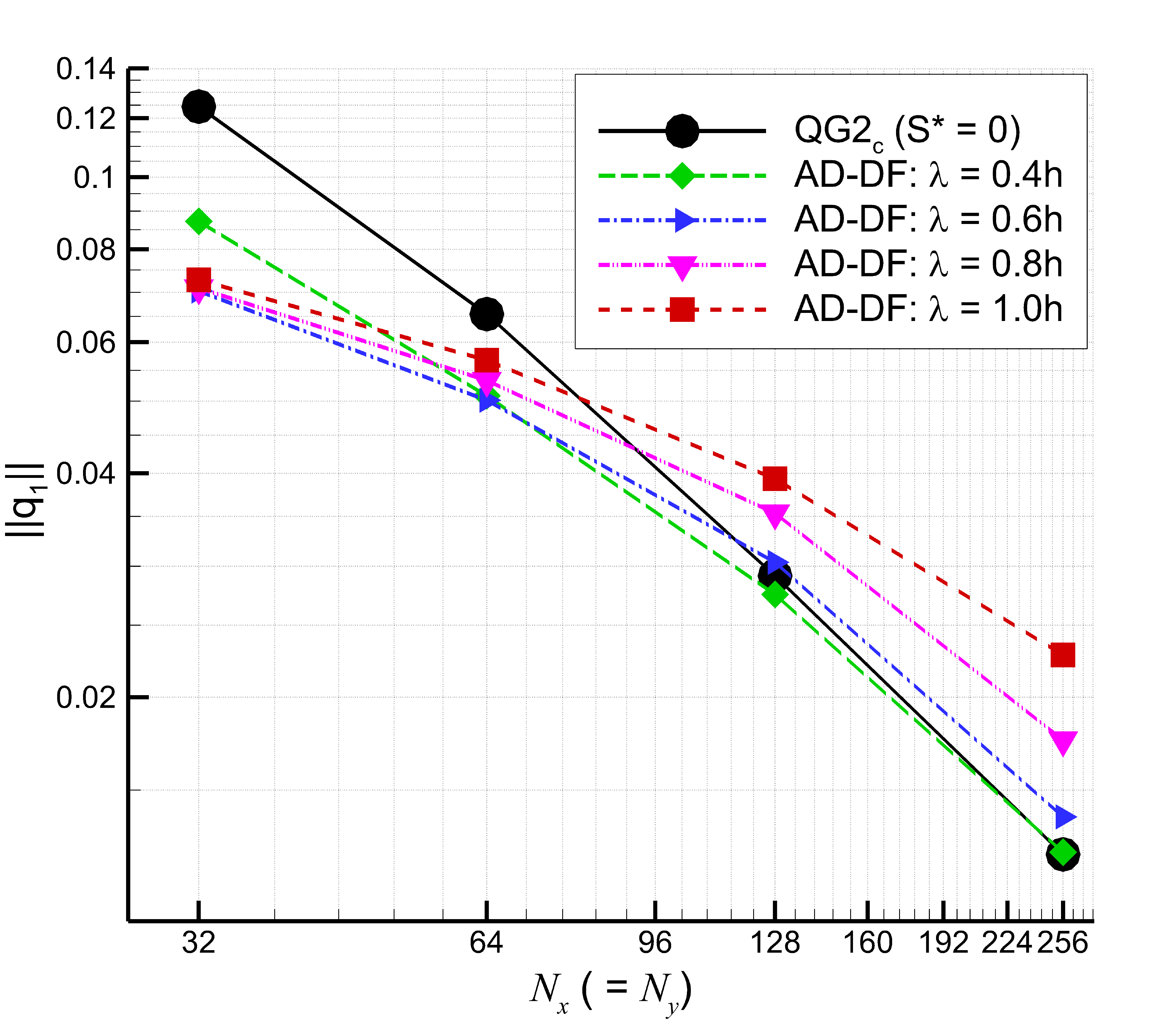}}
\subfigure[]{\includegraphics[width=0.5\textwidth]{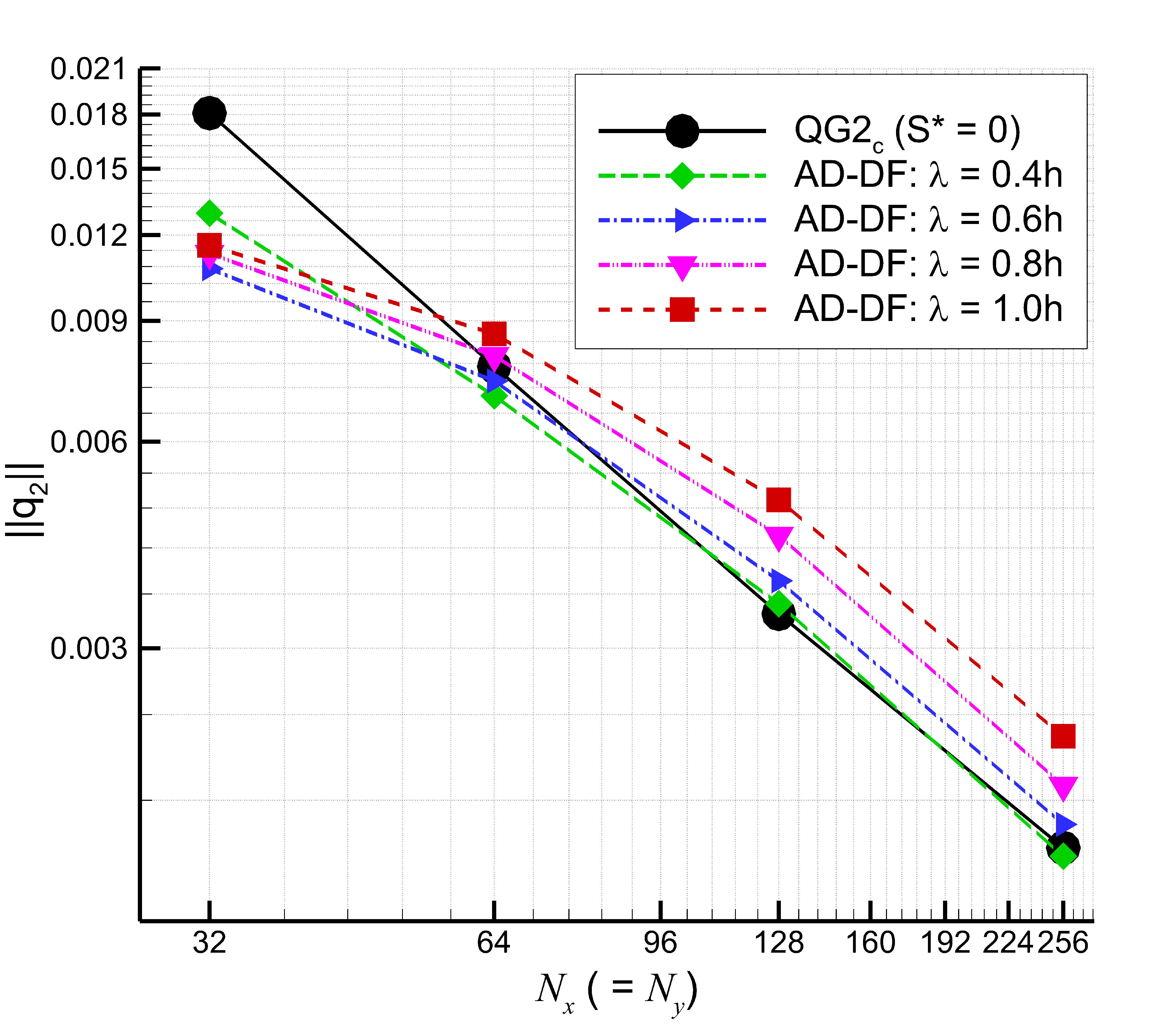}}
}
\caption{Experiment 1:  Log-log plot of the time-averaged $L^2$ norm of the error of the potential vorticity in the two layers, (a) $\| q_1 \|$ and (b) $\| q_2 \|$, for varying Helmholtz length $\lambda$ in the AD-DF model.
The results obtained with QG2$_c$ (the under-resolved numerical simulation without any subfilter-scale model) are also included for comparison purposes.
The reference solution used in the computation of the error is the numerical approximation obtained at a grid resolution of $512^2$.}
\label{fig:AD-DF-norms}
\end{figure}

\begin{figure}
\centering
\includegraphics[width=0.5\textwidth]{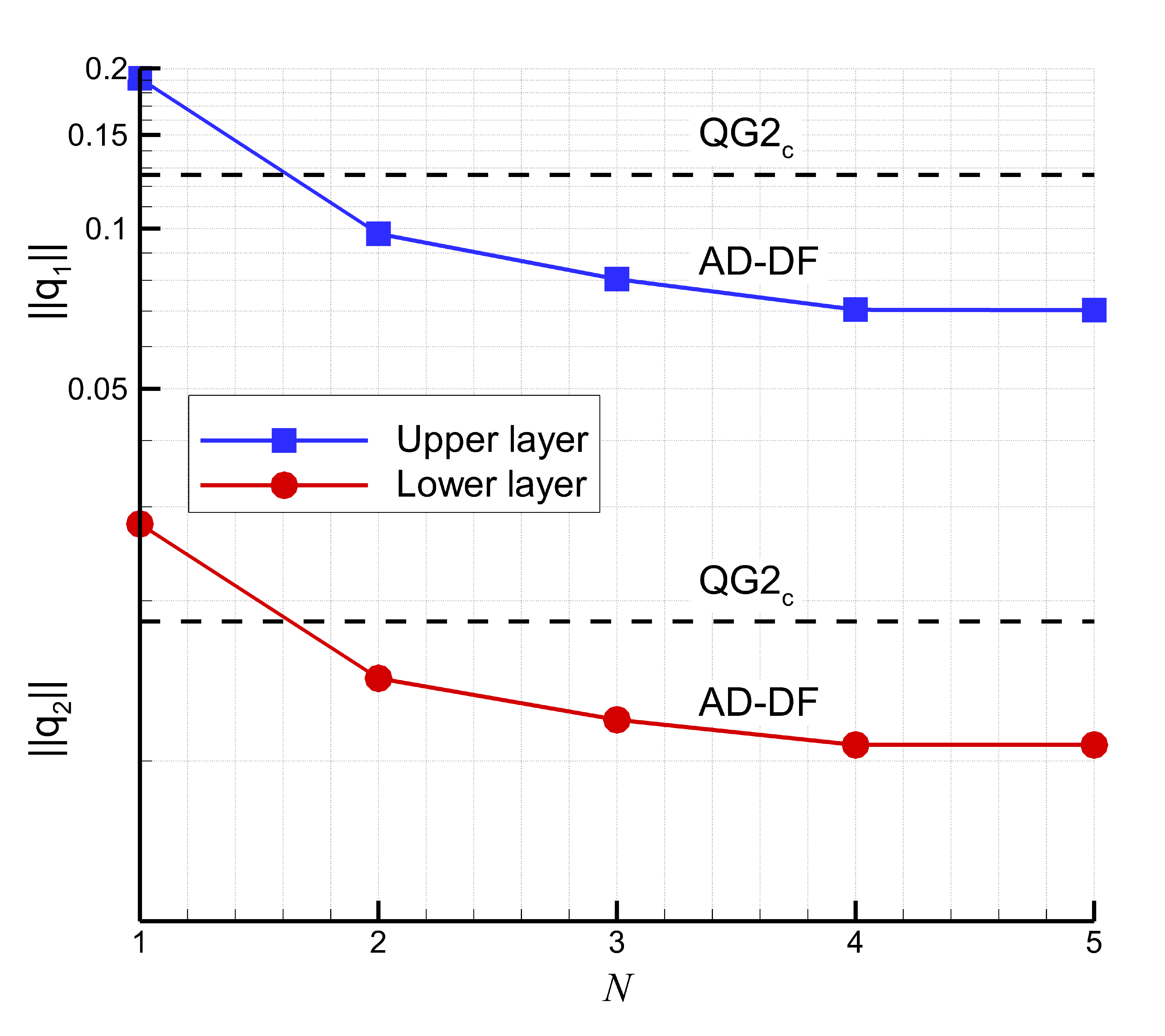}
\caption{
Experiment 1: The time-averaged $L^2$ norm of the error of the potential vorticity in the two layers, $\| q_1 \|$ and $\| q_2 \|$, for varying orders $N$ in the AD-DF model, at a grid resolution of $32^2$, and for a fixed parameter $\lambda=0.6 h$.
The results obtained with QG2$_c$ (the under-resolved numerical simulation without any subfilter-scale model) are also included (dashed lines) for comparison purposes.
The reference solution used in the computation of the error is the numerical approximation obtained at a grid resolution of $512^2$.}
\label{fig:AD-DF-N}
\end{figure}

Section~\ref{ss_adtf} clearly showed that, for a fixed value of the EV coefficient $\nu$, the AD-TF model can provide an accurate approximation of the mean flow field on a mesh that is significantly coarser than that used in a DNS.
Furthermore, it also showed that the AD-TF model is relatively insensitive with respect to changes in the smoothing parameter $\alpha$ used in the definition of the tridiagonal filter.
A natural question is whether the AD model is sensitive with respect to other choices in the input parameters, such as the spatial filter.
In this section, we numerically investigate the AD model equipped with a differential filter (given in Eq.~\eqref{eq:dishelm} and discussed in Section~\ref{ss_df}) instead of the tridiagonal filter used in Section~\ref{ss_adtf}.
The resulting LES model is denoted as AD-DF.

We start by performing a sensitivity study with respect to the model parameter $\lambda$ and the order $N$ in the AD-DF model, similar to the analysis performed in Section \ref{ss_adtf} for the AD-TF model.
For comparison purposes, we also include results for QG2$_c$ (theunder-resolved numerical simulation without any subfilter-scale model).
In order to quantify the results of the AD-DF model, we compute the error norms with respect to the DNS results having a resolution of $512^2$.
In both DNS and QG2$_c$ computations, the subfilter-scale term is set to zero: $S^{*}_{1} = S^{*}_{2} = 0$.

We first investigate the sensitivity of the AD-DF model with respect to the Helmholtz length, $\lambda$.
To this end, in Table~\ref{tab:L2-DF-lambda}, we fix the truncation order $N=5$, and present the time-averaged $L^2$ norm of the error of the potential vorticity in the two layers, $\| q_1 \|$ and $\| q_2 \|$, for varying grid resolutions, $N_x \times N_y$, and varying Helmholtz length, $\lambda$.
These results are also compared graphically in Fig.~\ref{fig:AD-DF-norms}.
Table~\ref{tab:L2-DF-lambda} and Fig.~\ref{fig:AD-DF-norms} yield the following conclusions.
At the $32^2$ and $64^2$ resolutions, all the $\lambda$ values yield similar results.
At the $128^2$ and $256^2$ resolutions, however, the values $\lambda = 0.4 h$ and $\lambda = 0.6 h$ yield the most accurate results; the values $\lambda = 0.8 h$ and $\lambda = 1.0 h$ yield inaccurate results.
In conclusion, the value $\lambda = 0.6 h$ appears to be optimal, since it yields the best results at the $32^2$ resolution in Table~\ref{tab:L2-DF-lambda} and Fig.~\ref{fig:AD-DF-norms}.
We also note that, for the values $\lambda = 0.4 h$ and $\lambda = 0.6 h$, the AD-DF model performs better than (or similar to) QG2$_c$ at all resolutions.
For the values $\lambda = 0.8 h$ and $\lambda = 1.0 h$, however, the AD-DF model is more accurate than QG2$_c$ at low resolutions ($32^2$ and $64^2$), but less accurate than QG2$_c$ at high resolutions ($128^2$ and $256^2$).
This behavior is natural, since, as explained in Section \ref{ss_df}, the higher values of $\lambda$ correspond to a higher level of numerical dissipation introduced by the DF.
A higher level of dissipation is beneficial to the numerical simulations at low resolutions, since it models some of the subgrid-scale effects.
At higher resolutions, however, the subgrid-scale effects become less important.
In this case, the dissipation introduced by the DF should also decrease.
This explains why, at higher resolutions, the lower values of $\lambda$ yield better results than the higher values of $\lambda$.

Next, we investigate the sensitivity of the AD-DF model with respect to the order $N$.
To this end, in Table~\ref{tab:L2-DF-N}, we fix the filtering parameter at $\lambda=0.6 h$ and the grid resolution  at $32^2$, and present the time-averaged $L^2$ norm of the error of the streamfunction, $\| \psi_1 \|$ and $\| \psi_2 \|$, and potential vorticity in the two layers, $\| q_1 \|$ and $\| q_2 \|$, for varying orders $N$ in the AD-DF model.
These results are also compared graphically in Fig.~\ref{fig:AD-DF-N}.
Based on the results in Table~\ref{tab:L2-DF-N} and Fig.~\ref{fig:AD-DF-N}, we conclude that the truncation order $N=4$ is the optimal value for the AD-DF model.
Indeed, increasing the value of $N$ from $1$ to $4$, results in a significant decrease in the error.
For $N = 5$, however, the decrease in the error is negligible.
Since increasing the value of $N$ implies more filtering operations in the computation of the subfilter-scale term and, thus, a higher computational time, the value $N=4$ yields the best results in terms of combined accuracy and efficiency.


The above sensitivity study clearly shows that, for a fixed value of the EV coefficient $\nu$, the AD-DF model can provide an accurate approximation of the mean flow field on a mesh that is significantly coarser than that used in a DNS.
It was also shown that the differential filter introduces a significant amount of numerical dissipation for higher values of $\lambda$.
The rest of the section is devoted to a careful numerical investigation of the amount of numerical dissipation in the AD-DF model by varying the EV coefficient $\nu$ in the model.

As mentioned in the introduction, the origin and modeling of the EV coefficient $\nu$ in the QG models is a thorny issue (the ``elephant in the room").
Indeed, Table~\ref{tab:vis} shows the wide range of values used for the EV coefficient $\nu$ over the years.
It is clear that no unique choice exists for $\nu$.
Instead, the value used in numerical simulations is dictated by the available computational resources.
To illustrate the importance of the particular value used for $\nu$ in practical computations, we carried out several high-resolution $128^2$ numerical simulations for various EV coefficients $\nu$.
\begin{figure}
\centering
\mbox{
\subfigure[]{\includegraphics[width=0.5\textwidth]{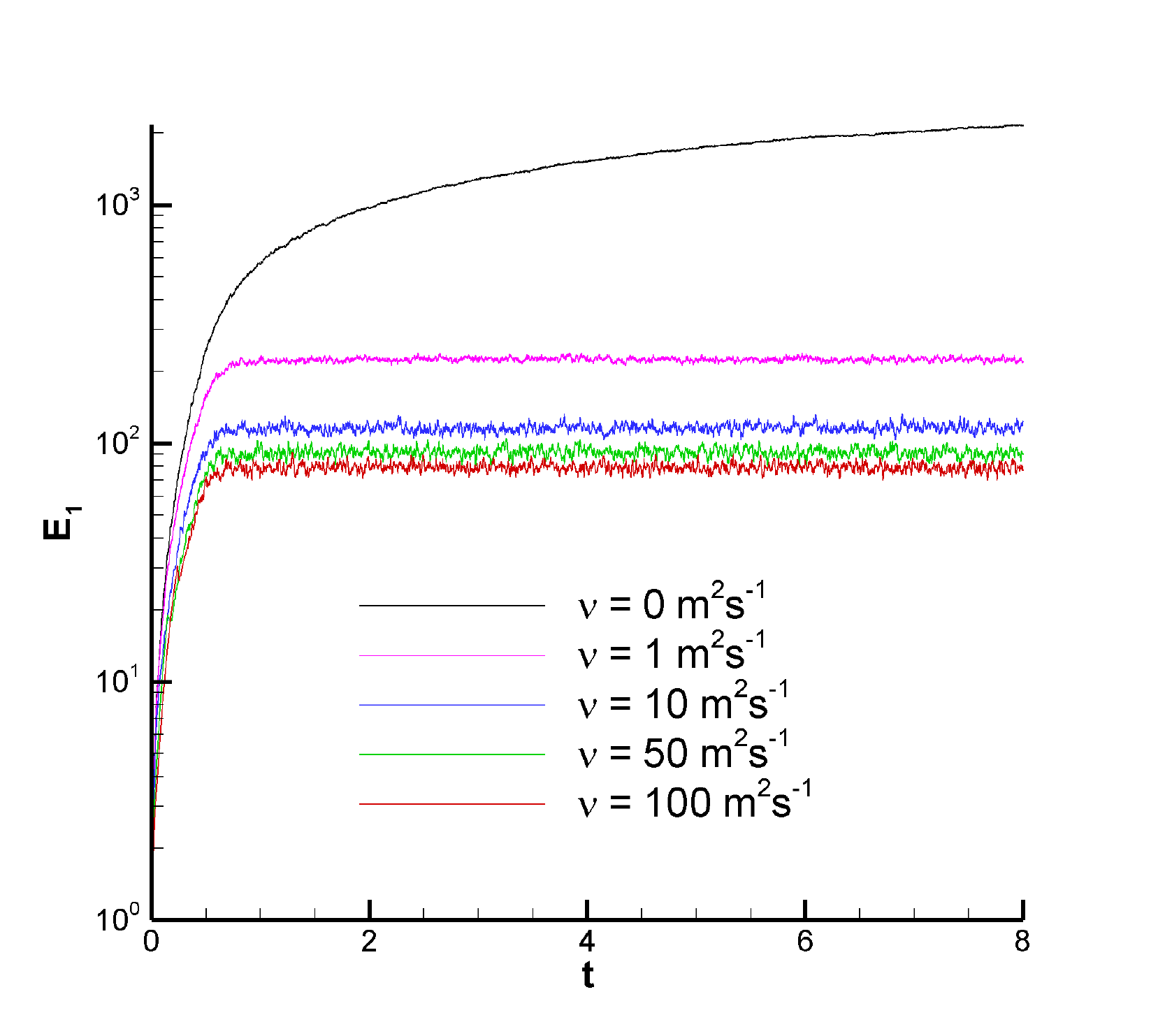}}
\subfigure[]{\includegraphics[width=0.5\textwidth]{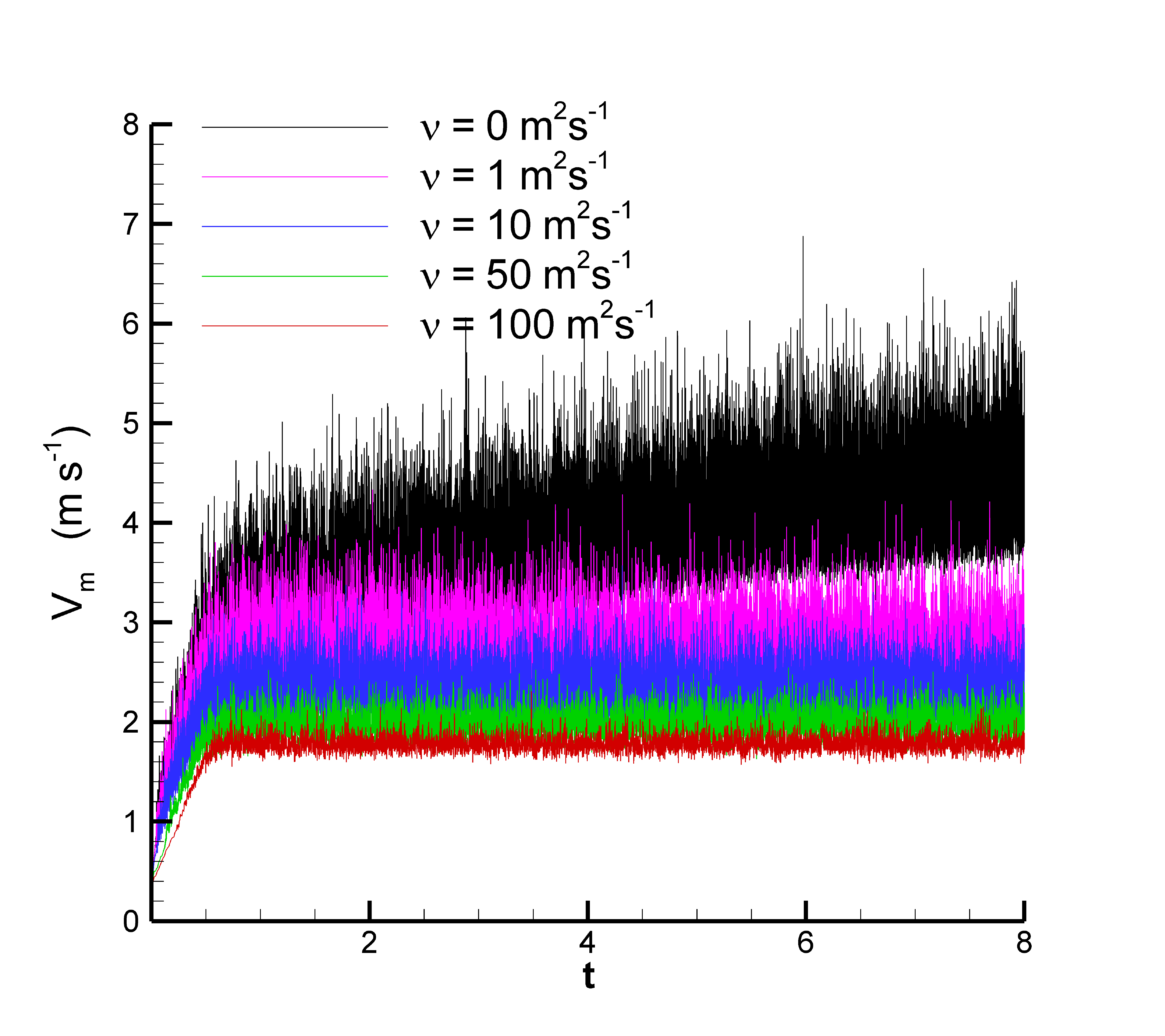}} }
\caption{
Experiment 1: Numerical simulation at a $128^2$ resolution for different values of the EV coefficient $\nu$. (a) Time history of basin integrated kinetic energy given by Eq.~\eqref{eq:26} for the upper layer, and (b) time series of the maximum speed $V_m$ in the field.
Note the sensitivity of the results with respect to $\nu$.
}
\label{fig:E1-vis}
\end{figure}
Fig.~\ref{fig:E1-vis}(a) shows the time series of the basin integrated kinetic energy for Experiment 1 for different values of $\nu$. The corresponding evolution of the maximum speed $V_m$ is also plotted in Fig.~\ref{fig:E1-vis}(b), in which we convert the dimensionless velocity to its dimensional counterpart to get a better physical insight. As seen from Fig.~\ref{fig:E1-vis}, after an initial transient spin-up process, the system with $\nu=100$ $m^2 s^{-1}$ reaches a statistically steady state at an average maximum speed of $1.78$ $m s^{-1}$ (having an upper bound of $2.05$ $m s^{-1}$ and a lower bound of $1.55$ $m s^{-1}$), which is close to the observed maximum zonal velocities of $2$ $m s^{-1}$ at $68^{\circ}W$ \citep{dijkstra2005nonlinear}.
Thus, Fig.~\ref{fig:E1-vis} illustrates the procedure used in choosing the EV coefficient $\nu$ in practical computations with the QG model:
The available computational resources dictate the numerical resolution that can be used; this, in turn, determines the EV coefficient $\nu$ that yields physical values for the computed flow fields (i.e., values that match those from observational data).
Using higher or lower values for $\nu$ can result in unphysical flow field data, as illustrated in Fig.~\ref{fig:E1-vis}.

\begin{figure}[!t]
\centering
\includegraphics[width=0.5\textwidth]{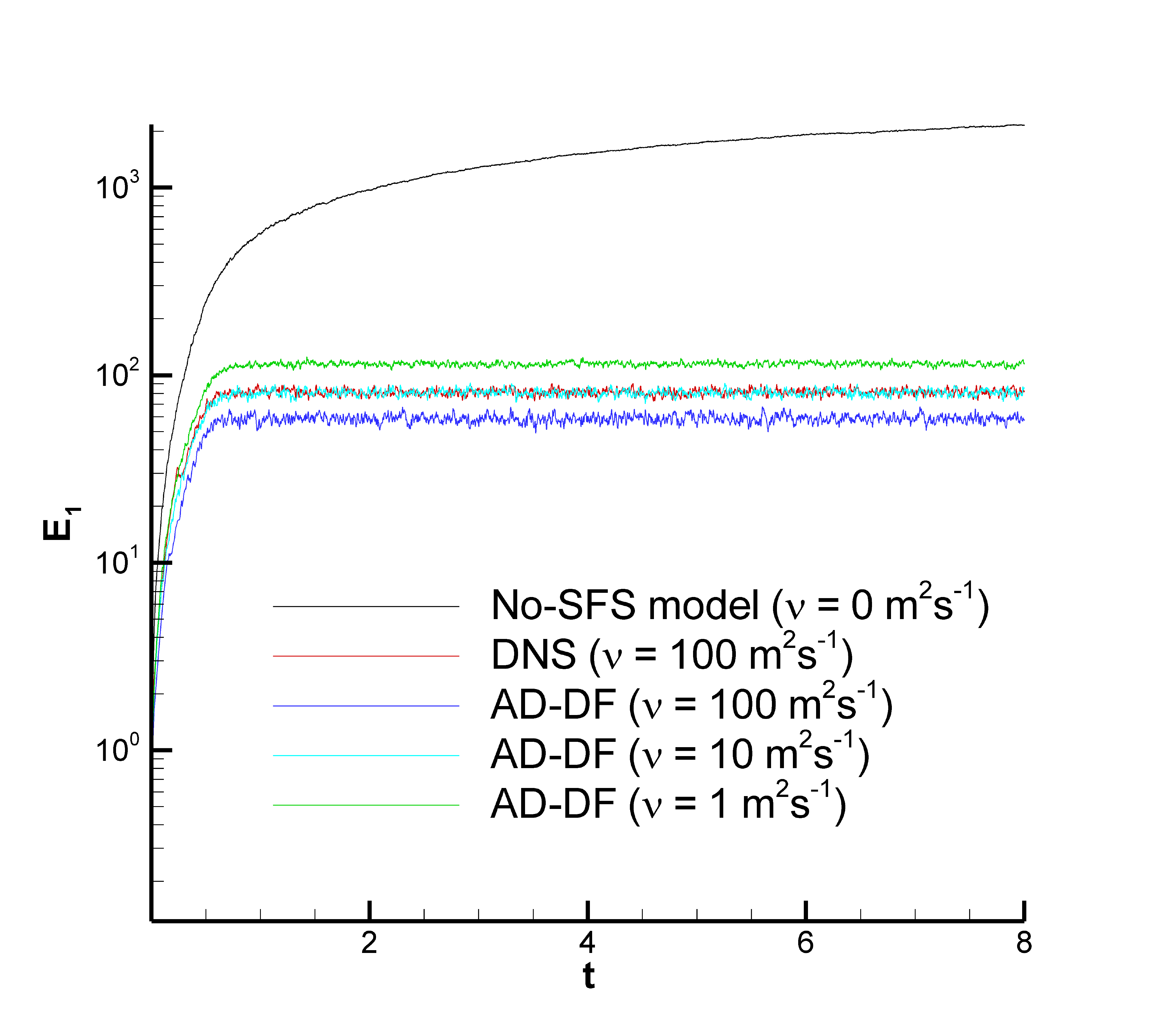}
\caption{
Experiment 1: Time histories of basin integrated kinetic energy for the upper layer obtained by the AD-DF model for $\mbox{Re}=580.97$  ($\nu=100$ $m^2 s^{-1}$), $\mbox{Re}=5809.7$ ($\nu=10$ $m^2 s^{-1}$), and $\mbox{Re}=58097$ ($\nu=1$ $m^2 s^{-1}$).
Results for the DNS and the No-SFS (under-resolved numerical simulation without any LES model) are also included.
Note that the AD-DF introduces numerical dissipation and produces the same mean flow field as the DNS on a coarser mesh and for a lower eddy viscosity coefficient $\nu$.}
\label{fig:E1-AD-hist}
\end{figure}

In order to measure the amount of numerical dissipation in the AD-DF model with $\lambda=2 h$, we run this model with EV coefficients that span three orders of magnitude.
The results for $\mbox{Re}=580.97$ ($\nu=100$ $m^2 s^{-1}$), $\mbox{Re}=5809.7$ ($\nu=10$ $m^2 s^{-1}$), and $\mbox{Re}=58097$ ($\nu=1$ $m^2 s^{-1}$) obtained with the AD-DF model are presented in Fig.~\ref{fig:E1-AD-hist}, which shows the time histories of the basin integrated kinetic energy given by Eq.~\eqref{eq:26} for the AD-DF model (for all three Reynolds numbers).
Results for the DNS and for the No-SFS run (the under-resolved numerical simulation without any LES model) for $\mbox{Re}=580.97$ ($\nu=100$ $m^2 s^{-1}$) (the Reynolds number used in Section \ref{ss_adtf}) are also included for comparison purposes.
As expected, No-SFS does yield a non-physical flow field with an unrealistically increasing energy level.
The kinetic energy of the AD-DF model for $\mbox{Re}=580.97$ ($\nu=100$ $m^2 s^{-1}$), on the other hand, is significantly lower than the kinetic energy of the DNS.
Thus, we conclude that the differential filter in the AD-DF model yields too much numerical dissipation.
Lowering the value of the eddy viscosity coefficient $\nu$ alleviates this problem.
Indeed, the AD-DF model with $\mbox{Re}=5809.7$ ($\nu=10$ $m^2 s^{-1}$) produces the same level of kinetic energy as the DNS for the $\mbox{Re}=580.97$ ($\nu=100$ $m^2 s^{-1}$).
Lowering even further the value of $\nu$ results in a kinetic energy level that is unrealistically high.
Based on the results in Fig. 20, we conclude that the differential filter in the AD-DF model introduces a wider range of numerical dissipation in the model.


We have already seen in Fig.~\ref{fig:E1-AD-hist} that the AD-DF model can successfully run on coarse meshes at a lower eddy viscosity coefficient $\nu$.
The transfer functions in Fig.~\ref{fig:tf} and Fig.~\ref{fig:df} clearly suggest that the AD-DF model should provide more dissipation than the AD-TF model.
The next natural step is to quantify how much effective dissipation is provided by each model.
We address this issue by performing numerical experiments for different values of the parameters $\alpha$ in the AD-TF model and $\lambda$ in the AD-DF model, and for various values of the eddy viscosity coefficient $0 \, m^{2}s^{-1} \leq \nu \leq 100 \, m^{2}s^{-1}$.
We note that, as expected, when the dissipation in the system is turned off (i.e., $\nu=0$ $m^{2}s^{-1}$), both the DNS and QG2$_c$ computations do not reach a quasi-stationary energy level; this is indicated in Table~\ref{tab:exp-32} by a dash symbol.
The AD truncation order is fixed to $N=5$.
The domain-integrated kinetic energy for the upper layer is presented in Table~\ref{tab:exp-32} for different values of $\nu$, $\alpha$, and $\lambda$.
The long time integrations are performed by using a coarse resolution of $32^2$ for all the runs.
The results obtained by the QG2$_c$, the under-resolved numerical simulation without any subfilter-scale model (i.e., $S^{*}_{i} =0$) at the same resolution of $32^2$ and the DNS results obtained at a resolution of $512^2$ are also included for comparison purposes.
Table~\ref{tab:exp-32} shows that the difference in mean kinetic energy level between the DNS and QG2$_c$ is quantitatively high for all values of $\nu$.
The reason is that the coarser resolution in QG2$_c$ does not effectively resolve the Munk scale.
Using the same coarse resolution of $32^2$, the AD-TF model with $0.05 \leq \alpha \leq 0.45$ predicts a more accurate energy level.
Decreasing $\alpha$ adds more numerical dissipation, which results in a decrease of the predicted energy level.
For the higher values of $\alpha$, the accuracy of the AD-TF model considerably degrades.
Thus, when the parameter $\alpha$ is between $\alpha=0.25$ and $\alpha=0.35$, the AD-TF model yields the most accurate results.
At the same coarse resolution of $32^2$, the AD-DF model with $0.4 h \leq \lambda \leq 2.0 h$ predicts a more accurate energy level than QG2$_c$.
Decreasing $\lambda$ adds a low level of numerical dissipation.
Increasing $\lambda$ provides a significant amount of numerical dissipation and decreases the predicted energy level.
The most  accurate results are obtained by the AD-DF model when the free parameter $\lambda$ lies between $\lambda=0.4 h$ and $\lambda=0.6 h$ for all values of $\nu$.
Overall, we conclude that, for a fixed value of the eddy viscosity coefficient $\nu$, we can obtain results close to the DNS results by tuning the modeling parameters $\alpha$ and $\beta$ in the AD-TF and AD-DF models, respectively.
Furthermore, the results in Table~\ref{tab:exp-32} show that the kinetic energy level predicted by the DNS for a given value of $\nu$ can be predicted by the AD-TF and AD-DF models with a lower value of $\nu$ when the model parameters $\alpha$ and $\beta$ are appropriately chosen.
Thus, as expected (see the transfer functions in Fig.~\ref{fig:tf} and Fig.~\ref{fig:df}), the AD-TF and AD-DF models do provide numerical dissipation to the system.
We note, however, that using the DF and TF without using the AD procedure does not provide a significant amount of numerical dissipation.
Indeed, running an under-resolved numerical simulation with $\nu = 0 \, m^{2}s^{-1}$ and using the DF to smooth out the potential vorticity and streamfunction values after each time-step yielded inaccurate results.
Finally, we emphasize that the numerical results in Table~\ref{tab:exp-32} should not interpreted as an argument for the superiority of the AD-LES models over standard EV models.
Instead, they simply show that the same kinetic energy level can be predicted in two different ways: by adjusting the eddy viscosity coefficient or by adjusting the parameters $\alpha$ and $\beta$ in the AD-TF and AD-DF models.
For completeness, in Table~\ref{tab:exp-128} we repeat the same numerical experiments as those displayed in Table~\ref{tab:exp-32}, but for a moderate resolution of $128^2$.
The same qualitative conclusions as those above can be drawn, except that the difference between DNS and QG2$_c$ in Table~\ref{tab:exp-128} is smaller due to the fairly well resolved Munk scales at this resolution.

\begin{table}[!t]
\centering
\caption{Experiment 1: Time-averaged basin-integrated kinetic energy of the upper layer, $E_1$, for varying modeling parameters, $\alpha$ and $\lambda$, varying eddy viscosity coefficients, $\nu$, and a fixed resolution of $32^2$.
The reference DNS solution obtained at a grid resolution of $512^2$ and the solution obtained by the QG2$_c$ without using any subfilter-scale model at a resolution of $32^2$ are also included for comparison purposes. The dash represents a nonstationary field.}
\label{tab:exp-32}
\begin{tabular}{lllllll}
\hline\noalign{\smallskip}
Method & $\nu=0$ $m^2s^{-1}$&  $\nu=1$ $m^2s^{-1}$& $\nu=5$ $m^2s^{-1}$& $\nu=10$ $m^2s^{-1}$& $\nu=50$ $m^2s^{-1}$& $\nu=100$ $m^2s^{-1}$\\
\noalign{\smallskip}\hline\noalign{\smallskip}
AD-TF: $\alpha=0.05$    & 68.744    & 61.655    & 52.597    & 48.790    & 33.884  &  27.695 \\
AD-TF: $\alpha=0.15$    & 69.620    & 67.875    & 60.054    & 55.825    & 41.881  &  38.532 \\
AD-TF: $\alpha=0.25$    & 85.855    & 83.992    & 78.974    & 71.994    & 53.641  &  48.478 \\
AD-TF: $\alpha=0.35$    & 107.444   & 106.521   & 94.917    & 87.655    & 83.846  &  74.202 \\
AD-TF: $\alpha=0.45$    & 369.895   & 323.502   & 245.234   & 210.642   & 145.469 &  119.488 \\
AD-DF: $\lambda=0.4 h$  & 220.331   & 212.155   & 164.394   & 139.083   & 96.006  &  84.913 \\
AD-DF: $\lambda=0.6 h$  & 135.768   & 117.392   & 90.414    & 77.151    & 50.566  &  42.623 \\
AD-DF: $\lambda=0.8 h$  & 85.939    & 86.124    & 64.816    & 51.566    & 31.910  &  26.804 \\
AD-DF: $\lambda=1.0 h$  & 71.628    & 75.752    & 49.740    & 40.499    & 23.448  &  18.161 \\
AD-DF: $\lambda=2.0 h$  & 49.906    & 43.736    & 27.501    & 21.936    & 9.556   &  7.164 \\
QG2$_c$ ($S^{*}_{i} =0$)& -         & 397.121   & 333.205   & 279.101   & 210.718 &  195.028 \\
DNS ($S^{*}_{i} =0$)    & -         & 128.698   & 121.196   & 117.014   & 96.593  &  81.609 \\
\noalign{\smallskip}\hline
\end{tabular}
\end{table}

\begin{table}[!t]
\centering
\caption{Experiment 1: Time-averaged basin-integrated kinetic energy of the upper layer, $E_1$, for varying modeling parameters, $\alpha$ and $\lambda$, varying eddy viscosity coefficients, $\nu$, and a fixed resolution of $128^2$.
The reference DNS solution obtained at a grid resolution of $512^2$ and the solution obtained by the QG2$_c$ without using any subfilter-scale model at a resolution of $128^2$ are also included for comparison purposes. The dash represents a nonstationary field.}
\label{tab:exp-128}
\begin{tabular}{lllllll}
\hline\noalign{\smallskip}
Method & $\nu=0$ $m^2s^{-1}$&  $\nu=1$ $m^2s^{-1}$& $\nu=5$ $m^2s^{-1}$& $\nu=10$ $m^2s^{-1}$& $\nu=50$ $m^2s^{-1}$& $\nu=100$ $m^2s^{-1}$\\
\noalign{\smallskip}\hline\noalign{\smallskip}
AD-TF: $\alpha=0.05$    & 101.317   & 95.878    & 84.092    & 76.945    & 60.018  &  53.374 \\
AD-TF: $\alpha=0.15$    & 129.586   & 124.196   & 105.293   & 94.670    & 69.004  &  59.290 \\
AD-TF: $\alpha=0.25$    & 164.488   & 159.169   & 135.967   & 118.318   & 79.629  &  66.723 \\
AD-TF: $\alpha=0.35$    & 217.861   & 218.466   & 173.868   & 147.645   & 90.977  &  74.544 \\
AD-TF: $\alpha=0.45$    & 336.068   & 393.889   & 238.706   & 190.372   & 105.839 &  83.325 \\
AD-DF: $\lambda=0.4 h$  & 240.249   & 187.409   & 125.260   & 102.714   & 69.399  &  60.189 \\
AD-DF: $\lambda=0.6 h$  & 128.823   & 105.917   & 78.761    & 68.464    & 54.023  &  48.134 \\
AD-DF: $\lambda=0.8 h$  & 80.713    & 67.224    & 52.469    & 47.975    & 41.991  &  37.782 \\
AD-DF: $\lambda=1.0 h$  & 57.065    & 47.336    & 38.009    & 35.994    & 32.924  &  30.316 \\
AD-DF: $\lambda=2.0 h$  & 26.376    & 19.411    & 15.607    & 15.116    & 14.149  &  14.166 \\
QG2$_c$ ($S^{*}_{i} =0$)& -         & 442.829   & 244.128   & 179.348   & 96.646  &  77.617 \\
DNS ($S^{*}_{i} =0$)    & -         & 128.698   & 121.196   & 117.014   & 96.593  &  81.609 \\
\noalign{\smallskip}\hline
\end{tabular}
\end{table}

\section{Conclusions}
\label{sec:cons}

A new approximate deconvolution large eddy simulation (AD-LES) model for the two-layer quasigeostrophic equations, a standard prototype of more realistic wind-driven ocean circulation, was introduced.
Two different ocean settings with eastward jet formations of different strengths were considered.
Two variants of the AD-LES model were proposed: one with a tridiagonal filter (AD-TF), and the other with a differential filter (AD-DF).
Both the AD-TF and the AD-DF models yielded accurate solutions, with physically relevant energy levels and realistic mean streamfunction and potential vorticity contour plots. A quantitative analysis of the effects of using AD-TF and AD-DF on the QG2 model was presented.
The two models also dramatically decreased the computational cost of the corresponding high-resolution numerical simulation, by using a mesh significantly coarser than the Munk scale.
We emphasize that the AD procedure plays an essential role in the success of the AD-LES modeling strategy.
Indeed, the underresolved numerical simulations without AD modeling on the same coarse mesh as that employed by the AD-LES models produced inaccurate results.
The AD-TF and AD-DF models, however, had different behaviors in terms of the numerical dissipation added to the system.
In fact, our numerical results showed that the AD-TF and AD-DF models can be employed successfully on meshes that are significantly coarser than the Munk scale {\it and} with an eddy viscosity coefficient that is dramatically lower than that used in the original two-layer quasigeostrophic equations by tuning the free parameters $\alpha$ and $\lambda$ appropriately.
We emphasize that the tuning of the AD-LES model parameters is essential in obtaining accurate results. We also note that this paper does not claim the superiority of the AD-LES method over other eddy viscosity type closure approaches since the underlying quasigeostrophic equations utilize an intrinsic eddy viscosity coefficient to account for large scale dissipation.
With this in mind, we also highlight that assessments and evaluations of various turbulence closure models for large eddy simulations of realistic oceanic basins are highly desirable, a topic we intend to further investigate in a future study.

\section*{Acknowledgements}
The authors thank the two reviewers whose comments and suggestions significantly improved this paper.
The authors greatly appreciate the support of the Institute for Critical Technology and Applied Science (ICTAS) at Virginia Tech via grant number 118709. The third author was also supported by the National Science Foundation via grant DMS-1025314 under the Collaboration in Mathematical Geosciences (CMG) initiative.

\bibliographystyle{elsarticle-harv}
\bibliography{ref}

\begin{thebibliography}{77}
\expandafter\ifx\csname natexlab\endcsname\relax\def\natexlab#1{#1}\fi
\expandafter\ifx\csname url\endcsname\relax
  \def\url#1{\texttt{#1}}\fi
\expandafter\ifx\csname urlprefix\endcsname\relax\def\urlprefix{URL }\fi

\bibitem[{Allen(1980)}]{allen1980models}
Allen, J.~S., 1980. {Models of wind-driven currents on the continental shelf}.
  Annu. Rev. Fluid Mech. 12, 389--433.

\bibitem[{Arakawa(1966)}]{arakawa1966computational}
Arakawa, A., 1966. {Computational design for long-term numerical integration of
  the equations of fluid motion: \uppercase{T}wo-dimensional incompressible
  flow. \uppercase{P}art \uppercase{I}}. J. Comput. Phys. 1~(1), 119--143.

\bibitem[{Awad et~al.(2009)Awad, Toorman, and Lacor}]{awad2009large}
Awad, E., Toorman, E., Lacor, C., 2009. {Large eddy simulations for
  quasi-2\uppercase{D} turbulence in shallow flows: \uppercase{A} comparison
  between different subgrid scale models}. J. Marine Syst. 77~(4), 511--528.

\bibitem[{Berloff et~al.(2009)Berloff, Kamenkovich, and
  Pedlosky}]{berloff2009mechanism}
Berloff, P., Kamenkovich, I., Pedlosky, J., 2009. {A mechanism of formation of
  multiple zonal jets in the oceans}. J. Fluid Mech. 628, 395--425.

\bibitem[{Berloff and McWilliams(1999)}]{berloff1999large}
Berloff, P.~S., McWilliams, J.~C., 1999. {Large-scale, low-frequency
  variability in wind-driven ocean gyres}. J. Phys. Oceanogr. 29, 1925--1949.

\bibitem[{Berselli et~al.(2006)Berselli, Iliescu, and
  Layton}]{berselli2006mathematics}
Berselli, L.~C., Iliescu, T., Layton, W.~J., 2006. {Mathematics of large eddy
  simulation of turbulent flows}. Springer Verlag.

\bibitem[{Berselli and Lewandowski(2011)}]{berselli2011convergence}
Berselli, L.~C., Lewandowski, R., 2011. Convergence of approximate
  deconvolution models to the mean \uppercase{N}avier-\uppercase{S}tokes
  equations. In: Annales de l'Institut Henri Poincare (C) Non Linear Analysis.
  Elsevier.

\bibitem[{Bryan(1963)}]{bryan1963numerical}
Bryan, K., 1963. {A numerical investigation of a nonlinear model of a
  wind-driven ocean}. J. Atmos. Sci. 20, 594--606.

\bibitem[{Campin et~al.(2011)Campin, Hill, Jones, and
  Marshall}]{campin2011super}
Campin, J.~M., Hill, C., Jones, H., Marshall, J., 2011. Super-parameterization
  in ocean modeling: \uppercase{A}pplication to deep convection. Ocean Modell.
  36, 90--101.

\bibitem[{Chang et~al.(2001)Chang, Ghil, Ide, and Lai}]{chang2001transition}
Chang, K.~I., Ghil, M., Ide, K., Lai, C. C.~A., 2001. {Transition to aperiodic
  variability in a wind-driven double-gyre circulation model}. J. Phys.
  Oceanogr. 31~(5), 1260--1286.

\bibitem[{Chen et~al.(2011)Chen, Gunzburger, and Ringler}]{chen2011scale}
Chen, Q., Gunzburger, M., Ringler, T., 2011. A scale-invariant formulation of
  the anticipated potential vorticity method. Mon. Wea. Rev. 139, 2614–--2629.

\bibitem[{Chen et~al.(2003)Chen, Ecke, Eyink, Wang, and
  Xiao}]{chen2003physical}
Chen, S., Ecke, R.~E., Eyink, G.~L., Wang, X., Xiao, Z., 2003. Physical
  mechanism of the two-dimensional enstrophy cascade. Phys. Rev. Lett. 91~(21),
  214501.

\bibitem[{Chow and Street(2009)}]{chow2009evaluation}
Chow, F.~K., Street, R.~L., 2009. Evaluation of turbulence closure models for
  large-eddy simulation over complex terrain: flow over \uppercase{A}skervein
  \uppercase{H}ill. J. Appl. Meteorol. Clim. 48~(5), 1050--1065.

\bibitem[{Chow et~al.(2005)Chow, Street, Xue, and Ferziger}]{chow2005explicit}
Chow, F.~K., Street, R.~L., Xue, M., Ferziger, J.~H., 2005. Explicit filtering
  and reconstruction turbulence modeling for large-eddy simulation of neutral
  boundary layer flow. J. Atmos. Sci. 62~(7), 2058--2077.

\bibitem[{Cummins(1992)}]{cummins1992inertial}
Cummins, P.~F., 1992. {Inertial gyres in decaying and forced geostrophic
  turbulence}. J. Mar. Res. 50~(4), 545--566.

\bibitem[{Cushman-Roisin and Beckers(2009)}]{cushman2009introduction}
Cushman-Roisin, B., Beckers, J.~M., 2009. Introduction to geophysical fluid
  dynamics: \uppercase{P}hysical and numerical aspects. Academic Press.

\bibitem[{DiBattista and Majda(2001)}]{dibattista2001equilibrium}
DiBattista, M.~T., Majda, A.~J., 2001. {Equilibrium statistical predictions for
  baroclinic vortices: \uppercase{T}he role of angular momentum}. Theor. Comp.
  Fluid Dyn. 14~(5), 293--322.

\bibitem[{Dijkstra(2005)}]{dijkstra2005nonlinear}
Dijkstra, H.~A., 2005. {Nonlinear physical oceanography}. Springer.

\bibitem[{Dijkstra and Ghil(2005)}]{dijkstra2005low}
Dijkstra, H.~A., Ghil, M., 2005. {Low-frequency variability of the large-scale
  ocean circulation: \uppercase{A} dynamical systems approach}. Rev. Geophys.
  43, 122--59.

\bibitem[{Domaradzki and Adams(2002)}]{domaradzki2002direct}
Domaradzki, J.~A., Adams, N.~A., 2002. {Direct modelling of subgrid scales of
  turbulence in large eddy simulations}. J. Turbul. 3~(24), 1--19.

\bibitem[{Duan et~al.(2010)Duan, Fischer, Iliescu, and
  {\"O}zg{\"o}kmen}]{duan2010bridging}
Duan, J., Fischer, P., Iliescu, T., {\"O}zg{\"o}kmen, T.~M., 2010. Bridging the
  \uppercase{B}oussinesq and primitive equations through spatio-temporal
  filtering. Appl. Math. Lett. 23~(4), 453--456.

\bibitem[{Dunca and Epshteyn(2006)}]{dunca2006stolz}
Dunca, A., Epshteyn, Y., 2006. {On the \uppercase{S}tolz-\uppercase{A}dams
  deconvolution model for the large-eddy simulation of turbulent flows}. SIAM
  J. Math. Anal. 37, 1890.

\bibitem[{Espa et~al.(2008)Espa, Carnevale, Cenedese, and
  Mariani}]{espa2008quasi}
Espa, S., Carnevale, G.~F., Cenedese, A., Mariani, M., 2008.
  {Quasi-two-dimensional decaying turbulence subject to the $\beta$ effect}. J.
  Turbul. 9~(36), 1--18.

\bibitem[{Fox-Kemper and Menemenlis(2008)}]{fox2008can}
Fox-Kemper, B., Menemenlis, D., 2008. {Can large eddy simulation techniques
  improve mesoscale rich ocean models?} in Ocean Modeling in an Eddying Regime,
  Geophys. Monogr. Ser. 177, edited by M. Hecht and H. Hasumi, 319--338.

\bibitem[{Gates(1968)}]{gates1968numerical}
Gates, W.~L., 1968. {A numerical study of transient Rossby waves in a
  wind-driven homogeneous ocean}. J. Atmos. Sci. 25, 3--22.

\bibitem[{Germano(1986)}]{germano1986differential}
Germano, M., 1986. Differential filters of elliptic type. Phys. Fluids 29,
  1757--1758.

\bibitem[{Ghil et~al.(2008)Ghil, Chekroun, and Simonnet}]{ghil2008climate}
Ghil, M., Chekroun, M.~D., Simonnet, E., 2008. {Climate dynamics and fluid
  mechanics: \uppercase{N}atural variability and related uncertainties}.
  Physica D 237~(14-17), 2111--2126.

\bibitem[{Gottlieb and Shu(1998)}]{gottlieb1998total}
Gottlieb, S., Shu, C.~W., 1998. {Total variation diminishing
  \uppercase{R}unge-\uppercase{K}utta schemes}. Math. Comput. 67~(221), 73--85.

\bibitem[{Griffa and Salmon(1989)}]{griffa1989wind}
Griffa, A., Salmon, R., 1989. {Wind-driven ocean circulation and equilibrium
  statistical mechanics}. J. Mar. Res. 47~(3), 457--492.

\bibitem[{Holland(1978)}]{holland1978role}
Holland, W.~R., 1978. {The role of mesoscale eddies in the general circulation
  of the ocean-numerical experiments using a wind-driven quasi-geostrophic
  model}. J. Phys. Oceanogr. 8~(3), 363--392.

\bibitem[{Holland and Lin(1975)}]{holland1975generation}
Holland, W.~R., Lin, L.~B., 1975. {On the generation of mesoscale eddies and
  their contribution to the oceanic general circulation. I. A preliminary
  numerical experiment}. J. Phys. Oceanogr. 5, 642--657.

\bibitem[{Holland and Rhines(1980)}]{holland1980example}
Holland, W.~R., Rhines, P.~B., 1980. {An example of eddy-induced ocean
  circulation}. J. Phys. Oceanogr. 10~(7), 1010--1031.

\bibitem[{Holm and Nadiga(2003)}]{holm2003modeling}
Holm, D.~D., Nadiga, B.~T., 2003. {Modeling mesoscale turbulence in the
  barotropic double-gyre circulation}. J. Phys. Oceanogr. 33~(11), 2355--2365.

\bibitem[{Iliescu and Fischer(2003)}]{iliescu2003large}
Iliescu, T., Fischer, P.~F., 2003. Large eddy simulation of turbulent channel
  flows by the rational large eddy simulation model. Phys. Fluids 15, 3036.

\bibitem[{Jiang et~al.(1995)Jiang, Jin, and Ghil}]{jiang1995multiple}
Jiang, S., Jin, F., Ghil, M., 1995. {Multiple equilibria, periodic, and
  aperiodic solutions in a wind-driven, double-gyre, shallow-water model}. J.
  Phys. Oceanogr. 25~(5), 764--786.

\bibitem[{Layton and Lewandowski(2006)}]{layton2006residual}
Layton, W., Lewandowski, R., 2006. Residual stress of approximate deconvolution
  models of turbulence. J. Turbul. 7, 1--21.

\bibitem[{Layton and Neda(2007)}]{layton2007similarity}
Layton, W., Neda, M., 2007. A similarity theory of approximate deconvolution
  models of turbulence. J. Math. Anal. Appl. 333~(1), 416--429.

\bibitem[{Layton and Rebholz(2012)}]{layton2012approximate}
Layton, W., Rebholz, L., 2012. Approximate Deconvolution Models of Turbulence:
  Analysis, Phenomenology and Numerical Analysis. Springer Verlag.

\bibitem[{Majda and Wang(2006)}]{majda2006non}
Majda, A., Wang, X., 2006. {Non-linear dynamics and statistical theories for
  basic geophysical flows}. Cambridge University Press.

\bibitem[{Maltrud and Vallis(1991)}]{maltrud1991energy}
Maltrud, M.~E., Vallis, G.~K., 1991. Energy spectra and coherent structures in
  forced two-dimensional and beta-plane turbulence. J. Fluid Mech. 228,
  321--342.

\bibitem[{Marshall et~al.(1997)Marshall, Hill, Perelman, and
  Adcroft}]{marshall1997hydrostatic}
Marshall, J., Hill, C., Perelman, L., Adcroft, A., 1997. Hydrostatic,
  quasi-hydrostatic, and nonhydrostatic ocean modeling. J. Geophys. Res. 102,
  5733--5752.

\bibitem[{McWilliams(2006)}]{mcwilliams2006fundamentals}
McWilliams, J.~C., 2006. {Fundamentals of geophysical fluid dynamics}.
  Cambridge University Press.

\bibitem[{Meacham(2000)}]{meacham2000low}
Meacham, S.~P., 2000. {Low-frequency variability in the wind-driven
  circulation}. J. Phys. Oceanogr. 30~(2), 269--293.

\bibitem[{Medjo(2000)}]{medjo2000numerical}
Medjo, T.~T., 2000. {Numerical simulations of a two-layer quasi-geostrophic
  equation of the ocean}. SIAM J. Numer. Anal. 37~(6), 2005--2022.

\bibitem[{Miller(2007)}]{miller2007numerical}
Miller, R.~N., 2007. {Numerical modeling of ocean circulation}. Cambridge
  University Press.

\bibitem[{Moin(2001)}]{moin2002fundamentals}
Moin, P., 2001. {Fundamentals of engineering numerical analysis}. Cambridge
  University Press.

\bibitem[{Munk(1950)}]{munk1950wind}
Munk, W.~H., 1950. {On the wind-driven ocean circulation}. J. Meteor. 7~(2),
  80--93.

\bibitem[{Nadiga and Margolin(2001)}]{nadiga2001dispersive}
Nadiga, B.~T., Margolin, L.~G., 2001. {Dispersive-dissipative eddy
  parameterization in a barotropic model}. J. Phys. Oceanogr. 31~(8),
  2525--2531.

\bibitem[{Najjar and Tafti(1996)}]{najjar1996study}
Najjar, F.~M., Tafti, D.~K., 1996. Study of discrete test filters and finite
  difference approximations for the dynamic subgrid-scale stress model. Phys.
  Fluids 8, 1076--1088.

\bibitem[{Nauw et~al.(2004)Nauw, Dijkstra, and Simonnet}]{nauw2004regimes}
Nauw, J.~J., Dijkstra, H.~A., Simonnet, E., 2004. {Regimes of low-frequency
  variability in a three-layer quasi-geostrophic ocean model}. J. Mar. Res.
  62~(5), 684--719.

\bibitem[{\"{O}zg\"{o}kmen et~al.(2009)\"{O}zg\"{o}kmen, Iliescu, and
  Fischer}]{ozgokmen2009large}
\"{O}zg\"{o}kmen, T., Iliescu, T., Fischer, P.~F., 2009. {Large eddy simulation
  of stratified mixing in a three-dimensional lock-exchange system}. Ocean
  Modell. 26~(3-4), 134--155.

\bibitem[{{\"O}zg{\"o}kmen and Chassignet(1998)}]{özgökmen1998emergence}
{\"O}zg{\"o}kmen, T.~M., Chassignet, E.~P., 1998. {Emergence of inertial gyres
  in a two-layer quasigeostrophic ocean model}. J. Phys. Oceanogr. 28~(3),
  461--484.

\bibitem[{{\"O}zg{\"o}kmen et~al.(2001){\"O}zg{\"o}kmen, Chassignet, and
  Rooth}]{özgökmen2001connection}
{\"O}zg{\"o}kmen, T.~M., Chassignet, E.~P., Rooth, C. G.~H., 2001. {On the
  connection between the \uppercase{M}editerranean outflow and the
  \uppercase{A}zores \uppercase{C}urrent}. J. Phys. Oceanogr. 31~(2), 461--480.

\bibitem[{Pedlosky(1987)}]{pedlosky1987geophysical}
Pedlosky, J., 1987. {Geophysical fluid dynamics}. Springer.

\bibitem[{Press et~al.(1992)Press, Teukolsky, Vetterling, and
  Flannery}]{press1992numerical}
Press, W.~H., Teukolsky, S.~A., Vetterling, W.~T., Flannery, B.~P., 1992.
  Numerical recipes in \uppercase{FORTRAN}: the art of scientific computing.
  Cambridge University Press.

\bibitem[{Rhines(1975)}]{rhines1975waves}
Rhines, P.~B., 1975. {Waves and turbulence on a beta-plane}. J. Fluid. Mech.
  69~(03), 417--443.

\bibitem[{Sagaut(2006)}]{sagaut2006large}
Sagaut, P., 2006. {Large eddy simulation for incompressible flows:
  \uppercase{A}n introduction}. Springer Verlag.

\bibitem[{Salmon(1998)}]{salmon1998lectures}
Salmon, R., 1998. {Lectures on geophysical fluid dynamics}. Oxford University
  Press.

\bibitem[{San and Staples(2012)}]{san2012high}
San, O., Staples, A.~E., 2012. High-order methods for decaying two-dimensional
  homogeneous isotropic turbulence. Comput. Fluids 63, 105--127.

\bibitem[{San and Staples(2013{\natexlab{a}})}]{san2012coarse}
San, O., Staples, A.~E., 2013{\natexlab{a}}. A coarse-grid projection method
  for accelerating incompressible flow computations. J. Comput. Phys. 233,
  480--508.

\bibitem[{San and Staples(2013{\natexlab{b}})}]{san2012stationary}
San, O., Staples, A.~E., 2013{\natexlab{b}}. Stationary two-dimensional
  turbulence statistics using a \uppercase{M}arkovian forcing scheme. Comput.
  Fluids 71, 1--18.

\bibitem[{San et~al.(2011)San, Staples, Wang, and Iliescu}]{san2011}
San, O., Staples, A.~E., Wang, Z., Iliescu, T., 2011. Approximate deconvolution
  large eddy simulation of a barotropic ocean circulation model. Ocean Modell.
  40, 120--132.

\bibitem[{Siegel et~al.(2001)Siegel, Weiss, Toomre, McWilliams, Berloff, and
  Yavneh}]{siegel2001eddies}
Siegel, A., Weiss, J.~B., Toomre, J., McWilliams, J.~C., Berloff, P.~S.,
  Yavneh, I., 2001. {Eddies and vortices in ocean basin dynamics}. Geophys.
  Res. Lett. 28~(16), 3183--3186.

\bibitem[{Smith et~al.(2002)Smith, Boccaletti, Henning, Marinov, Tam, Held, and
  Vallis}]{smith2002turbulent}
Smith, K.~S., Boccaletti, G., Henning, C.~C., Marinov, I., Tam, C.~Y., Held,
  I.~M., Vallis, G.~K., 2002. {Turbulent diffusion in the geostrophic inverse
  cascade}. J. Fluid Mech. 469, 13--48.

\bibitem[{Speich et~al.(1995)Speich, Dijkstra, and Ghil}]{speich1995successive}
Speich, S., Dijkstra, H., Ghil, M., 1995. {Successive bifurcations in a
  shallow-water model applied to the wind-driven ocean circulation}. Nonlinear
  Proc. Geoph. 2, 241--268.

\bibitem[{Stanculescu(2008)}]{stanculescu2008existence}
Stanculescu, I., 2008. Existence theory of abstract approximate deconvolution
  models of turbulence. Ann. Univ. Ferrara 54~(1), 145--168.

\bibitem[{Stolz and Adams(1999)}]{stolz1999approximate}
Stolz, S., Adams, N.~A., 1999. {An approximate deconvolution procedure for
  large-eddy simulation}. Phys. Fluids 11, 1699--1701.

\bibitem[{Stolz et~al.(2001{\natexlab{a}})Stolz, Adams, and
  Kleiser}]{stolz2001approximate}
Stolz, S., Adams, N.~A., Kleiser, L., 2001{\natexlab{a}}. {An approximate
  deconvolution model for large-eddy simulation with application to
  incompressible wall-bounded flows}. Phys. Fluids 13, 997--1015.

\bibitem[{Stolz et~al.(2001{\natexlab{b}})Stolz, Adams, and
  Kleiser}]{stolz2001approximatec}
Stolz, S., Adams, N.~A., Kleiser, L., 2001{\natexlab{b}}. {The approximate
  deconvolution model for large-eddy simulations of compressible flows and its
  application to shock-turbulent-boundary-layer interaction}. Phys. Fluids 13,
  2985--3001.

\bibitem[{Stolz et~al.(2004)Stolz, Adams, and Kleiser}]{stolz2004approximatei}
Stolz, S., Adams, N.~A., Kleiser, L., 2004. {The approximate deconvolution
  model for compressible flows: \uppercase{I}sotropic turbulence and
  shock-boundary-layer interaction}. in Advances in LES of Complex Flows, Fluid
  Mech. Appl. 65, edited by R. Friedrich and W. Rodi, 33--47.

\bibitem[{Stommel(1972)}]{stommel1972gulf}
Stommel, H., 1972. {The \uppercase{G}ulf \uppercase{S}tream: \uppercase{A}
  physical and dynamical description}. University of California Press.

\bibitem[{Sukoriansky et~al.(2007)Sukoriansky, Dikovskaya, and
  Galperin}]{sukoriansky2007arrest}
Sukoriansky, S., Dikovskaya, N., Galperin, B., 2007. {On the arrest of inverse
  energy cascade and the \uppercase{R}hines scale}. J. Atmos. Sci. 64~(9),
  3312--3327.

\bibitem[{Sura et~al.(2001)Sura, Fraedrich, and Lunkeit}]{sura2001regime}
Sura, P., Fraedrich, K., Lunkeit, F., 2001. {Regime transitions in a
  stochastically forced double-gyre model}. J. Phys. Oceanogr. 31~(2),
  411--426.

\bibitem[{Tanaka and Akitomo(2010)}]{tanaka2010alternating}
Tanaka, Y., Akitomo, K., 2010. {Alternating zonal flows in a two-layer
  wind-driven ocean}. J. Oceanogr. 66~(4), 475--487.

\bibitem[{Vallis(2006)}]{vallis2006atmospheric}
Vallis, G.~K., 2006. {Atmospheric and oceanic fluid dynamics:
  \uppercase{F}undamentals and large-scale circulation}. Cambridge University
  Press.

\bibitem[{Visbeck et~al.(1997)Visbeck, Marshall, Haine, and
  Spall}]{visbeck1997specification}
Visbeck, M., Marshall, J., Haine, T., Spall, M., 1997. Specification of eddy
  transfer coefficients in coarse-resolution ocean circulation models. J. Phys.
  Oceanogr. 27~(3), 381--402.

\bibitem[{Zhou and Chow(2011)}]{zhou2011large}
Zhou, B., Chow, F.~K., 2011. Large-eddy simulation of the stable boundary layer
  with explicit filtering and reconstruction turbulence modeling. J. Atmos.
  Sci. 68, 2142--–2155.

\end{thebibliography}

\end{document}